\documentclass[12pt,superscriptaddress]{iopart}

\usepackage{fancyhdr} 
\fancypagestyle{plain}{%
\fancyhead{} 
}

\usepackage{iopams}
\usepackage{bm}
\usepackage{graphicx}
\usepackage{showlabels}
\usepackage{amssymb}
\usepackage{hyperref}

\def\la{\label}
\def\be{\begin{equation}}
\def\beq{\begin{equation}}
\def\eeq{\end{equation}}
\def\ee{\end{equation}}
\def\bea{\begin{eqnarray}}
\def\eea{\end{eqnarray}}
\def\p{\partial}

\newcommand{\ii}{{{i}}}
\newcommand{\dd}{{{d}}}
\newcommand{\sd}{{{d}}}

\begin{document}

\title{Random Matrices in 2D, Laplacian Growth and Operator Theory}

\author{Mark Mineev$^1$, Mihai Putinar$^2$ and Razvan Teodorescu$^3$}
\address{$^1$M.S. P365, Los Alamos National Laboratory, Los Alamos, NM 87505}
\address{$^2$Mathematics Department, UCSB, Santa Barbara, CA 93106}
\address{$^3$Theoretical Division and the Center for Nonlinear Studies, Los Alamos, NM 87505}
\eads{\mailto{mariner@lanl.gov}, \mailto{mputinar@math.ucsb.edu},
\mailto{razvan@lanl.gov}}

\begin{abstract}
Since it was first applied to the study of nuclear interactions by Wigner and Dyson,
almost 60 years ago, Random Matrix Theory (RMT) has developed into a field of its own whithin applied mathematics, and is now essential to many parts of theoretical
physics, from condensed matter to high energy.
The fundamental results obtained so far rely mostly on the theory of random
matrices in one dimension (the dimensionality of the spectrum, or equilibrium
probability density). In the last few years, this theory has been extended
to the case where the spectrum is two-dimensional, or even fractal, with dimensions
between 1 and 2. In this article, we review these recent developments and indicate
some  physical problems where the theory can be applied. 
\end{abstract}

\submitto{\JPA}

\maketitle


\tableofcontents

\newpage

\section{Introduction} \label{sec:intro}

During the second half of last century and continuing through the present, random matrix theory has grown from a special method of theoretical physics, meant to approximate energy levels of complex nuclei \cite{Wigner1, Wigner2, Wigner3, Dyson1, Dyson2, Dyson3, Dyson4}, into a vast mathematical theory with many different application in physics, computer and electrical engineering. Simply describing all the developments and methods currently employed in this context would result in a monography much more extensive than this review. Therefore, we will only briefly mention topics which are themselves very interesting, but lie beyond the scope of this work. 

The applications of random matrix theory (RMT) into physics have been extended from the original subject, spectra of heavy nuclei, to descriptions of large $N$ $SU(N)$ gauge theory \cite{tHoft, IZPB}, 
critical statistical models in two dimensions \cite{Kazakov, Kostov, Serban}
disordered electronic systems \cite{Pastur, Wegner, Efetov, Zirnbauer, Fyodorov, Altshuler}, quantum chromo-dynamics (QCD) \cite{Verbaarschot, Akemann}, to name only a few. Non-physics applications range from communication theory \cite{MIMO} to stochastic processes out of equilibrium \cite{Schutz1, Schutz2} and even more exotic topics \cite{Cuernavaca}. 

A number of important results, both at theoretical and applied levels, were obtained from the connection between random matrices and orthogonal polynomials, especially in their weighted limit \cite{Szego,Saff-Totik,IZPB,BEH1,BEH2}. These works explored the relationship between the branch cuts of spectral (Riemann) curves of systems of differential equations and the support of limit measures for weighted orthogonal polynomials. Yet another interesting connection stemming from this approach is with the general (matrix) version of the Riemann-Hilbert problem with finite support \cite{Its, Deift}. 

In \cite{WZ03, Teodorescu04}, it was showed that such relationships also hold for the class of normal random matrices. Unlike in previous works, for this ensemble, the support of the equilibrium distribution for  the eigenvalues of matrices in the infinite-size limit, is two-dimensional, which allows to interpret it as a growing cluster in the plane. Thus, a direct relation to the class of models known as Laplacian Growth (both in the deterministic and stochastic formulations), was derived, with important consequences. In particular, this approach allowed to study formation of singularities in models of two-dimensional growth. 
Moreover, these results allowed to define a proper way of continuing the solution for singular Laplacian Growth, beyond the critical point. 

From the point of view of the dimensionality of the support for random matrix eigenvalues, it is possible to distinguish between 1-dimensional situations (which characterize 1 and 2-matrix models), and 2-dimensional situations, like in the case of normal random matrix theory. In fact, very recent results point to intermediate cases, where the support is a set of dimensional between 1 and 2. This situation is very similar to the description of disordered, interacting electrons in the plane, in the vicinity of the critical point which separates localized from de-localized behavior \cite{Efetov}. It is from the perspective of the dimensionality of support for equilibrium measure that we have organized this review. 

The paper is structured in the following way: after a brief summary of the main concepts in Section~\ref{first}, we explain the structure of normal random matrices in the limit of infinite size, in Section~\ref{second}. This allows to connect with planar growth models, of which Laplacian (or harmonic) Growth is a main representative. The following  two sections give a solid description of the physical (Section~\ref{third}) and mathematical (Section~\ref{fourth}) structure of harmonic growth. The discretized (or quantized) version of this problem is precisely given by normal random matrices, as we indicate in these sections. Next we present a general scheme for encoding shade functions
in the plane into linear data, specifically into a linear bounded
Hilbert space operator $T$ with rank one self-commutator
${\rm rank} [T^\ast,T] = 1$. This line of research goes back to
the perturbation and scattering theory of symmetric operators
(M. G. Krein's phase shift function) and to studies related to
singular integral operators with a Cauchy kernel type singularity.
Multivariate refinements of the "quantization scheme" we outline in
Chapter 5 lie at the foundations of both cyclic (co)homology
of operator algebras and of free probability theory. In view of the scope
and length of the present survey, we confine ourselves to only outline
the surprising link between quadrature domains and such Hilbert space
objects.

We conclude with an application of the operator formalism to the description of boundary singular points that are characteristic to Laplacian growth evolution, and a brief overview of other related topics.

\section{Random Matrix Theory in 1D} \la{first}
\subsection{The symmetry group ensembles and their physical realisations}

Following \cite{Mehta}, we reproduce the standard introduction of the symmetry-groups
ensemble of random matrices.  The traditional ensembles (orthogonal, unitary and symplectic) 
were introduced mainly because of their significance with respect to symmetries of hamiltonian 
operators in physical theories: time-reversal and rotational invariance corresponds to the orthogonal 
ensemble (which, for Gaussian measures, is naturally abbreviated GOE), while time-reversal alone and rotational invariance alone correspond to the symplectic and unitary ensembles, respectively (GSE and GUE for Gaussian measures). 

An invariant measure is defined for each of these ensembles, in the form 
\be
d \widetilde{\mu}(M) \equiv P(M) d \mu(M) \equiv 
Z^{-1} e^{-{\rm{Tr}}[W(M)]} d \mu(M), 
\ee 
where $M$ is a matrix from the ensemble, $Z$ is a normalization factor (partition function), Tr$[W(M)]$ is invariant under the symmetries on the ensemble, and $d \mu(M)$ is the appropriate flat measure for that ensemble: $\prod_{i \le j} d M_{ij}$ for orthogonal, $\prod_{i \le j}d {\mbox{\rm Re }} M_{ij} \prod_{i < j} d {\mbox{\rm Im }}M_{ij}$ for unitary, and $\prod_{i \le j}d M^{(0)}_{ij} 
\prod_{k=1}^3 \prod_{i < j} dM^{(k)}_{ij}$ for symplectic (where each matrix element is an element of the real Klein group, $M_{ij} = M^{(0)}_{ij} \cdot 1 + \sum_{k=1}^3 M^{(k)}_{ij} \cdot \mathbb{\sigma}_k $). Correspondingly, to each of these ensembles, a parameter $\beta$, indicating the number of independent real parameters necessary to describe the pair of values $M_{ij}, M_{ji}$, is introduced, with values $\beta = 1, 2, 4$ for orthogonal, unitary and symplectic ensembles, respectively.  

The invariance under transformations from the appropriate symmetry group leads to the following simplification of the measure: for any of these ensembles, the generic matrix $M$ can be diagonalized by a transformation $M = U^{-1} \Lambda U$, with $U$ from the same group, and $\Lambda = {\rm{diag}}(\lambda_1, \ldots, \lambda_N)$. The Jacobian of the transformation $M \to \Lambda, U$ (where $U$ is said to carry the ``angular" degrees of freedom of $M$) is $J = \prod_{i < j}|\lambda_i - \lambda_j|^{\beta}= |\Delta(\Lambda)|^{\beta}$, with $\Delta$ the Vandermonde determinant. The angular degrees of freedom can be integrated out (a trivial redefinition of the normalization factor), giving the simplified measure
\be
\rho(\lambda_1, \ldots, \lambda_N)\prod_{i=1}^N d\lambda_i = Z^{-1} e^{{\rm{Tr}}[W(\Lambda)]} 
|\Delta(\Lambda)|^{\beta} \prod_{i=1}^N d\lambda_i 
\ee 
For example, in the case of Gaussian measure $W(M) = -M^2$, the joint probability distribution function of eigenvalues, $\rho$, becames (up to normalization)
\be
\rho(\lambda_1, \ldots, \lambda_N)  = \exp \left [
-\sum_{i=1}^N \lambda_i^2 + \beta \sum_{i < j} \log |\lambda_i - \lambda_j|
\right ]
\ee
Clearly, this procedure is useful only if we are interested in computing expectation values of quantities which depend only of the distribution of eigenvalues, and not of the angular degrees of freedom. This is indeed the case for all situations of interest. 

The next standard transformation (which we discuss for the case of unitary ensemble, $\beta=2$) that is performed on the measure uses the well-known property of Vandermonde determinant $\Delta(\Lambda) = \det [\lambda_i^{j-1}]_{1 \le i, j \le N}$. Because of standard determinantal identities, this is equivalent with replacing each monomial $\lambda_i^{j-1}$ by a {\emph{monic}} polynomial of the same order, $P_{j-1}(\lambda_i) = \lambda_{i}^{j-1} + \ldots$. Finally, these polynomials may be chosen to be orthogonal with respect to the measure $e^{W(\lambda)}$, giving for the p.d.f. of eigenvalues the expression
\be
\rho(\lambda_1, \ldots, \lambda_N) = |\det [P_{j-1}(\lambda_i)e^{W(\lambda_i)/2}]|^2, 
\ee
which is simply the absolute value-squared of the wavefunction of the ground state for $N$ electrons in the external potential $W$.  As we shall see, this kind o physical interpretation may be generalized to the case of matrix ensembles with two-dimensional support of eigenvalues. 

\paragraph*{Generalizations of group ensembles} Recently, various generalizations were proposed 
in order to extend the theory for ensembles of matrices which are not associated with symmetry groups. 
In particular, ensembles of matrices which may be reduced to a tridiagonal form (instead of standard diagonal) by a transformation which eliminates ``angular" degrees of freedom, were introduced in \cite{Dumitriu}. As an interesting consequence, many results carry over to this case, while the parameter $\beta$ is allowed to take any positive real value. 

\subsection{Critical ensembles}

In this section we explain how, using properly chosen non-Gaussian measures, it is possible to construct ensembles of hermitian matrices (corresponding again to the unitary symmetry) which are in a sense, critical, i.e. for which a continuum limit ($N \to\infty$) may be defined. The discussion relies on the formulation based on orthogonal polynomials indicated above, and it follows (at a more elementary level) the general theory of Saff and Totik \cite{Saff-Totik}. 

\subsubsection{General formalism}

Let $d \mu(x) = e^{W(x)}d x  $ be a well-defined measure on the 
real axis, $W(x) \to - \infty$ as $|x| \to \infty$, and $P^{(1)}_n(x)$ the 
corresponding  family of orthogonal polynomials 
\be
\int_{-\infty}^\infty P^{(1)}_n(x) P^{(1)}_m(x) d \mu (x) = \delta_{nm}.
\ee
Orthonormal functions are obtained through $\psi_n(x) = P_n(x) e^{W(x)/2}$, 
which are orthogonal with respect to the flat measure on $\mathbb{R}$. We consider
 a deformation of this ensemble through a positive real 
parameter $ \lambda \ge 1$,  so that $d \mu_{\lambda}(x) = e^{\lambda W(x)} d x$  and
\be
\int_{-\infty}^\infty P^{(\lambda)}_n(x) P^{(\lambda)}_m(x) d \mu_{\lambda} (x) = \delta_{nm}.
\ee 
Clearly, if $W(x) $ is a monomial of degree $k$, the deformation amounts to 
a simple rescaling 
\be
P^{\lambda}_n(x) = \lambda^{1/2k} P^{(1)}_n(\lambda^{1/k}x) . 
\ee 
The first non-trivial example is a quartic polynomial of the type
\be
W(x) = -(x^2 + g x^4), \quad g> 0,
\ee
for which the deformation in not a simple rescaling. In this case, it is possible 
to consider a special limit $n \to \infty, \lambda \to \infty, \lambda \to n r_c$, where 
$r_c$ is a constant. As we will see, for a specific value of $r_c$, this limit yields 
a special asymptotic behavior of the orthonormal functions $\psi_n(x)$.  However, 
even for the simplest, trivial monomial (a Gaussian), which yields the Hermite polynomials, 
the asymptotic behavior of the orthogonal functions is non-trivial, in the sense that 
there are no known good approximations for the case $r_c = O(1)$.  

Generically, in this large $n, \lambda$ limit, we can ask where the wavefunction 
$\psi_n(x)$ will reach its maximum value, in the saddle point approximation:
\be
{\rm{max }}_{|x|}\partial_x |\psi_n(x)| = 0, 
\ee 
giving 
\be
\partial_x \left [ \sum_{i=1}^n \log (x-\xi_i) - n r_c  + W(x)\right ] = 0,
\ee
so that 
\be \label{saddle_point}
-r_c W'(x) = \frac{2}{n}\sum_{i=1}^n \frac{1}{x-\xi_i},  
\ee
where $\xi_i, i = 1, \ldots, n$ are the roots of the $n^{\rm th}$ polynomial.  

Let 
\be \label{cauchy}
\omega(z) = \frac{1}{n}\sum_{i=1}^n \frac{1}{\xi_i - z},
\ee
multiply (\ref{saddle_point}) by $(\xi_i - z)^{-1	}$ and sum over $i$, and obtain
\be \label{polynomialst}
\omega^2(z) - r_c W'(z) \omega(z) = 
- \frac{r_c}{n} \sum_{i=1}^n \frac{W'(z) -W'(\xi_i)}{z-\xi_i}.
\ee 
Equation (\ref{polynomialst}) can be solved in the large $n$ limit by  assuming 
that the roots will be distributed with density $\rho(\xi)$ on some
compact (possibly disconnected)  set $I \subset \mathbb{R}$. Defining
\be \label{definition}
R(z) = -\frac{4}{r_c} \int_I  \frac{W'(z) -W'(\xi)}{z-\xi} \rho(\xi) d \xi,
\ee
we obtain
\be \label{large}
\omega^2(z) - r_c W'(z) \omega(z) + \left ( \frac{r_c}{2}\right )^2 R(z) = 0.
\ee
The proper solution of (\ref{polynomialst}) (considering the behavior at $\infty$ of the function 
$\omega(z)$), is 
\be \label{solution}
\omega(z) = \frac{r_c}{2}\left [ W'(z) + \sqrt{(W'(z))^2 - R(z)}\right ],
\ee
and (since the function $\omega(z)$ is the Cauchy transform of the density $\rho(x)$), it
gives us the asymptotic distribution of zeros as
\be \la{ro}
\rho(x) = \frac{1}{2 \pi i} [\omega(x+ i 0) - \omega(x-i0)].
\ee
Finally, to obtain the asymptotic form of wave functions $\psi_n(x)$, we can write
\be \label{asymptotic}
n^{-1} \log \psi_n(x) \to 
\int \rho(\xi) \log (x- \xi)d \xi - \frac{r_c}{2} +W(x). 
\ee

\subsubsection{Continuum limit and integrable equations}

There are two related problems for the large $n$ limit of deformed ensembles described
in the previous section. The first is determination of the support of zeros $I$; the second is
the scaling behavior of the orthogonal functions $\psi_n(x)$. In general, the limiting support
$I$  may consist of several disconnected segments $I_k$, $I = \cup_{k=1}^{k=d} I_k$. In the simplest
case, it is just one interval $I = [a, b]\subset \mathbb{R}$. In this section we indicate how to determine 
this support as well as the density $\rho(x)$, and what this yields for the orthogonal functions. 

Let the function $W(x)$  be a polynomial of even degree $d$. From (\ref{definition}) we see that
$R(z)$ is a polynomial of degree $d-2$, and therefore solution (\ref{solution}) has generically 
$2(d-1)$ branch points. Thus, the function $\omega(z)$ typically has $d-1$ branch cuts, which 
constitute the disconnected support of distribution $\rho(z)$.  

We are interested in a special case, when $d-2$ of these cuts degenerate into double points, 
and there is a single interval $[a, b]$ which is the support of $\rho(z)$. This special case is called $critical$ and it provides new asymptotic limits for the orthogonal functions. We will also refer to
this solution as the ``single-cut" solution.

From the equation 
\be
\omega(x+ i0) - \omega(x-i0) = r_c W'(x),
\ee
we obtain for the single-cut solution
$$ 
\omega(z) = -\frac{r_c \sqrt{(z-a)(z-b)}}{2 \pi} \int_a^b \frac{W'(\xi)}{\sqrt{(b-\xi)(\xi-a)}} \frac{d \xi}{\xi -z}.
$$
The large $|z|$ behavior of this function is known from the continuum limit of (\ref{cauchy}), and it
implies the absence of regular terms in the Laurent expansion: 
\be
\omega(z) = -\frac{1}{z} + O(z^{-2}),
\ee
so that we impose the conditions
\be \label{cond1}
0 = \int_a^b \frac{W'(\xi)}{\sqrt{(b-\xi)(\xi-a)}} d \xi,
\ee
\be \label{cond2}
2 \pi = - r_c \int_a^b \frac{\xi W'(\xi)}{\sqrt{(b-\xi)(\xi-a)}} d \xi.
\ee

\paragraph*{Gaussian measure and the Hermite polynomials} Let $d=2$ and $-W(x) = ax^2, a > 0$. Then 
conditions (\ref{cond1},\ref{cond2}) give a symmetric support $[-b, b]$ where $b^2 = 2/(a r_c)$. 

More generally, using the saddle point equation for $|\psi_n(x)|$ at $x = a, b$ and (\ref{asymptotic}), we conclude that 
\be 
\frac{\log \psi_n (b + \zeta)}{n} = C_b - \frac{r_c}{2}\int_0^\zeta \sqrt{[W'(b+\eta)]^2 - R} d \eta
\ee
Since the integrand behaves like $\eta^{d-3/2}$, we obtain
\be
\psi_n(b+\zeta) = \psi_n(b) \exp \left [-\frac{nr_c}{2d-1} \zeta^{d-1/2} \right ].
\ee
We immediately conclude that for $d=2$ (Hermite polynomials), the asymptotic behavior 
is given  by the Airy function, $\exp z^{3/2}$.  The full scaling is achieved by considering 
the region around the end-point $b$, of order $\zeta = O(n^{-2/(2d-1)})$.Then we obtain 
\be
\psi_n(b+\tilde{\zeta}n^{-\frac{2}{2d-1}}) \sim \exp \left [ - \frac{r_c}{2d-1} \tilde{\zeta}^{d-1/2}\right ].
\ee

\subsubsection{Scaled limits of orthogonal polynomials and equilibrium measures}

The distribution of eigenvalues investigated in the previous sections illustrates the general 
approach developed by Saff and Totik \cite{Saff-Totik} for holomorphic polynomials 
orthogonal on curves in the complex plane. We sketch here the more general result because 
of its relevance to the main topic of this review. 

Given a set $\Sigma \in \mathbb{C}$ and a properly-defined measure on it $w(z)=e^{-Q(z)}$, 
we construct the holomorphic orthogonal polynomials $P_n(z)$, with respect to $w$. We then 
pose the question of finding the ``extremal" measure  (its support $S_w$ and density $\mu_w$),
such that the $F$-functional $F(K) \equiv \log {\rm{cap}} (K) - \int Q d \omega_K$, with ${\rm{cap}}(K)$ and 
$\omega_K$ the capacity, respectively the equilibrium measure of the set $K$, is maximized by
$S_w$. Furthermore, $\mu_w$ satisfies energy and capacity constraints on $S_w$.  

The remarkable fact noticed in \cite{Saff-Totik} is that if the extremal value $F(S_w)$ is approximated 
by the weighted $monic$ polynomials $\tilde{P}_n(z)$ as $(||w^n \tilde{P}_n||_\Sigma^*)^{1/n} \to \exp (-F_w)$ (where we use the weak star norm), then the asymptotic zero distribution of $\tilde{P}_n$ gives the support $S_w$. Hence, (\ref{ro}) may be interpreted as giving both the support of the extremal 
measure (labeled $\rho$ in this formula), as well as its actual density. 

The extremal measure has the physical interpretation of the ``smallest" equilibrium measure which 
gives a prescribed logarithmic potential at infinity. According to the concept of ``sweeping" (or ``balayage", see \cite{Saff-Totik}), the extremal measure is obtained as a limit of the process, under the constraints  
imposed on the total mass and energy of the measure. As we have shown in this chapter, for the case of 1D measures, this extreme case is given by weighted limits of orthogonal polynomials.

\section{Random Matrix Theory in higher dimensions} \la{second}

In this chapter, we show how to generalize the concepts of equilibrium measure, extremal measure, 
and their relations to orthogonal polynomials and ensembles of random matrices, in the case of 
two-dimensional support. The applications of this theory to planar growth processes will be discussed in the following two  chapters.

\subsection{The Ginibre-Girko ensemble}

We begin with a brief discussion on the oldest and simplest ensemble of random matrices with planar support. The ensemble of complex, $N \times N$ random matrices with identical, independent, 
zero-mean Gaussian-distributed entries, was first studied by J. Ginibre in 1965 \cite{Ginibre}, and 
then it was generalized for non-zero mean Gaussian by Girko in 1985 \cite{Girko}. Consider $N \times N$ random matrices with eigenvalues 
$z_k \in \mathbb{C}$, and joint p.d.f. 
\be
dP_N \sim \prod_{1 \le i < j \le N} |z_i - z_j|^2 \prod_{1 \le k \le N} \mu_N(z_k),
\ee 
where $\mu_N(z_k) = e^{-N|z_k|^2} d {\rm {Re}} z_k d {\rm {Im}} z_k$. Then,  in the large $N$ limit, 
the measure 
$
\frac{1}{N}\sum_k \delta(z-z_k) 
$
converges weakly to the uniform measure on the unit disk. This is known as the Circular Law. If the exponent of the pure Gaussian is perturbed by a quadratic term, the result holds for a corresponding 
elliptical domain, giving the Elliptical Law. The same limiting curves (circular and elliptical) describe the 
graph of the distribution of $real$ eigenvalues for Hermitian ensembles, with pure and perturbed 
Gaussian measures. In that case, the laws are known as Wigner-Dyson \cite{Wigner1, Dyson1}
 and Marchenko-Pastur, respectively (although the last one was originally derived for covariance
 matrices built from sparse regression matrices \cite{MarPas}). 
  
Extensions and exceptions from the circular and elliptical laws were found by relaxing the
conditions of the theorems. In particular, deviations from uniformity for angular statistics 
in the case of Gaussian measure were derived in \cite{Rider}, while the case of heavy-tails
distributions was investigated in \cite{Sosh, Zakh} and subsequent publications.

\subsection{Normal matrix ensembles}

A special case of matrices with complex eigenvalues is given by $normal$ 
matrices. 
A matrix $M$ is called normal if it commutes with its
Hermitian conjugate: $[M, M^{\dag}]=0$, so that both $M$ and $M^{\dag}$
can be diagonalized simultaneously.  The statistical weight of the normal matrix ensemble is given through a general potential $W(M,M^{\dag})$ \cite{Zaboronsky}:
\be
\label{ZN}
e^{\frac{1}{\hbar}{\rm tr} \, W(M, M^{\dag})} \sd\mu (M).
\ee
Here $\hbar$ is a parameter, and the measure of integration
over normal matrices
is induced by the flat metric
on the space
of all complex matrices ${\sd}_C M$, where
${\sd}_C M = \prod_{ij}\sd\, {\rm Re} \, M_{ij} \sd\, {\rm Im} \, M_{ij}$.
Using a standard procedure,
one passes to the joint
probability distribution
of eigenvalues of normal matrices $z_1,\dots,z_N$, where $N$ is the 
size of the matrix:
\be
\label{mean}
\frac{1}{N! \tau_N}|\Delta_N (z)|^2 \,\prod_{j=1}^N
e^{\frac{1}{\hbar}W(z_j,\bar z_j)} {\sd}^2 z_j
\ee
Here
$\sd^2 z_j \equiv \sd x_j \, \sd y_j$ for $z_j =x_j +iy_j$,
$\Delta_N(z)=\det (z_{j}^{i-1})_{1\leq i,j\leq N}=
\prod_{i>j}^{N}(z_i -z_j)$
is the Vandermonde determinant, and
\be
\label{tau}
\tau_N = \frac{1}{N!}\int
|\Delta_N (z)|^2 \,\prod_{j=1}^{N}e^{\frac{1}{\hbar}
W(z_j,\bar z_j)} d^2z_j
\ee
is a normalization factor, the partition function
of the matrix model (a $\tau$-function).

A particularly important  special  case arises
if the potential $W$ has the form
\be
\label{potential}
W=-|z|^2+V(z)+\overline{V(z)},
\ee
where $V(z)$ is a holomorphic function in a domain which
includes the support of eigenvalues (see also a comment in the end of
Section~\ref{F1}
about a proper definition of the ensemble with this potential).
In this case, a normal matrix ensemble gives the same distribution as
       a general complex matrix ensemble.
A general complex matrix can be decomposed as $M=U(Z+R)U^\dagger$,
where $U$ and  $Z$
are  unitary and diagonal matrices, respectively, and $R$
is an upper triangular matrix. The distribution (\ref{mean})
holds for the elements of the diagonal matrix $Z$
which are eigenvalues of $M$.
Here we mostly focus on
the special potential (\ref{potential}),  and also assume
that the field
\be\la{a}
A(z)=\p_z V(z)
\ee
is a globally defined meromorphic function.

\subsection{Droplets of eigenvalues}
\begin{figure} \begin{center}
\includegraphics*[width=5cm]{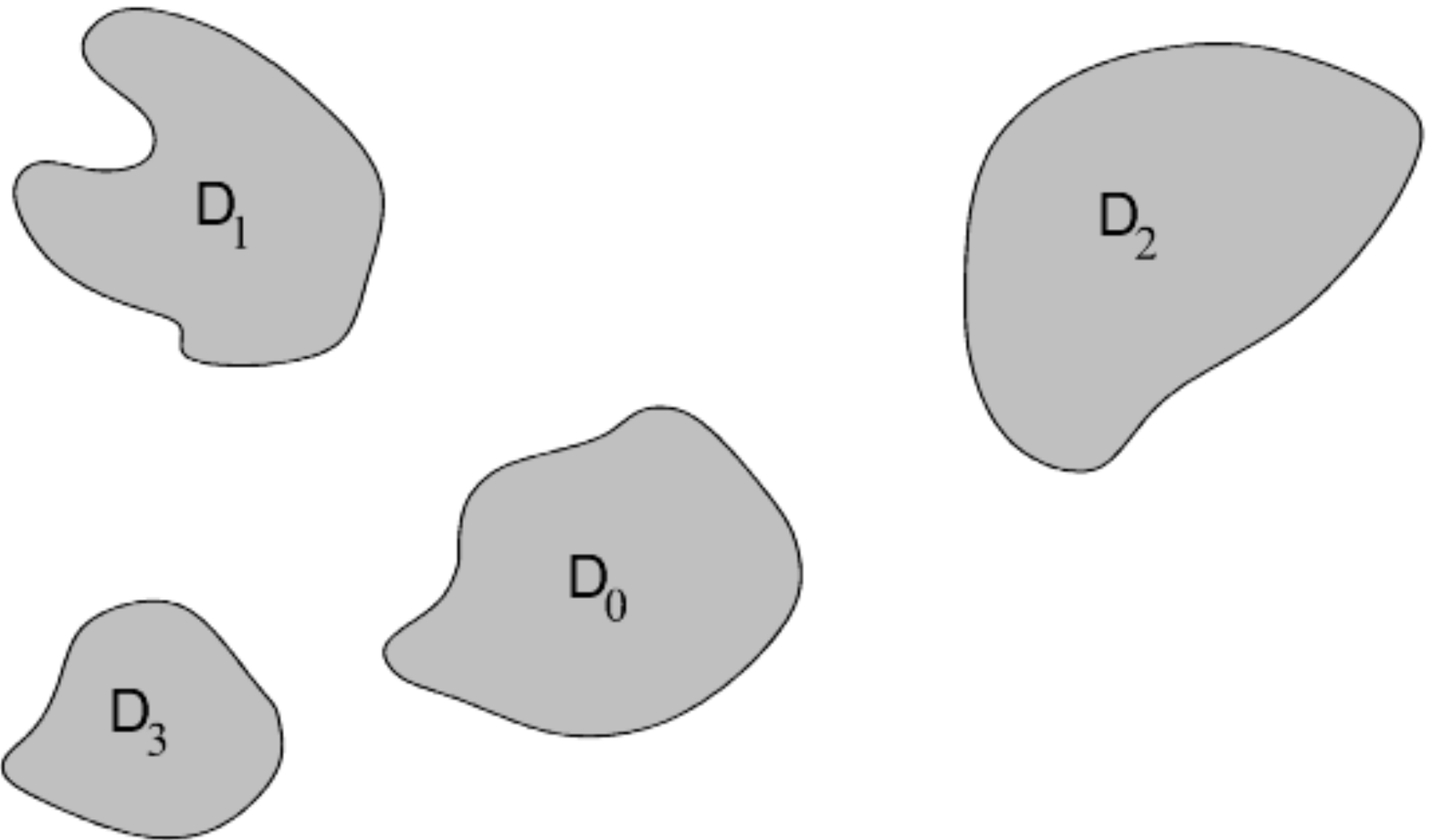}
\includegraphics*[width=5cm]{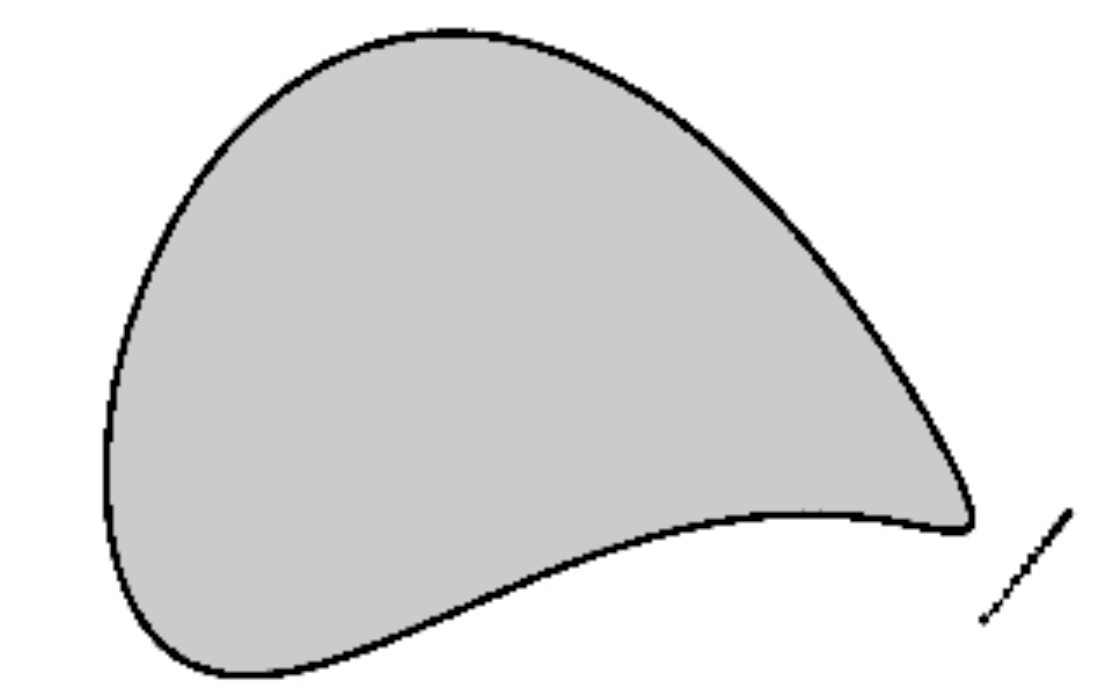}
\caption{A support of eigenvalues consisting of four disconnected components (left).
The distribution of eigenvalues for potential $V(z) = - \alpha \log (1- z/\beta) - \gamma z $. 
(right)
}
\label{droplets}
\end{center} \end{figure}

     In the large $N$ limit ($\hbar\to 0$,
$N\hbar$ fixed), the eigenvalues of matrices from the
ensemble densely occupy  a connected domain $D$
       in the complex plane, or, in general, several
disconnected domains.
This set (called the support of eigenvalues)
has sharp edges
(Figure~\ref{droplets}). We refer to the connected components
$D_\alpha$ of the domain $D$ as {\it droplets}.

For algebraic domains (the definition follows)
the eigenvalues are distributed with the density
$\rho=-\frac{1}{4 \pi}\Delta W$,
where $\Delta =4\p_z \p_{\bar z}$ is the 2-D Laplace operator
\cite{WZ03}.
For the potential (\ref{potential}) the density is uniform.
The shape of the support
of eigenvalues is the main subject of this chapter. 
For example, if the
potential is Gaussian \cite{Ginibre},
\be\la{e}
A(z)=2t_2 z,
\ee
the domain is an ellipse.  If $A$ has one simple pole,
\be
\la{J1}
A(z)=-\frac{\alpha}{z-\beta}-\gamma
\ee
the droplet (under certain conditions discussed below) has the profile
of an aircraft wing given by the Joukowsky map
(Figure~\ref{droplets}). If $A$ has one double pole
(say, at infinity),
\be\la{hyp}
A(z)=3t_3 z^2,
\ee
the droplet is a hypotrochoid. If $A$ has two or more simple poles, there may be
more than one droplet. This support and density represent the equilibrium solution 
to an electrostatic problem, as we will indicate in a later section. 

\subsection{Orthogonal polynomials and distribution of eigenvalues}\la{F1}

Define the exact $N$-particle wave function (up to a phase), by
\be
\label{psi}
\Psi_N(z_1,\dots,z_N)=
\frac{1}{\sqrt{N! \tau_N}}\Delta_N (z) \,e^{\sum_{j=1}^{N}
\frac{1}{2\hbar}W(z_j, \bar z_j)}.
\ee
The joint probability distribution (\ref{mean})
is then equal to $|\Psi(z_1,\dots,z_N)|^2$.

Let the number of eigenvalues (particles)
increase while the potential  stays fixed. If the
support of eigenvalues is
simply-connected, its  area
grows as $\hbar N$.  One can describe the
evolution of the  domain through the density of particles
\begin{equation}\la{51}
\rho_N(z)=N \int|\Psi_{N}(z,z_1,z_2,\dots,z_{N-1})|^2
d^2z_1\dots d^2 z_{N-1},
\ee
where $\Psi_N$ is given by (\ref{psi}).

We introduce a set of orthonormal one-particle functions
on the complex plane as matrix elements of transitions
between $N$ and $(N+1)$-particle states:
\begin{equation}\la{5}
\frac{\psi_N(z)}{\sqrt{N+1}}= \int\Psi_{N+1} (z,z_1,z_2,\dots,z_N)
\overline{\Psi_{N}(z_1,z_2,\dots,z_N)} d^2z_1\dots d^2 z_N
\end{equation}
Then the rate of the density change is
\begin{equation}\la{1}
\rho_{N+1}(z)-\rho_{N}(z)=|\psi_N(z)|^2.
\end{equation}
The proof of this formula
is based on the representation of the $\psi_n$ through
holomorphic biorthogonal polynomials $P_n(z)$. Up to a phase
\begin{equation}\label{O3}
\psi_n (z)=
e^{\frac{1}{2\hbar}W(z, \bar z )}P_n (z),\quad
\quad P_n (z)=\sqrt{\frac{\tau_n}{\tau_{n+1}}} z^n +\ldots
\end{equation}
The polynomials $P_n(z)$ are
biorthogonal on the complex plane
with the weight $e^{W/\hbar}$:
\begin{equation}\la{29}
\int e^{W/\hbar}P_n(z)\overline{P_m(z)}d^2 z=
\delta_{mn}.
\end{equation}
The proof of these formulae
is standard in the theory of orthogonal
polynomials. Extension to the biorthogonal case adds no
difficulties.

We note that, with
   the choice of potential (\ref{potential}), the integral
representation (\ref{29}) has only a formal meaning,
since the integral diverges
unless the potential is Gaussian.
A proper definition of the  wave functions goes through
recursive relations (\ref{L1}, \ref{M}) which follow from the integral
representation.
The same comment applies to the $\tau$-function (\ref{tau}).
The wave function is not  normalized everywhere
in the complex plane. It may diverge at the poles of the
vector potential field.

\subsection{Wavefunctions, recursions and integrable hierarchies}
 
 In order to illustrate the mathematical connection between this theory and 
 equivalent formulations which we present in Chapter 5, it is necessary to 
 make a digression through the formalism of infinite, integrable hierarchies. 
 In particular, we choose the case of the Kadomtsev-Petviashvilii (KP) 
 hierarchy, and follow the notations in \cite{Dikey}. 
 
\subsubsection{Pseudo-differential operators} We denote by $\mathcal{A}$ the algebra constructed from differential polynomials of the type $P = \p^n + u_{n-2}\p^{n-2} + \ldots + u_1 \p + u_0$, where $\p = \p / \p z$ is a differential symbol with respect to some (complex) variable $z$, and $u_0, u_1, \ldots, u_{n-2}$ (note: $u_{n-1}$ can be always set to zero) are generically smooth functions in $z$ and (if necessary) other variables $t_1, t_2, \ldots $. On this algebra, we define the ring of pseudo-differential operators 
$\mathcal R$, consisting of (formal) operators defined by the infinite series
\beq 
L =  \sum_{-\infty}^n c_k \p^k, \quad \p^{-1} \equiv \int dz, 
\eeq
where coefficients are again smooth functions, and the negative powers in the expansions contain $integral$ operators. For any such operator, we denote by $L_+$ the purely differential part and by $L_-$ the remainder of the series:
\beq
L_+ \equiv \sum_{0}^n c_k \p^k, \quad L = L_+ + L_-.
\eeq

Let 
\beq \la{KP_op}
\mathcal L = \p + u_0 \p^{-1} + u_1\p^{-2} + \ldots
\eeq
be a pseudo-differential operator such that $\mathcal{L}_+ = \p$. Then, introducing the infinite set of $times$ ${\bm {t}} = t_1, t_2, \ldots ,$ such that all coefficients $u_k, c_k $ above are generically functions of $\bm t$, the KP hierarchy has the form
\beq
\frac{\p \mathcal{L}}{\p t_k} = [\mathcal{L}^k_+, \mathcal{L}], \quad k = 1, 2, \ldots
\eeq
More explicitly, we note that the hierarchy consists of the differential equations satisfied by the {\emph{coefficients}} of the operator $\mathcal L$. As a consequence of the compatibility of all the equations in the hierarchy, we have the {\emph{zero-curvature equations}} 
\beq
[\p_{t_k} - \mathcal{L}^k_+ , \p_{t_p} - \mathcal{L}^p_+] = 0, \quad \forall t_k, t_p.
\eeq

\subsubsection{Level reductions}

The KP hierarchy contains many other known integrable hierarchies, particularly the KdV hierarchy, as reductions to a certain $level$ $n$ in the hierarchy. For example, assume that the operator $\mathcal L$ satisfies the constraint 
\beq
\mathcal{L}^2_- = 0,
\eeq 
i.e. it is the square root of a {\emph{differential}} operator $L$ of order 2:
\beq
\mathcal{L} = L^{1/2}, \quad L = \p^2 + 2u_0.
\eeq
Then it follows that for all even powers $n=2m$, $\mathcal{L}^n_{+} = \mathcal{L}^n$, so that 
$[\mathcal{L}^n_{+}, \mathcal{L}] = 0$, so there is $no$ dependence on the even times $t_2, t_4, \ldots$. 
This sub-hierarchy is called $level-2$ KdV, because the first non-trivial zero-curvature equation of the hierarchy is the famous Korteweg-de Vries equation:
\beq
\mathcal{L}^3_+ \equiv P = \p^3 + \frac{3}{2}\left [ u_0 \p + \p u_0 \right ], \quad
\frac{\p L}{\p t_3} = [P, L] \Rightarrow u_{t_3} = 6 uu_z + u_{zzz},
\eeq
where we have used $u_0 = u$ for clarity. 

This formulation of the KdV equation makes use of the notion of $Lax$ pair $L, P$, which is central to the {\emph{inverse scattering method}} for solving nonlinear integrable differential equations The idea is quite physical: assume that the operators $L, P$ act on a wavefunction $\psi(x,t)$ such that 
\beq
L\psi = \lambda \psi, \quad \frac{\p \psi}{\p t} = P \psi,
\eeq
where eigenvalues $\lambda$ form the spectrum of $L$. Then applying the Lax pair equation to the eigenvalue equation, we obtain $\p \lambda / \p t = 0$, i.e. the evolution under these equations leaves the spectrum invariant. This allows to construct the initial state from the final state, hence the inverse scattering appellation. 

\subsubsection{Tau functions and Baker-Akhiezer function}

At the level of systems of PDE, the $\tau-$function and the Baker-Akhiezer function are introduced, by analogy with the Lax par formulation indicated above, in the following way: 

\paragraph*{Baker-Akhiezer function} Consider the function $\psi(z,t_1, t_2, \ldots)$ satisfying
\beq
\mathcal{L} \psi = z \psi, \quad \frac{\p \psi}{\p t_k} = \mathcal{L}^k_+ \psi, \quad \forall k \ge 1.
\eeq
This is the Baker-Akhiezer function of the KP hierarchy. 

\paragraph*{Fundamental property of the Baker-Akhiezer function} 

\begin{quotation}
Let $\phi = 1 + \sum_0^\infty k_i \p^{-i-1}$ be the ``dressing" operator defined such that 
$\mathcal{L} = \phi \p \phi^{-1}$. Also, introduce the function $g(z, t_1, \ldots) = \exp [\sum_{1}^\infty t_k z^k]$. Then the Baker-Akhiezer function satisfies:
$$
\psi = \hat{k}(z) g(z, t_1, \ldots), \quad \hat{k}(z) = 1 + \sum_{0}^\infty k_i z^{-i-1},
$$
where $\hat{k}$ is the ``scalar" analog of the dressing operator $\phi$. 
\end{quotation}

\paragraph*{Tau function} Using the notation introduced above, we have the following property: 

\begin{quotation}
There exists a function $\tau(z, t_1, \ldots)$ such that 
$$
\psi(z, t_1, \ldots) = g \cdot \frac{\tau(z, t_1-\frac{1}{z}, t_2 - \frac{1}{2z^2}, \ldots)}{\tau(z, t_1, t_2, \ldots)} 
= g \cdot \frac{\exp \left [\sum_1^\infty -\frac{1}{kz^k}\frac{\p }{\p t_k} \right ] \tau(z, t_1, \ldots)}{\tau(z, t_1, \ldots)}
$$
\end{quotation}
Now let us consider the generalized overlap function
$$
\psi_N(z,\bar w) = \tau_{N}^{-1}
\int\Psi_{N+1} (z,z_1,z_2,\dots,z_N)
\overline{\Psi_{N+1}(w,z_1,z_2,\dots,z_N)} d^2z_1\dots d^2 z_N,
$$
and expand for $|z|, |w| \to \infty$. We obtain 
\be
\psi_N(z, \bar w) =  \frac{(z\bar w)^{N}}{\tau_{N}} 
\exp \left [\sum -\frac{1}{z^k\bar w^p}\frac{\p }{\p a_{kp}} \right ] 
\tau_N,
\ee
where $a_{kp}$ is the corresponding interior bi-harmonic moment. Therefore, we may regard 
the $\tau$-function and the scaled wavefunction introduced earlier as canonical objects describing 
an integrable hierarchy. This fact will be illustrated in more detail in the next section. 

\subsection{Equations for the wave functions and the
spectral curve}\la{1A}
In this section we specify the potential to be
of the form (\ref{potential}).
It is convenient to modify the
exponential factor of the wave function.
Namely, we define
\be \la{chin}
\psi_n (z)=
e^{-\frac{|z|^2}{2\hbar}+\frac{1}{\hbar}V(z)}P_n (z),\quad\mbox{and}\quad\chi_n (z)=
e^{\frac{1}{\hbar}V(z)}P_n (z),
\ee
where the holomorphic
functions $\chi_n(z)$
are orthonormal in the complex plane
with the weight
$e^{-|z|^2/\hbar}$.
Like traditional orthogonal polynomials, the biorthogonal polynomials
$P_n$ (and the corresponding wave functions)
obey a set of differential equations with
respect to the argument $z$, and recurrence relations with respect to
the degree $n$. Similar equations
for  two-matrix models are discussed in  numerous papers
(see, e.g., \cite{Aratyn}).

We introduce the $L$-operator (the Lax operator)
as multiplication by $z$ in the
basis $\chi_n$:
\begin{equation}\la{L1}
L_{nm}\chi_m(z)=z\chi_n(z)
\end{equation}
(summation over repeated indices is implied).
Obviously, $L$ is a lower triangular matrix with one
adjacent upper diagonal, $L_{nm}=0$ as $m>n+1$.
Similarly, the differentiation
$\p_z$ is represented by an
upper triangular matrix with one adjacent lower diagonal.
Integrating by parts the matrix elements of the $\p_z$, one finds:
\begin{equation}\la{M}
(L^{\dag})_{nm}\chi_m =
\hbar\p_z\chi_n,
\end{equation}
where $L^{\dag}$
is the Hermitian conjugate operator.

The matrix elements of $L^{\dag}$ are
$(L^{\dag})_{nm}=\bar L_{mn}=A(L_{nm})+
\int e^{\frac{1}{\hbar}W}\bar P_m(\bar z)\p_z P_n(z) d^2z$, where
the last term is a lower triangular matrix.  The latter can be written
       through negative powers of the Lax operator.
Writing $\p_z\log P_n(z)=\frac{n}{z}+\sum_{k>1}v_k(n)z^{-k}$,
one represents $L^{\dag}$ in the form
\begin{equation}\la{M1}
L^{\dag}=A(L)+(\hbar n) L^{-1}+\sum_{k>1}v^{(k)}L^{-k},
\end{equation}
where $v^{(k)}$ and $(\hbar n)$ are
diagonal matrices with elements $v_n^{(k)}$
and $(\hbar n)$.
The coefficients $v_{n}^{(k)}$ are determined by the condition that
lower triangular matrix elements of $A(L_{nm})$ are cancelled.

In order to emphasize the structure of the operator $L$, we
write it in the basis of the
shift operator \footnote{The shift operator
$\hat w$  has no
inverse. Below $\hat w^{-1}$ is understood as a shift to the left
defined as  $\hat w^{-1}\hat w=1$.
Same is applied to the operator $L^{-1}$. To avoid a possible
confusion, we emphasize that although $\chi_n$
is a right-hand eigenvector of $L$, it is not a right-hand
eigenvector of $L^{-1}$.}
$\hat w$ such that $\hat w f_n =f_{n+1}\hat w$ for any
sequence $f_n$. Acting on the wave function, we have:
$$\hat w
\chi_n=\chi_{n+1}.$$
In the $n$-representation, the operators $L$, $L^{\dag}$
acquire the form
\begin{equation}\la{M11}
L=r_n \hat w+\sum_{k\geq 0} u_{n}^{(k)} \hat w^{-k},\quad
L^{\dag} = \hat w^{-1} r_n+
\sum_{k\geq 0} \hat w^{k} \bar u_{n}^{(k)}.
\end{equation}
Clearly, acting on $\chi_n$, we have the commutation
relation (``the string equation")
\begin{equation}\la{string}
[L, \, L^{\dag}]=\hbar.
\end{equation}
This is the compatibility condition of Eqs. (\ref{L1}) and (\ref{M}).

Equations (\ref{M11}) and (\ref{string})
completely determine the coefficients
$v_n^{(k)}$, $r_n$ and $u_n^{(k)}$. The first one connects the coefficients
to the parameters of the potential.
The second equation is used to determine how the coefficients
       $v_n^{(k)}$, $r_n$ and $u_n^{(k)}$ evolve with $n$. In particular,
the diagonal part of it reads
\be\la{Area}
n\hbar=r_n^2-\sum_{k\geq 1}\sum_{p=1}^k|u_{n+p}^{(k)}|^2.
\ee
Moreover, we note that all the coefficients can be expressed
through the $\tau$-function (\ref{tau}) and its derivatives
with respect to parameters of the potential.
This representation is particularly simple for $r_n$:
$
r_n^2 =\tau_n \tau_{n+1}^{-2}\tau_{n+2}
$.

\subsubsection{Finite dimensional reductions}
If the vector potential $A(z)$ is a rational function,
the coefficients $u_{n}^{(k)}$ are not all independent.
The number of independent coefficients
equals the number of independent parameters of the potential.
For example, if the holomorphic part
of the potential, $V(z)$, is a polynomial of degree $d$,
the series
(\ref{M11}) are truncated at $k= d-1$.

In this  case the semi-infinite system of linear equations (\ref{M})
and the recurrence relations (\ref{L1})
can be cast in the form of a set of finite dimensional equations whose
coefficients are rational functions of  $z$, one
system for every $n>0$.
The system of differential equations generalizes the
Cristoffel-Daurboux second order
differential equation
valid for orthogonal polynomials. This fact has been observed in
recent papers \cite{BEHdual,Eynard03}
for biorthogonal polynomials emerging in the
Hermitian two-matrix model
with a polynomial potential. It is
applicable to our case (holomorphic biorthogonal
polynomials) as well.

In a more general case,
when $A(z)$ is a general rational function with $d-1$ poles
(counting multiplicities), the series (\ref{M11})
is not truncated. However, $L$ can be represented
as a ``ratio",
\beq\label{LKK}
L=K_{1}^{-1}K_{2}=M_2 M_{1}^{-1},
\eeq
where the operators $K_{1,2}$, $M_{1,2}$ are
polynomials in $\hat w$:
\beq\label{KK}
K_1 =\hat w^{d-1}+\sum_{j=0}^{d-2}A_{n}^{(j)} \hat w^j\,,
\quad
K_2 =r_{n\! +\! d\! -\! 1}\hat w^{d}+\sum_{j=0}^{d-1}B_{n}^{(j)} \hat w^j
\eeq
\beq\label{MM}
M_1 =\hat w^{d-1}+\sum_{j=0}^{d-2}C_{n}^{(j)} \hat w^j\,,
\quad
M_2 =r_{n}\hat w^{d}+\sum_{j=0}^{d-1}D_{n}^{(j)} \hat w^j
\eeq
These operators obey the
relation
\beq\label{KMKM}
K_1 M_2 =K_2 M_1.
\eeq
It can be proven that
the pair of operators $M_{1,2}$ is uniquely determined
by $K_{1,2}$ and vice versa. We note that the
reduction (\ref{LKK}) is a difference analog of the
``rational" reductions of the Kadomtsev-Petviashvili
integrable hierarchy considered in \cite{Krichev-red}.

The linear problems (\ref{L1}), (\ref{M}) acquire the form
\beq\label{L1a}
(K_2\chi )_n =z \, (K_1 \chi )_n\,, \quad
(M_{2}^{\dag} \chi )_n =\hbar \p_z (M_{1}^{\dag}\chi ) _n.
\eeq
These equations are of {\it finite order}
(namely, of order $d$), i.e., they connect values of $\chi_n$
on $d+1$ subsequent sites of the lattice.

The semi-infinite set $\{\chi_0,\chi_1,\dots\}$ is
then a ``bundle" of
$d$-dimensional vectors
$${\underline\chi}(n)= (\chi_n,\chi_{n+1},\dots,
\chi_{n+d-1})^{{\rm t}}$$
(the index ${\rm t}$
means transposition, so
${\underline\chi}$ is a column vector).
The dimension of the vector is the number of poles
of $A(z)$ plus one.
Each vector obeys a closed $d$-dimensional linear
differential equation
\begin{equation}
\la{M2}
\hbar\p_z{\underline\chi}(n)={\mathcal L}_n (z){\underline\chi}(n),
\end{equation}
where the $d\times d$ matrix
${\mathcal L}_n$ is a ``projection" of the operator $L^{\dag}$
onto the $n$-th $d$-dimensional space. Matrix elements of the
${\mathcal L}_n$ are rational functions of
$z$ having the same poles as $A(z)$ and also a pole at
the point $\overline{A(\infty )}$. (If $A(z)$ is a polynomial,
all these poles accumulate to a multiple pole at infinity).

We briefly describe the procedure of
constructing the finite dimensional matrix
differential equation. We use the first
linear problem in (\ref{L1a})
to represent the shift operator
as a $d\times d$ matrix ${\mathcal W}_n (z)$ with
$z$-dependent coefficients:
\be\la{W}
{\mathcal W}_n(z){\underline\chi}(n)={\underline\chi}(n \! +\! 1).
\ee
This is nothing else than rewriting the scalar linear problem
in the matrix form.
Then the matrix ${\mathcal W}_n (z)$ is to be substituted into the
second equation of (\ref{L1a}) to
determine ${\mathcal L}_n(z)$ (examples follow).
The entries of ${\mathcal W}_n(z)$ and ${\mathcal L}_n (z)$
obey the Schlesinger
equation, which follows from compatibility of
(\ref{M2}) and (\ref{W}):
\be\la{WW}
\hbar\p_z {\mathcal W}_n=
{\mathcal L}_{n+1} {\mathcal W}_n
-{\mathcal W}_n {\mathcal L}_n.
\ee

This procedure has been realized explicitly for
polynomial potentials in
recent papers \cite{BEHdual,Eynard03}.
We will work it out in detail for our three examples:
${\underline\chi}(n)=(\chi_n,\,\chi_{n+1})^{{\rm t}}$ for the ellipse
(\ref{e}) and
the aircraft wing (\ref{J1}) and
${\underline\chi}(n)=(\chi_n,\,\chi_{n+1},\chi_{n+2})^{{\rm t}}$
for the hypotrochoid (\ref{hyp}).

\subsection{Spectral curve}\la{Curve1}

According to the general theory of
linear differential equations, the
semiclassical (WKB) asymptotics of solutions to
Eq. (\ref{M2}), as $\hbar \to 0$,
is found by solving the eigenvalue problem for
the matrix ${\mathcal L}_n (z)$ \cite{Wasow}.
More precisely, the basic object of the WKB approach is the
spectral curve \cite{Wasow} of the matrix ${\mathcal L}_n$, which is
defined, for every integer $n>0$, by the secular equation
$\det ({\mathcal L}_n (z)-\tilde z) = 0$
(here $\tilde z$ means $\tilde z \cdot {\bf 1}$, where
${\bf 1}$ is the unit $d\times d$ matrix).
It is clear that the left hand side of the secular
equation is a polynomial in $\tilde z$ of degree $d$.
We define the spectral curve by an equivalent equation
\begin{equation}\la{qc}
f_n (z,\tilde z)=a(z)\det ({\mathcal L}_n (z)-\tilde z) = 0,
\end{equation}
where the factor $a(z)$ is added
to make $f_n(z,\tilde z)$ a polynomial in $z$ as well.
The factor $a(z)$ then has zeros at the points where
poles of the matrix function ${\mathcal L}(z)$ are located.
It does not depend on $n$.
We will soon see that the degree of
the polynomial $a(z)$ is equal to $d$.
Assume that all poles of $A(z)$ are simple, then zeros of
the $a(z)$ are
just the $d-1$ poles of $A(z)$ and another simple zero
at the point $\overline{A(\infty)}$. Therefore,  we conclude that
the matrix ${\mathcal L}_n (z)$ is rather special.
For a general $d\times d$ matrix function with the same
$d$ poles, the factor $a(z)$ would be of degree $d^2$.

Note that the matrix ${\mathcal L}_n (z) -\bar z$ enters
the differential equation
\begin{equation}\la{M2'}
\hbar\p_{z}|{\underline\psi}(n)|^2=\bar {\underline\psi}(n)
({\mathcal L}_n (z)- \bar z)
{\underline\psi}(n)
\end{equation}
for the squared amplitude
$|{\underline \psi}(n) |^2={\underline \psi}^{\dag}(n)
{\underline \psi}(n) =
e^{-\frac{|z|^2}{\hbar}} |{\underline \chi}(n)|^2$
of the vectors
${\underline \psi}(n)$ built from the orthonormal
wave functions (\ref{O3}).

The equation of the curve can be interpreted as a
``resultant" of the non-commutative polynomials
$K_2 -zK_1$ and $M_{2}^{\dag}-\tilde z M_{1}^{\dag}$
(cf. \cite{BEHdual}). Indeed, the point $(z, \tilde z)$
belongs to the curve if and only if the linear system
\beq\label{linsys}
\left \{ \begin{array}{l}
(K_2 c)_k =z (K_1 c)_k \quad \quad n-d\leq k \leq n-1
\\ \\
(M_{2}^{\dag} c)_k =\tilde
z (M_{1}^{\dag} c)_k \quad \quad n\leq k \leq n+d-1
\end{array}\right.
\eeq
has non-trivial solutions. The system contains $2d$ equations
for $2d$ variables $c_{n-d}\, , \ldots , c_{n+d -1}$.
Vanishing of the
$2d \, \times \, 2d$ determinant yields the equation of
the spectral curve.
Below we use this method to find the equation of the curve
in the examples.
It appears to be much easier than the determination of the
matrix ${\mathcal L}_n(z)$.

The spectral curve (\ref{qc}) possesses an important property:
it admits an antiholomorphic involution.
In the coordinates $z, \tilde z$ the involution reads
$(z, \tilde z)\mapsto (\overline{\tilde z}, \bar z)$.
This simply means that the secular equation
$\det (\bar {\mathcal L}_n (\tilde z)-z) = 0$
for the matrix $\bar {\mathcal L}_n (\tilde z)\equiv
\overline{{\mathcal L}_n (\overline{\tilde z})}$
defines the same curve.
Therefore, the polynomial $f_n$
takes real values for $\tilde z =\bar z$:
\be\la{anti1}
f_n(z,\bar z)=\overline{f_n ( z, \bar z)}.
\ee
Points of the real section of the curve
($\tilde z =\bar z$) are
fixed points of the involution.

The curve (\ref{qc})
was discussed in recent papers \cite{BEHdual,Eynard03}
in the context of Hermitian two-matrix models with
polynomial potentials. The dual realizations of the curve
pointed out in \cite{BEHdual} correspond to the
antiholomorphic involution in our case.
The involution can be proven along the lines of these
works. The proof is rather technical and we omit
it, restricting ourselves to the
examples below. We simply note that the involution relies on the fact
that the squared modulus of the wave function is real.

We will give a concrete example for the construction of the 
spectral curve, after a brief but necessary detour through the continuum
limit of this problem. 

\subsubsection{Schwarz function}

The polynomial $f_n(z, \bar z)$ can be
factorized in two ways:
\be\la{h1}f_n(z,\bar z)=a(z)(\bar z-S_n^{(1)}(z))
\dots (\bar z-S_n^{(d)}(z)),
\ee
where $S_n^{(i)}(z)$ are eigenvalues of the matrix
${\mathcal L}_n (z)$, or
\be\la{ah1}
f_n(z,\bar z)=\overline{a(z)}( z-\bar S_n^{(1)}(\bar z))\dots ( z-\bar
S_n^{(d)}(\bar z)),
\ee
where $\bar S_n^{(i)}(\bar z)$ are eigenvalues of the matrix
$\bar {\mathcal L}_n (\bar z)$.
One may understand them as different branches of a
multivalued function $S(z)$ (respectively, $\bar S(z)$)
on the plane (here we do not indicate the dependence on $n$,
for simplicity of the notation).
It then follows that
$S(z)$ and $\bar S(z)$
are mutually inverse functions:
\begin{equation}\la{anti11}
\bar S(S(z))=z.
\end{equation}

An algebraic function  with this property
is called {\it the Schwarz function}.
By the equation $f(z, S(z))=0$, it defines a complex curve
with an antiholomorphic involution. An upper bound
for genus of this curve is $g=(d-1)^2$, where $d$ is the
number of branches of the Schwarz function.
The real section of this curve is a set of all fixed
points of the involution. It consists of a number
of contours on the plane (and possibly a number of
isolated points, if the curve is not smooth).
The structure of this set is known to be
  complicated. Depending on
coefficients of the polynomial, the number of disconnected
contours in the real section may vary from $0$ to $g+1$.
If the contours divide the complex curve into two
disconnected ``halves", or sides (related by the involution), then
the curve can be realized as
the {\it Schottky double} \cite{Gustafsson90} of one of
these sides. Each side is a Riemann surface with a boundary.

Let us come back to
equation (\ref{M2}). It has $d$ independent solutions.
They are functions on the spectral curve.
One of them is a physical solution
corresponding to biorthogonal polynomials.
The physical solution defines the ``physical sheet"
of the curve.

The Schwarz function on the physical sheet
is a particular root, say $S^{(1)}_n(z)$,   of the
polynomial $f_n(z, \tilde z)$ (see (\ref{h1})).
It follows from (\ref{M1}) that
this root  is
selected by the requirement that
it has the same poles and residues as the potential $A$.

\subsubsection {The Schottky
double}\label{S}

The Schwarz function describes more than just the boundary of
clusters of eigenvalues.
Together with other sheets
it defines a Riemann surface.
If the   potential $A(z)$ is meromorphic,
the Schwarz function is an algebraic function. It
satisfies a polynomial equation $f(z, S(z))=0$.

\begin{figure} \begin{center}
      \includegraphics*[width=5cm]{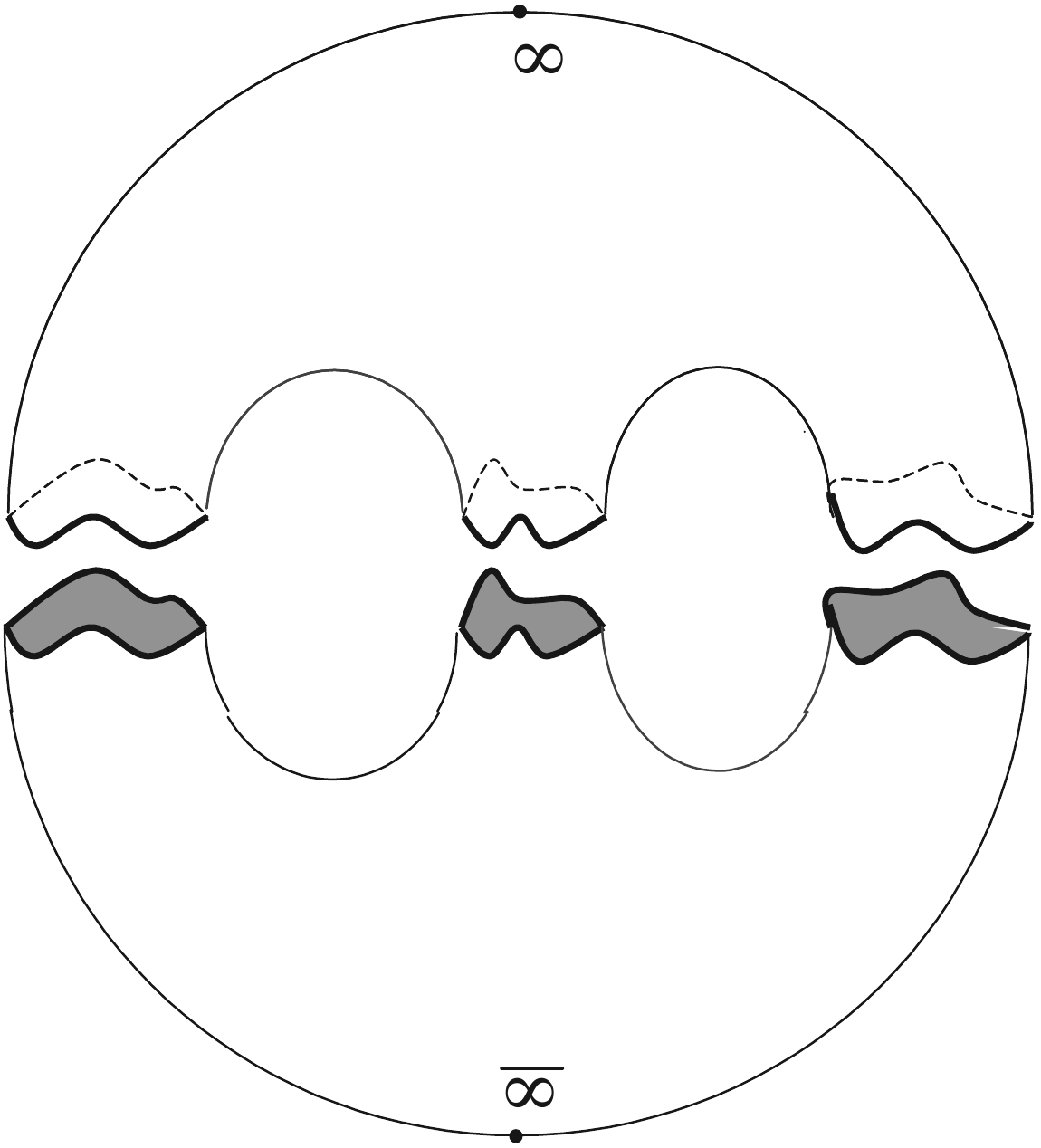}
      \caption{\label{Schottky} The Schottky double. A Riemann surface with
boundaries
     along the droplets (a front side) is
  glued to its mirror image (a back side).}
      \end{center} \end{figure}

The function  $f(z,\tilde z)$, where $z$
and $\tilde z$ are treated as two independent complex arguments, defines
a Riemann surface with antiholomorphic involution (\ref{anti1}).
If the involution divides the surface into two disconnected
parts, as explained above,
the Riemann surface is
       the {\it Schottky double} \cite{Gustafsson90} of
one of these parts.

There are two complementary ways to describe
this surface. One is through the algebraic covering
(\ref{h1}, \ref{ah1}). Among $d$ sheets we distinguish
a {\it physical} sheet.
The physical sheet is
selected by the condition that  the differential $S(z)dz$ has the same
poles and residues  as
the differential of the potential $A(z)dz$.
It may happen that the condition
$\bar z=S^{(i)}(z)$ defines a planar curve (or
several curves, or a set of isolated points)
for branches
other than the physical one. We refer to the interior of these
   planar curves as {\it virtual} (or
unphysical) droplets
     situated on sheets other than physical.

Another way emphasizes the antiholomorphic involution. Consider a
meromorphic function $h(z)$ defined
on a Riemann surface with boundaries. We call this surface the front side.
The Schwarz reflection principle extends   any meromorphic function on
the front side
     to a meromorphic function on the Riemann surface without
boundaries. This is done
     by adding another copy of the Riemann surface with
boundaries (a back side),
glued to the front side along the boundaries, Figure~\ref{Schottky}.
The value of the
     function $h$ on the mirror point on the back side
is $h(\overline{S(z)})$.
     The copies are glued along the boundaries:
$h(z)=h(\overline{S(z)})$
if the point $z$ belongs to the boundary.
The same extension rule applies to differentials.
Having a meromorphic differential $h(z)dz$ on the
front side, one extends it to  a
meromorphic differential
$h(\overline{S( z)})d\overline{S(z)}$ on the back side.

This definition can be applied to  the Schwarz function itself.
We say that the Schwarz
     function on the double is $S(z)$ if the point is
on the front side, and
     $\bar z$ if the point belongs to the back side
(here we understand $S(z)$
as a function defined on the complex curve, not just on the physical sheet).

The number of sheets of the curve is the number of poles (counted
with their multiplicity)
of the function $A(z)$ plus one.
Indeed, poles of $A$   are  poles of
     the Schwarz function on the front side of the double. On the
back side, there is also a pole at
infinity. Since $S(z=\infty)=A(\infty)$, we have
$\bar S(\bar z=A(\infty))=\infty$.
Therefore, the factor $a(z)$ is a polynomial
with zeros at the poles of $A(z)$
and at $\overline{A(\infty)}$, and
$$d\equiv\mbox{number of sheets} =
\mbox{number of poles of $A$ + 1}.$$
The front and  back sides meet at planar curves $\bar z=S(z)$.
These curves are
  boundaries of the droplets. We repeat  that not all droplets
are physical.
Some of them may belong to unphysical sheets, Figure~\ref{torus}.

Boundaries of droplets, physical and virtual, form a subset of the
$\bf a$-cycles on the curve. Their number
 cannot exceed the genus of the
curve plus one:
$$\mbox{number of droplets} \le g+1.$$

The sheets meet along cuts located inside droplets.
The cuts that belong to physical droplets show up on  unphysical sheets.
On the other hand, some
     cuts show up on the physical sheet (Figure~\ref{torus}).
They correspond to droplets situated on
     unphysical sheets.
\begin{figure} \begin{center}
      \includegraphics*[width=5cm]{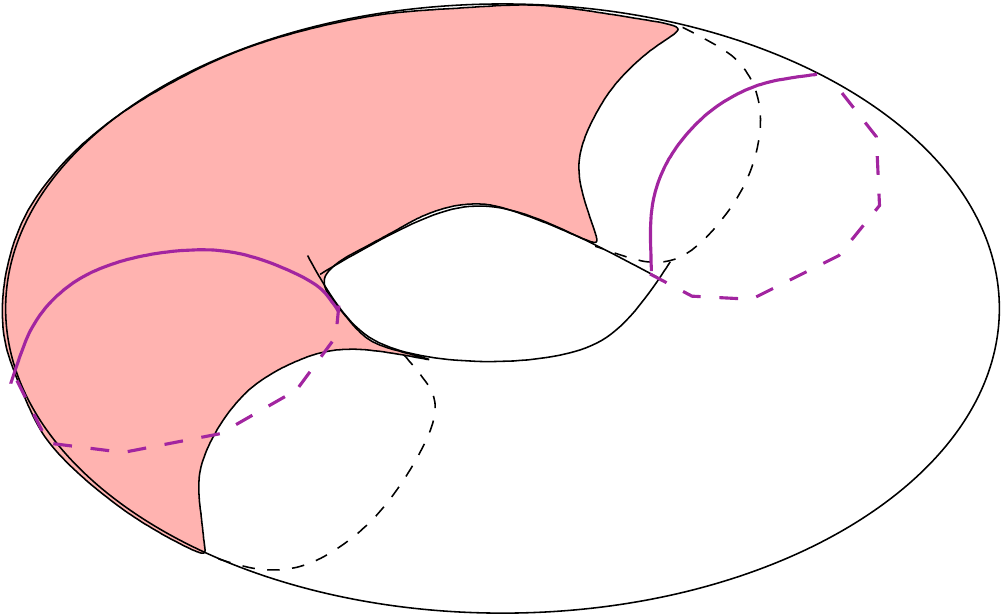}
      \caption{\label{torus}Physical and unphysical droplets on a  torus.
The physical sheet (shaded) meets the unphysical sheet
along the cuts. The cut situated inside the
unphysical droplet appears on the physical sheet.
The boundaries of the droplets (physical and virtual)
belong to different sheets.
This torus is the Riemann surface corresponding to the ensemble
with the potential $V(z) = - \alpha \log (1- z/ \beta) - \gamma z $.}
\end{center} \end{figure} 

The Riemann-Hurwitz theorem computes the genus of the curve as
$$ g=\mbox{half
the number of
branching points} - d+1. $$

With  the help of the Stokes formula, the numbers $\{ \nu_{\alpha} \}$
are identified with  areas of
the droplets:
$|\nu_a|=\frac{1}{2\pi  \hbar}\int_{D_a} d^2z$.
   For a nondegenerate curve,
   these numbers are not necessarily positive. Negative
numbers correspond to droplets located on unphysical sheets. In this
case, $\{\nu_a\}$
do not correspond to the number of eigenvalues located inside each droplet,
as it is the case for algebraic domains, when all
filling numbers are positive.

\subsubsection{Degeneration of the spectral curve} \la{alg}
Degeneration of the complex curve gives the most interesting physical
aspects of growth.
There are several levels of degeneration. We briefly discuss them below.

\paragraph*{Algebraic domains and double points}
A special  case occurs when the Schwarz function on the physical sheet
is meromorphic. It
     has no other singularities than poles of $A$.
This is the case of algebraic domains .
They appear in the
semiclassical case.
This situation occurs if  cuts
     on the physical sheet, situated outside physical droplets, shrink
     to  points, i.e., two or more
branching points merge. Then  the physical sheet meets other
     sheets along cuts situated inside physical droplets only and also at
some points on their exterior ({\it double points}).
     In this case the Riemann surface degenerates. The genus
is given by the number of physical
droplets only. The filling factors  are all positive.

\begin{figure} \begin{center}
      \includegraphics*[width=5cm]{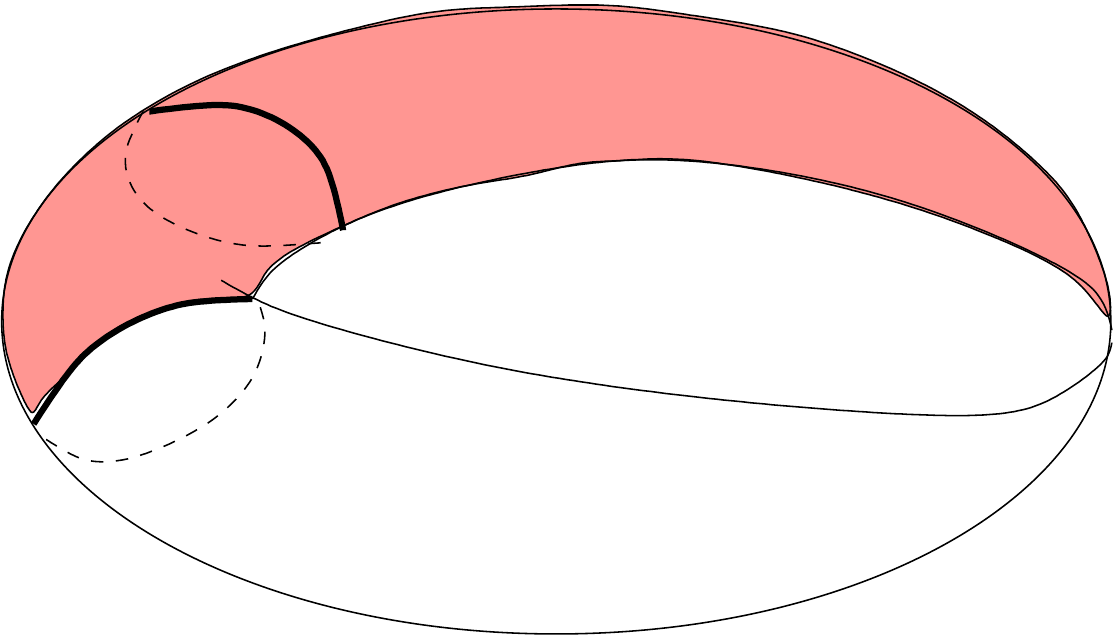}
      \caption{\label{degtorus}Degenerate torus corresponds to the
algebraic domain for the Joukowsky map.}
\end{center} \end{figure}

In the case of algebraic domains, the physical branch of the
Schwarz function
     is a well-defined
meromorphic function. Analytic continuations of $\bar z$
     from different disconnected parts of the boundary
give the same result. In this case, the
     Schwarz function can be written through the Cauchy transform of the
physical droplets:
\begin{equation}\label{28}
S(z)=A(z)+\frac{1}{\pi}\int_{D}\frac{d^2 \zeta}{z-\zeta}.
\end{equation}

Although algebraic domains occur in physical problems
such as Laplacian growth,
their semiclassical evolution is limited.
Almost all algebraic domains will be broken in a  growth process.
Within a finite
     time (the area of the domain) they degenerate further into
critical curves.
The Gaussian potential (the Ginibre-Girko ensemble),
which leads to a single
droplet of the form of an ellipse is a known exception.

\paragraph*{Critical degenerate curves}

Algebraic domains appear as
a result of merging of simple branching points on the physical
sheet. The double points are located
outside physical droplets.
Remaining branching points belong to
the interior of
physical droplets. Initially, they survive
in the degeneration process.
However, as known in the theory of Laplacian growth,
   the process necessarily leads to a further
degeneration. Sooner  or later, at least one of the
interior branching points  merges with one of the double points in
the exterior.  Curves degenerated in this
manner are called
{\it critical}. For the genus one and three
this degeneration is discussed below.

Since interior branching points
can only merge with exterior branching points
on the boundary of the
droplet, the boundary develops a cusp, characterized by a pair $p, q$
of mutually prime integers. In local coordinates
around such a cusp, the curve looks like $x^p\sim y^q$. The fact that the growth of algebraic domains always leads to critical
curves is known
in the theory of Laplacian growth   as
finite time singularities. 

The degeneration process seems to be a feature of the semiclassical
approximation. Curves treated beyond this approximation never
degenerate.

\subsubsection{Example: genus one curve}\la{J}

The potential is
$V(z) = - \alpha \log (1- z/\beta) -
\gamma z,\quad A(z)=-\frac{\alpha}{z-\beta}-\gamma$.
There is one pole at $z=\beta$  on the first (physical) sheet.
At $z=\infty$ on the first sheet
$S(z)\to-\gamma+\frac{n\hbar-\alpha}{z}$.
Therefore, the Schwarz function has another pole
at the point $-\bar\gamma$
on another sheet.  All the poles are simple.
According to the general arguments of
    Sec.~\ref{S},  the number of sheets is 2,
the number of branching points is 4. The genus is 1.
The curve has the form
$$
f(z, \bar z) = z^2 \bar z^2 +k_1
z^2\bar z +\bar k_1 z\bar z^2  + k_2 z^2 + \bar k_2 \bar z^2 +
k_3 z\bar z
+k_4z+
\bar k_4\bar z +h =0.
$$
The points at infinity  and  $-\bar\gamma$ belong to
the second
sheet of the algebraic covering. Summing up,
\[ S(z)=
\left \{\begin{array}{rll}
-\frac{\alpha}{z-\beta}& \mbox{as} & z\to\beta_1, \\
(-\gamma+\frac{n\hbar-\alpha}{z} ) & \mbox{as} & z\to\infty_1, \\
\frac{n\hbar-\bar\alpha}{z+\bar\gamma}& \mbox{as}
& z\to -\bar\gamma_2, \\
(\bar\beta-\frac{\bar\alpha}{z})& \mbox{as} & z\to\infty_2.
\end{array}
\right. \]
where, by 1 and 2  we indicate the sheets.

Poles and residues of the Schwarz function
determine   all the coefficients of the curve
$f(z,\bar z)=a(z) (\bar z-S^{(1)}(z))(\bar z-S^{(2)}(z))=
\overline{a(z)} (z-\bar S^{(1)}(\bar z))(z-\bar
S^{(2)}(\bar z))$ except one.
The behavior at $\infty$ of $z, \bar z$
gives $k_1=\gamma -\bar \beta, \, k_2=-\gamma \bar \beta$.
Hereafter we choose the origin by
setting  $\gamma=0$.
The equation of the curve then reads $f_n(z, \bar z) = 0$, where
$f_n(z, \bar z) $ is given by
\be\la{8}
z^2 \bar z^2 -
z^2\bar z \bar\beta  -z\bar z^2\beta  +
\left ( |\bar\beta|^2 +\alpha+\bar\alpha-n\hbar
\right ) z\bar z
+z\bar\beta(n\hbar-\alpha)+
\bar z\beta(n\hbar-\bar\alpha)
+h_n
\ee
The free term $h_n$ is to be determined by
filling factors of the two
droplets $\nu_1$ and $\nu_2=n-\nu_1$.
A detailed analysis shows
that the droplets belong to different
     sheets (Figure~\ref{torus}).
Therefore, $\nu_2$ is negative.

A boundary of a physical  droplet is given by the
equation
$\bar z=S^{(1)}(z)$ (Figure~\ref{droplets}).
The second droplet
belongs to the unphysical sheet. Its
boundary is given by
     $\bar z=S^{(2)}(z)$. The explicit form of both branches is
$$S^{(1,2)}=\frac{1}{2}\bar\beta-
\frac{\beta(n\hbar-\bar\alpha)+(\alpha+\bar\alpha-n\hbar)z\mp
\sqrt{(z-z_1)(z-z_2)(z-z_3)(z-z_4)}}{2(z-\beta)z},$$
where the branching points $z_i$ depend on $h_n$.

If the filling factor of the physical droplet is equal to $n$, the cut
inside the unphysical droplet is of the order of
   $\sqrt{\hbar}$. Although it never vanishes, it
  shrinks to a double point
$z_3=z_4=z_*$ in a semiclassical limit.
The sheets meet at the double
point $z_*$ rather than along the cut:
$\sqrt{(z-z_1)(z-z_2)(z-z_3)(z-z_4)}\to
(z-z_*)\sqrt{(z-z_1)(z-z_2)}$.
In this case, genus of the curve reduces to zero and
the exterior of the physical droplet becomes an algebraic domain.
This condition determines $h$,
and also the position of the double point (Figure~\ref{degtorus}).
The double point is a saddle point for the
level curves of  $f(z,\bar z)$.
If all the parameters are real, the double point is stable in
$x$-direction and unstable in
$y$-direction. 

If this solution is chosen,
the exterior of the physical droplet can be mapped
to the exterior of the unit disk by the Joukowsky map
\be\la{222}
z(w)=rw+u_0 + \frac{u}{w-a},\quad |w|>1,\quad |a|<1.
\ee
The inverse map
is given by the branch $w_1(z)$ (such that $w_1\to\infty$ as
$z\to\infty$)
of the double valued function
$$
w_{1,2}(z) = \frac{1}{2r} \left [
z-u_0 + ar \pm \sqrt{(z -z_1)(z-z_2) } \right ],\quad
z_{1,2}=u_0+ar\mp 2\sqrt{r(u+au_0)}.
$$

     The function
\be\la{23}
\bar z(w^{-1})=rw^{-1}+\bar u_0 + \frac{\bar u}{w^{-1}-\bar a}
\ee
     is a meromorphic function of
$w$ with two simple poles at $w=0$ and $w=\bar a^{-1}$. Treated as a
function of $z$, it  covers
the $z$-plane twice. Two branches of the Schwarz function are
$S^{(1,2)}(z)=\bar z(w_{1,2}^{-1}(z))$. On the physical sheet,
$S^{(1)}(z)=\bar z(w_1(z))$ is the analytic continuation
of $\bar z$ away from the boundary. This function is
meromorphic outside the droplet. Apart from
a cut between the branching points
$z_{1,2}$, the sheets also meet at
the double point $z_*=-\bar\gamma+a^{-1}re^{2i\phi}$,
where $S^{(1)}(z_*)=S^{(2)}(z_*), \phi = \arg (ar+\frac{u\bar a}{1-|a|^2})$.

Analyzing singularities of the Schwarz function,
   one  connects parameters  of
the conformal map with the deformation parameters:
\be\la{27}
\left \{
\begin{array}{rcl}

     \gamma  & = &  \frac{\bar u}{\bar a} - \bar u_0,  \\
n\hbar -\bar\alpha  & = &  r^2 -\frac{ur}{a^2},  \\

     \beta & = &  \frac{r}{\bar a} + u_0  + \frac{u \bar a}{1-|a|^2}
\end{array}\right.
\ee
$$
\mbox{Area of the droplet}\sim n\hbar=r^2-\frac{|u|^2}{(1-|a|^2)^2}.
$$

A critical degeneration occurs when the double point merges with a
branching point located inside the droplet
($z_*=z_2$) to form a triple point $z_{**}$.
This may happen on the boundary only.
At this point, the boundary has a $(2,\,3)$
cusp.
   In local coordinates, it is
$x^2\sim y^3$. This is a critical point of the conformal map:
$w'(z_{**})=\infty$.
A critical point
inevitably results from the evolution
at some finite critical area.

     A direct way to  obtain the
complex curve from the conformal map is the following. First,
rewrite (\ref{222}) and (\ref{23}) as
\be\la{22}\left
\{
\begin{array}{lcl}
z-u_0+ar & = & rw+a(z+\bar\gamma)w^{-1}\\
\bar z-\bar u_0+\bar a r & = & rw^{-1}+\bar a(\bar z+\gamma)w,
\end{array}
\right.
\ee
and treat $w$ and $1/w$ as independent variables.
Then impose the condition $w\cdot w^{-1} = 1$. One obtains
$$
\left |\det
\left [
\begin{array}{cc}
z- u_0 + ar & a(z+\gamma) \\
\bar z -\bar u_0 + \bar a r & r
\end{array}
\right ]\right |^2
=\left (\det
\left [
\begin{array}{cc}
r & a(z+\bar\gamma) \\
     \bar a( \bar z+\gamma) & r
\end{array}
\right ]\right ) ^2.
$$
This gives the equation of the curve and in particular $h$,
in terms of $u,\, u_0,\,r,\,a$ and eventually through the deformation
parameters $\alpha,\,\beta,\,\gamma$ and $t$.

The semiclassical analysis gives a guidance for the form of the
recurrence relations.
Let us use an ansatz for the $L$-operator, which resembles
   the conformal map (\ref{222}):
$$
L=r_n \hat w +u_{n}^{(0)} + (\hat w -a_n )^{-1} u_n,
$$
so that
\begin{eqnarray}
(\hat w -a_n ) L = (\hat w -a_n ) r_n \hat w +
(\hat w - a_n ) u^{(0)}_n + u_n, \la{701} \\
L^{\dag}(\hat w^{-1} -\bar a_n ) =
\hat w^{-1} r_n (\hat w^{-1} -\bar a_n ) +
\bar u^{(0)}_n (\hat w^{-1} -\bar a_n ) + \bar u_n \, \la{7111},
\end{eqnarray}
where $\hat w$ is the shift operator $n\to n+1$.

Now we follow the procedure of the previous section. Since the
potential has only
    one pole,  ${\mathcal L}_n$ can be cast into
    $2\times 2$ matrix form. Let us
    apply the lines (\ref{701}, \ref{7111})
    to an eigenvector $(c_n, c_{n+1})$ of a yet unknown operator
${\mathcal L}_n$, and set
   the eigenvalue to be $\tilde z$:
\be\la{94}\left
\{
\begin{array}{ccc}
(z+r_{n-1} a_{n-1}-u^{(0)}_{n})c_{n} & = & r_{n} c_{n+1} +
a_{n-1}(z+\bar \gamma_{n-1})c_{n-1} \\
(\tilde z +  r_{n}\bar a_{n} -
\bar u^{(0)}_{n+1} )c_{n} & = &
\bar a_{n+1}(\tilde z + \gamma_{n+1}) c_{n+1}  + r_{n} c_{n-1}.
\end{array}\right.
\ee
We have defined $\bar \gamma_n = \frac{u_n}{a_n} - u^{0}_n$.
The equations are compatible if $c_{n-1}$ and $c_{n+1}$ found
through $c_n$
    differ by the shift $n\to n+2$. We have
\be \nonumber
c_{n+1} = \frac{c_n}{d_n} \det \left | \begin{array}{cc}
z+r_{n-1} a_{n-1} -u^{(0)}_{n}&  a_{n-1}(z+\bar \gamma_{n-1}) \\
\tilde z + r_n\bar a_{n} - \bar u^{(0)}_{n+1} & r_{n}
\end{array} \right | = c_n \frac{\widetilde{{\mathcal D}}_{n}}{d_n},
\ee
\be \nonumber
c_{n-1} = \frac{c_n}{d_n} \det \left | \begin{array}{cc}
r_{n} & z + r_{n-1} a_{n-1} -u^{(0)}_{n}  \\
\bar a_{n+1}( \tilde z +
\gamma_{n+1}) & \tilde z + r_n\bar a_{n} - \bar u^{(0)}_{n+1}
\end{array} \right | = c_n \frac{{\mathcal D}_n}{d_n},
\ee
where
\be \nonumber
d_n =
\det \left | \begin{array}{cc}
r_{n}  & a_{n-1}(z+\bar \gamma_{n-1})   \\
    \bar a_{n+1} (\bar z +\gamma_{n+1}) & r_{n}.
\end{array} \right |
\ee
This  yields  the curve
\be\la{811}
\widetilde{\mathcal D}_n\cdot {\mathcal D}_{n+1}=d_nd_{n+1}.
\ee
Comparing the two forms of the curve
(\ref{8}) and (\ref{811}),
    we obtain the  conservation laws of growth:
\be \nonumber
\gamma =\gamma_n = \frac{\bar u_n}{\bar a_n}-\bar u^{0}_n,
\ee
\be\nonumber
\beta  = \frac{r_n}{\bar a_{n+1}} + u^{(0)}_{n+1} + \frac{u^{(0)}_{n+1}
a_n \bar a_{n+1}}{1-a_{n}\bar a_{n+1}},
\ee
\be \nonumber
     n \hbar -\bar \alpha= r_n r_{n+1} - \frac{r_{n+1}u_{n+1}}{a_{n}a_{n+1}} .
\ee
They are the quantum version of (\ref{27}).

\subsection{Continuum limit and conformal maps}

The geometrical meaning of the complex curve (\ref{qc}) is
 straightforward: at fixed shape parameters $t_k$ and area parameter $\hbar$, 
increasing $n$ yields 
growing domains that represent the support of 
the corresponding $n \times n$ model. 
A remarkable feature of
this process is that it preserves the external harmonic moments of the domain  $\mathbb{D}_n$, 
\beq \la{moments}
t_k(n) = t_k(n-1), \quad t_k(n) = -\frac{1}{\pi k}\int_{\mathbb{C} \setminus \mathbb{D}_n} \frac{d^2 z}{z^k}, \quad k \ge 1.
\eeq
The only harmonic moment which changes in this process is the normalized area  $t_0 = \frac{1}{\pi} \int d^2 z,$
and it increases in increments of $\hbar$ (hence the meaning of $\hbar$ as quantum of area).  We may say that the growth 
of the NRM ensemble consists of increasing the area of the domain by multiples of $\hbar$, while preserving all the other
external harmonic moments.  The continuum version of this process, known as $Laplacian$ $Growth$, is a famous problem of complex analysis. It arises in the two-dimensional hydrodynamics of  two non-mixing fluids, one inviscid and the other viscous, upon neglecting the effects of surface tension, where it is known as the Hele-Shaw problem. The following chapters discuss this classical problem in great detail.  

As we will see, Laplacian growth can be restated simply as a problem of finding the uniform equilibrium measure, subject to constraints on the total mass, and the asymptotic expansion of the logarithmic potential at infinity. As long as a classical solution exists, the machinery of NRM does not seem necessary. However, Laplacian growth (as a class of processes), is characterized by finite-time singularities. In that case, the only way to reformulate the problem is similar to the Saff-Totik approach to the extremal measure, and is deeply related to weighted limits of orthogonal polynomials in the complex plane. 

\section{Laplacian Growth} \la{third}

\subsection{Introduction}

Laplacian growth (LG) is defined as the motion of a planar 
domain, whose boundary velocity is a gradient of the Green
function of the same domain (also called a {\it harmonic
measure}). This deceivingly simple process appears to be connected 
to an impressive number of non-trivial physical and mathematical
problems \cite{Oxf98,MWZ}. As a highly unstable, dissipative,
non-equilibrium, and nonlinear phenomenon, it is famous for
producing different universal patterns \cite{ST, Ristroph}.

Numerous non-equilibrium physical processes of apparently
different nature are examples of Laplacian growth: 
viscous fingering \cite{ST}, slow freezing of fluids (Stefan
problem) \cite{Pelce}, growth of snowflakes \cite{Nakaya}, crystal
growth, amorphous solidification \cite{Langer}, electrodeposition
\cite{Gollub}, bacterial colony growth \cite{BenJacob},
diffusion-limited aggregation (DLA) \cite{DLA81}, motion of a
charged surface in liquid Helium \cite{Zubarev}, and secondary
petroleum production \cite{Bear}, to name just a few.

A major consequence from the current development of the subject is a
discovery of a new and unexpectedly fruitful mathematical structure, which
is capable to predict and explain key physical observations in
regimes, totally inaccessible by any other available mathematical
method.

The first section of this chapter is a brief history of physics
covered by the Laplacian growth.  The second section addresses in detail the exact time-dependent solutions of the Laplacian Growth Equation,  and the last  is a detailed presentation of the analytic and algebraic-geometric structure of Laplacian growth.  

\subsection{Physical background}

\paragraph*{Darcy's law}

In 1856, while completing a hydrological study for the city of Dijon,  H. Darcy noticed that the rate of flow (volume per unit time) $Q$
through a given cross-section, $t_0$, is (a) proportional to 
$t_0$, (b) inversely proportional to the length, $L$, taken between
positions of efflux and influx, and (c) linearly proportional to
pressure difference, $\Delta p$, taken between the same two
levels. In short,
\begin{equation}
\la{Darcy}
Q =  - \frac{k\,t_0}{L}\,\Delta p ,
\end{equation}
where $k$ is a positive constant.  As one can see, Darcy's observation 
coincides with Ohm's law, upon identifying $Q,\,\Delta p$, and $k$ as the total current
through the cross-section $t_0$, the electric potential difference,
and the electrical conductivity, respectively.  Rewriting (\ref{Darcy}) in
a differential form, as for Ohm's law,
we obtain
\begin{equation}
\la{d'arcy}
{\bf v} = -k \nabla p,
\end{equation}
where ${\bf v}$ is the velocity vector field of fluid particles,
properly coarse-grained to assure its smoothness over
infinitesimally small volumes.  Here the kinetic coefficient, $k$,
(the same as in (\ref{d'arcy})), is called a (hydraulic) conductivity 
and can depend on position.  The equation (\ref{d'arcy})
constitutes the Darcy's law in a differential form. For 
homogenous $k$,
\begin{equation} \la{DARCY}
{\bf v} = \nabla (- k p),
\end{equation}
Darcy's law merely states that a flow through uniform porous
media (sand in Darcy's experiments) is purely
potential (no vortices), where the pressure field, $p$, is a
velocity potential up to a constant factor.  Assuming constant $k$ 
and the fluid  incompressible, $\nabla \, {\bf \cdot
\, \bf v} = 0$, we find that pressure $p$ is a harmonic function,
\begin{equation}
\la{LaplaceEq}
\nabla^2 p = 0
\end{equation}
As seen from purely dimensional considerations, the
conductivity $k$ equals 
\begin{equation}
\la{conductivity}
k = C\frac{d^2}{\mu},
\end{equation}
where $d$ is the average linear size of a pore in cross-section, $\mu$ 
is the dynamical viscosity of the fluid under consideration, and the dimensionless coefficient, $C$, is usually small and media-dependent.  (It is of the order of the
density of voids in a given porous medium).

It follows from (\ref{d'arcy}) and (\ref{conductivity}), that if $\mu$
is negligibly small (an almost inviscid liquid), pressure gradients are
also negligibly small, regardless of how fast fluid moves (but still much slower than the velocity of sound in this liquid in order to assure
incompressibility assumed earlier).

\paragraph*{Laplacian growth in porous media}

Assume that a fluid with a viscosity $\mu_1$ occupying a domain
$D_1(t)$ at the moment $t$ pushes another fluid with a viscosity
$\mu_2$ occupying the domain $D_2(t)$ at the same time $t$ through
a uniform porous media. Then the Laplace equation will hold for
both pressures $p_1$ and $p_2$ corresponding to domains $D_1$ and
$D_2$ respectively:
\begin{equation}
\la{LMuskat}
\nabla^2 p_i = 0 \qquad {\rm{in}} \,\,D_i(t),
\end{equation}
where $i = 1,2$. At the interface $\Gamma(t)$, where two fluids
meet (but do not mix), their normal velocities coincide because of
continuity and equal to the normal component $V_n$ of the velocity
of the boundary, $\Gamma(t) = \partial D_1 = - \partial D_2$:
\begin{equation}
\la{vMuskat}
{\bf v_1}|_n = {\bf v_2}|_n = V_n \qquad \rm{at} \,\,\Gamma(t).
\end{equation}
The pressure field $p$ at the interface $\Gamma(t)$ (by the Laplace law)
 has a jump equal to the mean local curvature
$\kappa$ multiplied by the {\it {surface tension}} $\sigma$:
\begin{equation}
\la{pMuskat}
p_1 - p_2 =  \sigma \kappa \qquad \rm{at} \,\,\Gamma(t).
\end{equation}
Unless the local curvature is very high, this surface tension
correction is usually very small, and so is often neglected. If to
supplement the last three equations by boundary conditions at
external walls or/and at infinity (they may include sources/sinks
of fluids either extended or point-like), then the free boundary
problem of finding $\Gamma(t)$ by initially given $D_1$ and $D_2$
is completely formulated.

The process described by (\ref{LMuskat}, \ref{vMuskat}, \ref{pMuskat}) is typical
for various geophysical systems, for instance for petroleum production,
where a less viscous fluid (usually water) pushes a much more viscous
one (oil) toward production wells. This process is
very unstable and most initially smooth water/oil fronts will quickly break
down and become fragmented. 

\paragraph*{The Hele-Shaw cell}

In 1898,  H.S. Hele-Shaw proposed an interesting
way to observe and study two-dimensional fluid flows by using two
closely-placed parallel glass plates with a gap between them
occupied by the fluid under consideration \cite{Hele-Shaw}.   This simple
device appears to be very useful in various investigations and is
now called a Hele-Shaw cell after its inventor.
Remarkably, a viscous fluid, governed in 3D by the Stokes law,
\begin{equation}
\la{Stokes}
\mu \nabla^2 {\bf v} = \nabla p,
\end{equation}
after being trapped in a gap of a width $b$, between the plates of a Hele-Shaw cell, 
obeys Darcy's law (\ref{d'arcy}) with a conductivity equal to $k = b^2/(12\mu)$. The
derivation of the formula
\begin{equation}
\la{HSdarcy}
{\bf v} = -\frac{b^2}{12\mu}\nabla p,
\end{equation}
which is to be understood as a 2D vector field in a plane parallel
to the Hele-Shaw cell plates, is rather trivial and results from
the averaging of (\ref{Stokes}) over the dimension perpendicular to the
plates  \cite{Lamb, LL}).  Thus, displacement of viscous fluid by the
(almost) inviscid one in a Hele-Shaw cell became a major
experimental tool to investigate a 2D Laplacian growth.  Various
versions of 2D Laplacian growth in a Hele-Shaw cell corresponding to different
geometries are shown in Figure~\ref{fig:LG}.
\begin{figure}
\includegraphics[width=12cm]{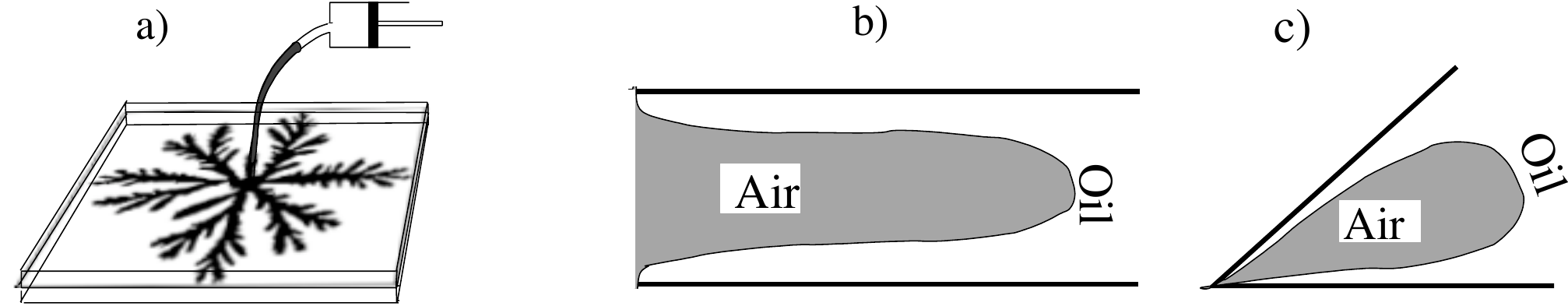}
\caption{\label{fig:LG} Laplacian Growth in a Hele-Shaw cell for the
radial {\it a}), channel {\it b}), and wedge {\it c}) geometries. }
\end{figure}

\paragraph*{Idealized Laplacian growth}

In 1945, Polubarinova-Kochina \cite{PK} and Galin \cite{Galin}
simultaneously, but independently, derived a nonlinear
integro-differential equation for an oil/water interface in 2D
Laplacian growth, after neglecting surface tension, $\sigma$, and
water viscosity, $\mu_{water} = 0$.  Assuming for simplicity a
singly connected oil bubble, occupying a domain $D(t)$ surrounded
by water and having a sink at the origin, $0\in D(t)$, we will
obtain this equation starting from the system
\begin{eqnarray} \la{darcy}
\left\{
\begin{array}{l}
\nabla^2 p = \rho\quad {\rm in} \, D(t),\\
p = 0 \quad {\rm at~ the~interface}, \Gamma(t) = \partial D(t)
\label{eq2}
\\
V_n =- \partial_n p \quad \rm{at~the~interface}, \Gamma(t),
\end{array}
\right.
\end{eqnarray}
where $\rho$ and $\partial_n$ are density of sources and the
normal derivative respectively.   This system is a reduction of
(\ref{LMuskat}, \ref{vMuskat}) after simplifications mentioned
above and using the fact that the normal boundary velocity, $V_n$,
equals to the normal components of the fluid velocity at the boundary,
which is $- \partial_n p$ by virtue of the Darcy law (\ref{d'arcy}).
Here and below the conductivity $k$ is scaled to one.  The density
of sources, $\rho$, in this case equals $\rho(z) = -\delta^2(z)$,
which corresponds to a sink of unit strength located at the origin.

\paragraph*{The Laplacian growth equation}

Coming back to the derivation, we apply the conformal map
from the unit disc in the complex plane $w = \exp(-p + i\phi)$,
where the (stream) function $\phi(x,y)$ is harmonically conjugate
to $p(x,y)$, into the domain $D(t)$ in the "physical" complex
plane $z = x+iy$, and zero maps to zero with a positive
coefficient.  Denoting the moving boundary as $z(t,l)$, where $l$
is the arclength along the interface, one obtains
\be
\la{LG}
V_n = {\rm Im}(\bar z_t z_l) = -\partial_n p = \partial_l \phi,
\ee
It is trivial to see that the chain of three equalities in (\ref{LG})
represent respectively the definition of  $V_n$ in terms of a moving complex
boundary, $z(t,l)$ (the first equality), the kinematic identity
expressed by the last equation in the system (\ref{darcy}) (the second
one), and the Cauchy-Riemann relation (the last one) between $p$ and
$\phi$. After reparamerization, $l \to \phi$, we arrive to the
equation
\begin{equation}
\la{LGE}
{\rm Im}(\bar z_t z_{\phi}) = 1.
\end{equation}
which  possesses
many remarkable properties, as will be seen below.  The equation
(\ref{LGE}) is usually referred as the Laplacian growth equation
(LGE) or the Polubarinova-Galin equation. In \cite{PK,Galin} it was noticed a fully unexpected
feature of the equation (\ref{LGE}): the boundary, $z(t,\phi)$,
taken initially as a polynomial of $w = \exp(i\phi)$, will remain 
a polynomial of the same degree with time-dependent coefficients
during the course of evolution, so new degrees of freedom,
describing the moving boundary, will not appear.

An even more remarkable observation concerning the equation (\ref{LGE}),
belongs to Kufarev \cite{Kuf}, who found that a boundary taken as
a rational function with respect to $w = \exp(i\phi)$ will stay as
such during the evolution.  Moreover, he managed to integrate this
dynamical system explicitly, and found first integrals of motion
associated with moving poles and residues of the conformal map,
$z(t, \exp(i\phi))$, describing the boundary.  The authors
\cite{PK,Galin,Kuf} have however noticed that all the solutions obtained
are short-lived, both because of instability and due to the finite volume
of $D(t)$, which is destined to shrink, because of a sink(s) located
inside.  We will address these interesting observations in detail
in the second section of this chapter.

\paragraph*{LGE in the evolutionary form}

It is of help to present (\ref{LGE}) in the evolutionary form, defined
as the dynamical system, where the time derivative constitutes the LHS
and does enter the RHS.  For this purpose we rewrite (\ref{LGE}) as
$$
{\bar z}_t z_{\phi} = i + t_0,
$$
where $t_0$ is real.  Dividing both sides by $|z_{\phi}|^2$, we will
obtain
$$
\frac{\bar z_t}{\bar z_{\phi}} = \frac{i+t_0}{|z_{\phi}|^2}
$$
Taking conjugate form both sides and multiplying by $i$, we will
have
$$
i\frac{z_t}{z_{\phi}} = \frac{1+it_0}{|z_{\phi}|^2}
$$
The LHS is the analytic function outside the unit disk in the
$w$-plane.  In accordance with the last equation, the real part of
this analytic finction along the unit circumference equals
$|z_{\phi}|^{-2}$.  To recover the analytic function from the
boundary value of its real part at the unit circle is a well-known
procedure involving either the Hilbert transform or the Schwarz
integral.  The result is
\be
\la{eLGE}
i z_t = -
z_{\phi}\,\int_0^{2\pi}\,\frac{e^{is}+e^{i\phi}}{e^{is}-(1+\epsilon)e^{i\phi}}\,
\frac{1}{|z_s|^2}\,\frac{ds}{2\pi},
\ee
where an infinitesimally small positive $\epsilon$ indicates
correct limiting value of the integral while approaching the unit
circumference.  This useful formula was obtained by Shraiman and
Bensimon in 1984 \cite{bs84}.  This  expression for (\ref{LGE})
in the evolutionary form reveals the nonlocal nature of Laplacian
growth due to the integral in the RHS.

The equation (\ref{eLGE}) helps to prove a beautiful statement, that
every singularity of the function $z(t,w)$ moves toward the unit circle
from inside, or in other words the radial component of the 2D velocity
of any singularity of the conformal map is positive.  To prove the claim,
we replace $\phi$ in (\ref{eLGE}) by $W$, defined earlier as $W = -p + i\phi$.  Then after we notice that
$$
-\frac{z_t(t, e^W)}{z_{W}(t, e^W)} = \left [ \frac{dW}{dt}\right ]_{z = const}
$$
and that near a singular point $w=a$ we can replace $W = \log(w) =
\log(a)$, we can rewrite the real part of (\ref{eLGE}) as
\begin{equation}
\la{out0}
\frac{d\,\log\,|a|}{d\,t} = \left [ \frac{1}{|z_w|^2} \right ]_{w = a} > 0.
\end{equation}
Thus, we proved that each singular point of the conformal map
moves toward the unit circle from inside, so the origin is a
repellor for this dynamical system, and the unit circumference
is an attractor.

\paragraph*{Diffusion limited aggregation}

The physics section of the survey cannot be completed without
mentioning a fascinating discovery by T.A. Witten and L.M. Sander,
who observed \cite{DLA81} in 1981 that a cluster on a 2D square
lattice, grown by subsequent attaching to it a Brownian diffusive
particles, eventually becomes a self-similar fractal (see Figure~\ref{dla})
with a robust universal fractal (Haussdorff) dimension given by
\begin{equation}
D_0 = \lim_{\epsilon \to 0}\,\frac{\log
N(\epsilon)}{\log(1/\epsilon)} = 1.71 \pm .01,
\end{equation}
where $1/\epsilon$ is a linear size of a cluster measured by a
small ``yard stick'', $\epsilon$, and $N(\epsilon)$ is a minimal
number of (small) boxes with a side $\epsilon$, which covers the
cluster.
\begin{center}
\begin{figure}
\includegraphics[width=5cm]{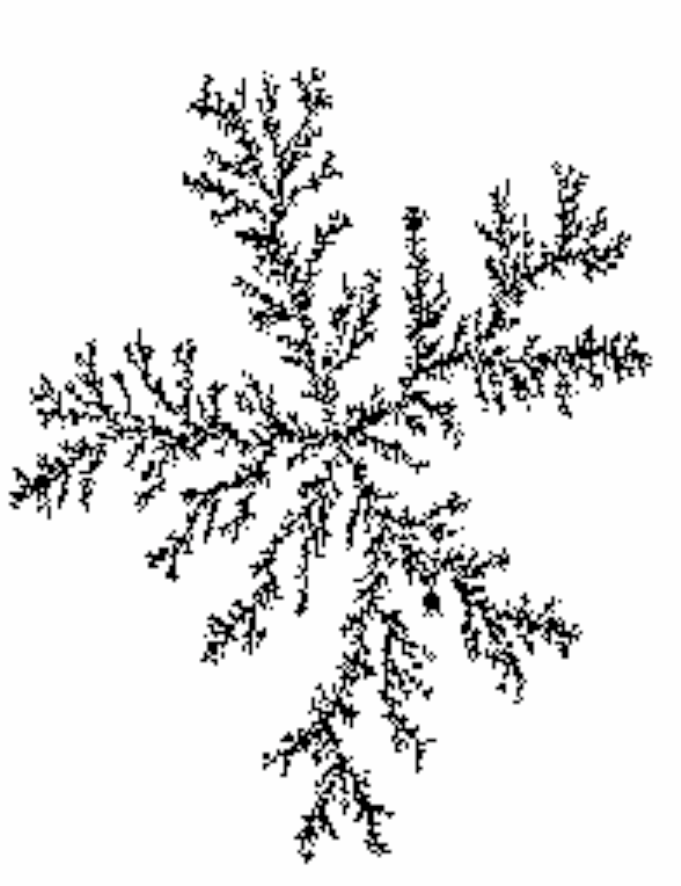}
\caption{\label{dla} A DLA cluster, $n=100 000$.}
\end{figure}
\end{center}
Remarkably, this fractal appeared to be self-similar
after appropriate statistical averaging.  This means that its
higher multi-fractal dimensions, $D_q$, defined as
\begin{equation}
D_q = \lim_{\epsilon \to 0}\,\frac{\log( \sum_{i}^N
p_i^q)}{\log(1/\epsilon)},
\end{equation}
where $p_i$ stands for a portion of a tiny box of a size of
$\epsilon$, covered by the cluster under consideration, appear to
be equal to each other, and to $D_0$, which is 1.71, as indicated
above.  Later, these findings were significantly clarified and
refined in many respects, but the major challenge: how to
calculate the universal dimension defined above still is an open
question (see a relatively recent review \cite{Halsey} and references therein).

This problem is tightly connected to the
Laplacian growth.  Until very recently there were numerous claims
that the DLA process is drastically different from the Laplacian
growth, and even statements appeared that the DLA and fractals
grown in Laplacian growth belong to different universality classes
\cite{Procaccia}.  However, the recent experiments by Praud and Swinney
\cite{HLS-1} made crystal clear that the multi-fractal spectrum of
a cluster grown in a viscous fingering process in a Hele-Shaw cell
(that is a Laplacian growth)
coincides with the DLA
spectrum up to the margin accuracy of  $1\%$, which is the maximal
accuracy in these measurements.  Thus, despite of its discrete
and a stochastic nature, DLA can be understood by a
continuum and deterministic Laplacian growth  (\ref{darcy}).

\paragraph*{Related problems}

Below is a list of physical problems  connected with Laplacian growth.

First, there is the so-called "singular" Laplacian growth,
where a growing domain consists of needles with zero areas and
divergent curvature at the tip.  The mathematical description for
this dynamics should be reformulated, since the gradient of
pressure $p$ diverges near moving needle tips, so the boundary
velocity should be replaced by an appropriately regularized law.
Interesting works by Derrida and Hakim \cite{DH}, and by Peterson
\cite{Peterson} in this direction deserve special attention.

There is also a considerable amount of works in so-called
nonlinear mean-field dynamics, where a phase field is involved, which
gradually changes from unity in one of moving phases toward zero
inside the second one \cite{CahnHill,Gollub-Langer}.  Many of these
processes, including dynamics of
miscrostructure \cite{Khachaturyan} in materials, growth of
bacterial colonies in nutritional environment \cite{Matsushita},
and spinodal decomposition \cite{SpinDec}, governed
by the time-dependent Ginzburg-Landau and the Cahn-Hilliard
equations, are reduced to the Laplacian growth interface dynamics
in a special singular limit, when the phase field degenerates to a
step-function, thus becoming a characteristic function of a moving
domain with a well defined boundary \cite{caginalp,Langer-1}.  This
is certainly worth to mention, both because it significantly
enriches a physical process by introducing an additional field
(the phase field) and since this is conceptually related to a
random matrix approach to Laplacian growth, addressed in the survey, and where a distribution of
eigenvalue support will play a role of a mean-field phase,
introduced in this paragraph.

Let us also mention several more ``selection puzzles", which belong
to the Laplacian growth in various settings:  
selection of a shape of a separated inviscid bubble, observed by
Taylor and Saffman in a viscous flow in a rectangular Hele-Shaw
cell \cite{TS} from a continuous family of possible solutions (not
to be confused with the Saffman-Taylor fingers family described in
\cite{ST}); selection of a so-called ``skinny'' finger in a
Hele-Shaw cell accelerated by a tiny inviscid bubble near the nose
of a finger \cite{Libch}; and prediction of the periodicity
for the so-called side-branching structure in dendritic
growth \cite{Glicksman}. These phenomena have
the same (or almost the same) mathematical description.

Another important comment about physics of Laplacian growth is
that  Darcy's law (\ref{d'arcy}) is invalid near walls of a Hele-Shaw
cell, including proximity to both parallel plates. This is because
averaging of the Stokes flow, $\mu \nabla^2 {\bf v} = \nabla p$,
given by (\ref{Stokes}) will no longer bring us to (\ref{HSdarcy}), due to 
boundary
layer effects. This apparent difficulty gives rise to the study of
an interface dynamics with a Stokes flow, which is an extension of
the Hele-Shaw (Darcy's) flow.  The Stokes flow also contains
remarkable physics and beautiful
mathematics \cite{Hopper,Rich,Crowdy}, which is still yet to
be fully understood.  

\subsection{Exact solutions}

\paragraph*{Cardioid}

Consider the equation of
motion for the droplet boundary under Laplacian growth 
\begin{equation}
\label{eq5} {\rm Im}(\bar z_t z_{\phi}) = Q,
\end{equation}
where $2\pi Q$ stands for a rate of a source (sink). Here,
$z(t,e^{i\phi})$ is conformal inside the unit circle, $|w|<1$, $0
\to 0$, and $w = e^{i\phi}$ in the equation.  When one
tries to solve (\ref{eq5}), the solution,
\begin{equation}
\la{eq6}
z = r(t)e^{i\phi},
\end{equation}
comes to mind first, as the simplest one.  It describes initially
circular droplet centered at the origin, which uniformly grows
(shrinks) while continuing to be a circle. Indeed, substituting
(\ref{eq6}) into (\ref{eq5}) one obtains
\begin{equation}
\la{eq7}
r(t) = \sqrt{2(|Q|T + Q t)},
\end{equation}
where a constant of integration, $T$ stands for an initial time.
When $Q<0$ (suction), the circle shrinks to a point at $t = T$,
and the solution (\ref{eq6}) ceases to exist after $T$.  Could
one find any other exact solutions, less trivial than given by
(\ref{eq6})?

Remarkably, the answer is yes, despite of nonlinearity of the
Laplacian growth equation (\ref{eq5}).  Let's add to (\ref{eq6})
an initially small quadratic correction,
\begin{equation}
\la{eq8}
z = r(t)e^{i\phi} + a(t)e^{2i\phi}.
\end{equation}
The domain bounded by the curve described by (\ref{eq8}), named {\it a cardiod}, is 
connected if $|a| < r/2$.  Substituting (\ref{eq8}) into (\ref{eq5}) one obtains
two coupled nonlinear first order ODEs w.r.t. $r$ and $a$:
\begin{eqnarray}
\left\{
\begin{array}{l}
r\,\dot r + 2\,a\,\dot a = Q\\
\dot a\,r + 2\,a\,\dot r= 0,
\end{array}
\right.
\end{eqnarray}
with an easily found solution
\begin{eqnarray}
\left\{
\begin{array}{l}
a r^2 = a_0\\
r^2 + 2 a^2 = 2(|Q|t_0 + Q t)
\end{array} \right.
\end{eqnarray}
with $a_0$ and $t_0$ as constants of integration.  If $Q>0$
(injection), the cardiod will grow becoming more and more like a
circle during the evolution.  If instead $Q<0$ (suction) the
cardioid (\ref{eq8}) shrinks, deforms and ceases to exist after $t^* =
t_0 + 3 a_0^{2/3}/( \sqrt[3]{16}Q)$. This happens when the
critical point of the conformal map given by (\ref{eq8}) reaches the unit
circle from outside. Then the cardioid ceases to be analytic and
earns a needle-like cusp (a point of return with infinite
curvature).  This cusp is called type 3/2 (alternatively (2,3)-cusp) because in
local Cartesian coordinates it is described by the
equation $y^2  \sim x^3$.  We will
see later that this kind of cusps is typical for those solutions
of Laplacian growth which cease to exist in finite time.

\paragraph*{Polynomials}

As a generalization, we are going to prove now that all
polynomials of $w$, which describe boundaries of analytic domains
when $|w|=1$, are solutions of (\ref{eq5}).
Assume a droplet
is initially described by a trigonometric polynomial (with all
critical points lying outside the unit disk, because its interior
conformally maps onto a droplet):
\begin{equation}
\la{eq10}
z = \sum_{k=1}^N\,a_k e^{ik\phi}.
\end{equation}
Substituting (\ref{eq10}) into (\ref{eq5}), one obtains $N$ coupled ODE's for
time-dependent coefficients $a_k$, and remarkably there are no
other degrees of freedom which appear during the evolution.  In other
words, the evolving droplet will continue to be described by the
polynomial (\ref{eq10}), with coefficients, $a_k$, changing in time in
accordance with these ODE's:
\begin{equation}
\la{eq11}
\sum_{k=1}^{N-n}\,[k\,a_k\,{\dot{\bar a}}_{k+n} + (k+n)\dot a_k
\,{\bar a}_{k+n}] = Q\delta_{n,0}\qquad n = 0, 1, \ldots ,N-1.
\end{equation}
Moreover, (\ref{eq11}) can be integrated explicitly.  Indeed, we notice
first that the equation for $k=N-1$, namely
\begin{equation}
a_1\,{\dot{\bar a}}_N + N\dot a_1 \,{\bar a}_N = 0,
\end{equation}
is trivially solved with the answer
\begin{equation}
\la{N}
{\bar a}_N\,a_1^N = C_N,
\end{equation}
where $C_N$ is the constant of integration. Substituting (\ref{N})
into the $(N-2)^{nd}$ equation, which has a form
\begin{equation}
\la{N-1}
a_1\,{\dot{\bar a}}_{N-1} + (N-1){\dot a_1} \,{\bar a}_{N-1} +
a_2\,{\dot{\bar a}}_N + N{\dot a_2} \,{\bar a}_N = 0,
\end{equation}
we notice that the LHS of (\ref{N-1}) is
proportional to a full derivative from the expression
$$ a_1^{N-1}\,{\bar a_{N-1}} + N\,C_N\frac{a_2}{a_1^2}$$
and is zero in accordance with the RHS of (\ref{N-1}).  Thus we obtain
\begin{equation}
a_1^{N+1}\,{\bar a_{N-1}} + N\,C_N\,a_2= C_{N-1}\,a_1^2,
\end{equation}
where $C_{N-1}$ is a constant of integration.  Knowing $a_{N-1}$
and $a_N$ in terms of $a_1$ and $a_2$ we can easily integrate the
third equation from the end of the system (\ref{eq11}), namely the
$(N-3)^{rd}$ equation.  The result is

\begin{equation}
a_1^{N+2}\bar a_{N-2} + (N-1)C_{N-1}a_2a_1^2 + NC_Na_3a_1 -
N(N+1)\frac{C_Na_2^2}{2} = C_{N-2}a_1^4.
\end{equation}

Continuing in this way, we obtain an explicit dependence of ${\bar
a_k}$ as a linear combination of constants of motion, $C_k$, with
coefficients which are polynomial forms w.r.t. $a_1, a_2, \ldots$.
The equation for $n=0$ from (\ref{eq11}) already constitutes the full
derivative and, as such, is trivially integrated:
\begin{equation}
\sum_{k=1}^N k|a_k|^2 = 2(C_0 + Qt),
\end{equation}
where $C_0$ is a constant of integration.  Here the LHS is a
(scaled) area of the droplet, and the equation states that the
area changes linearly in time.  In other words, we integrated the
system (\ref{eq11}), and the solutions are polynomial forms with respect
to $a_k$, linear w.r.t. integrals of motion, $C_k$, explicitly
obtained.

As in the case of cardioid, in case $Q>0$ the dynamics is stable
and the droplet becomes eventually more and more round since all
$a_k$ decay in time, but $a_1$ in contrary, grows, as one can
easily verify by looking to the system (\ref{eq11}).  If $Q<0$, 
then the droplet shrinks and the solution ceases to exist in finite 
time.  This happens because a critical point(s) hits a unit circle 
from outside manifesting a break of analyticity by making a cusp 
(of a 3/2 kind in general case).  Except such rare cases as a circle
centered at the location of sink, the solution stops to exist
prior to the formation of a cusp, because of the droplet 
being completely sucked by the sink.

The fact that a finite time singularity (a cusp) is generic follows
directly from (\ref{N}): since the conformal radius, $a_1$, should
decrease as the area shrinks, then the coefficient, $a_N$, grows
in time by virtue of (\ref{N}),  eventually bringing the system
to a cusp.

Now consider the external Laplacian growth, where an inviscid
bubble, surrounded by a viscous fluid grows (shrinks) because of a
source (sink) at infinity.  Then we map conformally the exterior of
the unit disk in the $w$-plane to the exterior of a bubble
(viscous region) in the physical $z$-plane with a simple pole and
positive residue (which is a conformal radius) at infinity.

Here an analogy of the polynomial ansatz (\ref{eq11}) will be the formula
\begin{equation}
\la{neg}
z = \sum_{k=-1}^N\,a_k e^{-ik\phi},
\end{equation}
where $a_{-1} = r$ is the conformal radius, that is the radius of
a circle perturbed by the rest of $a_k$'s.  This case is also
integrable in a way, very similar to the interior case shown above
\cite{Mineev90}.  One can also see that for an unstable LG, that is a
growing bubble in the exterior problem, a finite time cusp is
unavoidable.  Indeed, the system of ODEs for the ansatz (\ref{neg}) will
look the same as (\ref{eq11}), but with values of $n$ extended from $-1$
to $N+1$.  Thus one can easily see that $a_N = C_N r^{N-1}$.  This
means that the highest harmonic will grow faster than a conformal
radius, which should eventually break domain's analyticity through
a cusp \cite{Howison91}.

The area, $t_0$, of the growing bubble in this case equals
\be \la{area}
t_0 = |r|^2 - \sum_{k > 0} k |a_k|^2.
\ee

Consider the simplest non-trivial example for (\ref{neg}), 
which describes a shape with three-fold symmetry,
\be \la{ellipse}
z(w) = rw + \frac{a}{w^2},
\ee
we have
\be
a = 3C r^2,
\ee
where $C$ is a constant of integration, and the scaled area of the droplet
identified with time $t$ in this case, is:
\be
t_0=t = |r|^2 - 18|C|^2 |r|^4.
\ee
Clearly, this polynomial in $|r|^2$ has a global maximum at $r_c$ solving $36 |r_c|^2 |C|^2 = 1$. We call the corresponding value of the area $t_c = |r_c|^2/2$ {\emph{critical}}, and conclude that the dynamics will lead to finite-time singularities for any initial condition $t_3 \ne 0$ Figure \ref{critical}.

\begin{figure}[ht!]
\begin{center}
\includegraphics*[width=7cm]     {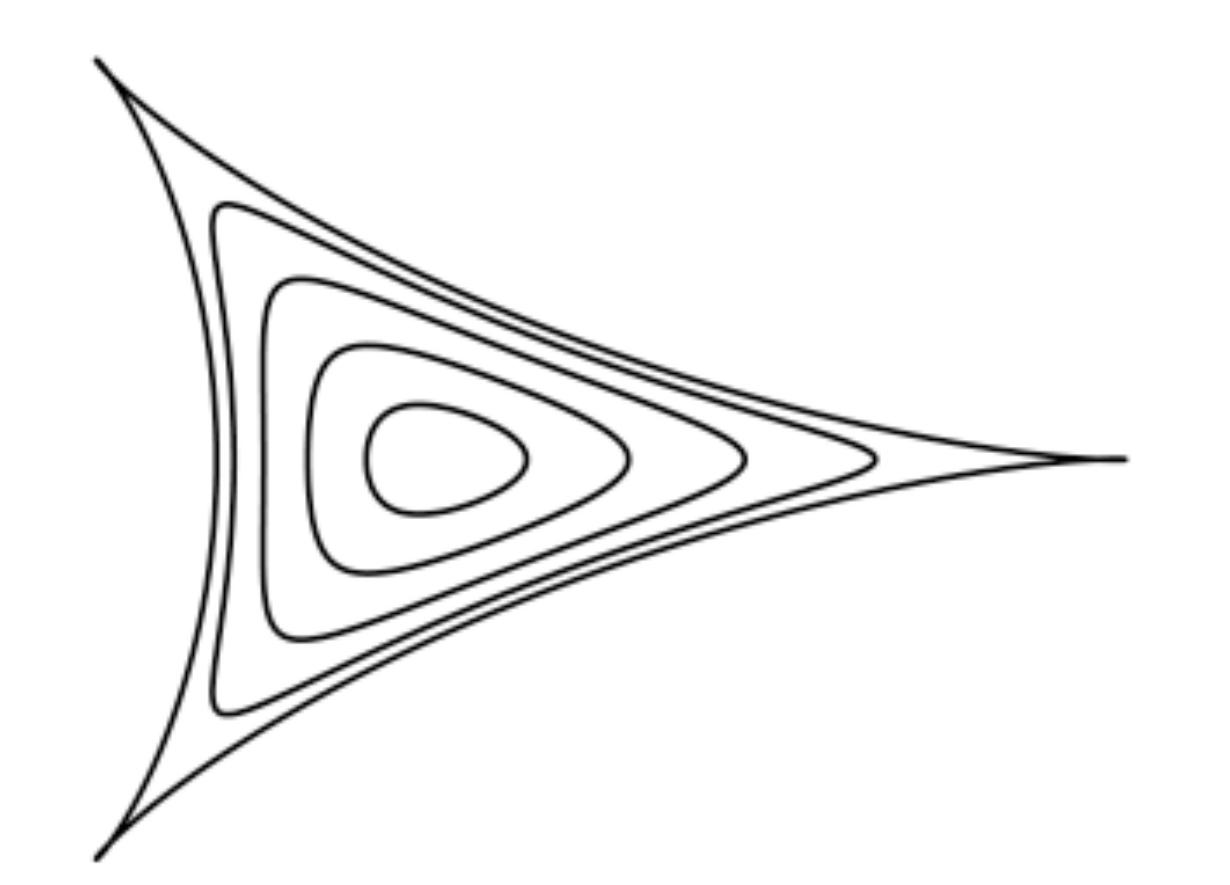}
\caption{A Hele-Shaw droplet approaching the critical area.}
\label{critical}
\end{center}
\end{figure}

In summary, we have shown that the Laplacian growth is integrable
for polynomial (time-dependent) conformal mappings both in
interior and exterior problem, and in both cases  
(growing of a bubble in the exterior and suction of a
droplet for the interior problem) finite time singularities in the
form of the 3/2-cusps are unavoidable.  This is caused by an
ill-posedness of the Laplacian growth without regularized factors,
such as surface tension.

\paragraph*{Rational functions}

Kufarev \cite{Kuf} found a class of
rational solutions of the equation (\ref{eq5}) with simple poles. Let us
show that all rational conformal maps from exterior (interior) of
the unit disk to the exterior (interior) of a domain D are
solutions of the LGE (\ref{eq5}). We will include in our proof multiple
poles for the sake of generality. Specifically, we claim that the
expression
\begin{equation}
\la{rat}
z = rw +
\sum_{k=1}^N\sum_{l=1}^{P_k}\,\frac{A_{kl}}{(w-a_k)^{p_{kl}}},
\end{equation}
where $p_{kl} \in \mathbb{N}$, solves (\ref{eq5}).  Indeed, after
substitution of (\ref{rat}) into (\ref{eq5}), putting $w=e^{i\phi}$, 
we obtain a double sum, which we can decompose to elementary fractions with
respect to $(e^{i\phi}-a_k)^{p_{kl}}$ by using repeatedly the identity
\begin{equation}
\frac{1}{(e^{i\phi}-a_k)(e^{-i\phi}-{\bar a_l})} =
\frac{1}{1-a_k\,{\bar a_l}}\, \left (1 + \frac{a_k}{e^{i\phi} - a_k} -
\frac{{\bar a_l}}{e^{-i\phi}- {\bar a_l}} \right )
\end{equation}
Equating coefficients prior to all independent modes to zero and
sum of all constants (the zeroth mode) to $Q$ in accordance with
(\ref{eq5}), we see, after some algebra, that all the
expressions are full derivatives, and after integration we obtain
the following equations:
\begin{eqnarray}
\la{rat.sol}
\left\{
\begin{array}{l}
r^2-\sum_{k=1}^N\sum_{l=1}^{P_k}\,A_{kl}[{\bar z}(1/{\bar
a_k})]^{(p_{kl}-1)}/(p_{kl}-1)! = C + Qt \\
z(1/{\bar a_k}) = \beta_k\qquad k=1,2,\ldots,N\\
A_{kl} = \alpha_{kl}[z_w(1/w)]^{p_{kl}}|_{w={\bar a_k}}\qquad
k=1,2,\ldots,N; \quad l=1, \ldots,P_k,
\end{array}
\right.
\end{eqnarray}
where $C$, $\alpha_{kl}$, and $\beta_k$ are constants of
integration.  It is possible to show that in unstable LG, that is
a an exterior problem with growth or an interior problem with
shrinking, all the solutions (\ref{rat}) blow up in finite time by
forming cusps, generally of the 3/2 kind.  

Another interesting class of rational solutions was also found by
Kufarev \cite{Kuf} in case there are several sources instead of
one, located at $z_k$ with rates $Q_k$, and $k=1,2,\ldots,N$.  In
this case the velocity potential (scaled pressure) diverges near
$z_k$ logarithmically with coefficients $Q_k$:
\begin{equation}
-p = Q_k\,\log(z-z_k) + {\rm regular\,\,terms} \qquad ({\rm
when}\,\,z \to z_k).
\end{equation}
In this case the Laplacian growth equation has a form
\begin{equation}
{\rm Im}(\bar z_t z_{\phi}) = {\rm Re}\sum_{k=1}^M
\frac{Q_k}{1-b_k(t)e^{i\phi}},
\end{equation}
where $b_k(t)$ are time dependent inverse conformal pre-images of
sources locations, $z_k$, so that
\begin{equation}
\la{A}
z_k = z({\bar b_k}^{-1}) \qquad k=1, \ldots , M.
\end{equation}
In this case, the most general rational solution has a form
\begin{equation}
\la{ist}
z = rw +
\sum_{k=1}^N\sum_{l=1}^{P_k}\,\frac{A_{kl}}{(w-a_k)^{p_{kl}}} +
\sum_{k=1}^M\frac{B_k}{w-b_k}
\end{equation}
The result of integration  is then  given by (\ref{rat.sol}), where summations
incorporate the last sum in (\ref{ist}), the equation (\ref{A}), and
\begin{equation}
B_k = C_k \,t\, [z_w(1/w)]|_{w={\bar b_k}} \qquad k=1,2,\ldots,M,
\end{equation}
where $C_k$ are additional constants of motion.  It is worth to
mention that even if the initial configuration $z(0,e^{i\phi})$
does not include poles at $b_k$ (knows nothing about sources $Q_k$
at $z_k$), the solution earns terms with simple poles at $b_k$
immediately from the start, as one can see from the last equation.

Thus, the singularities of any solution can be split into those
imposed by the source location ($b_k$ in our case) and  those
determined by initial configuration, that are $a_k$.

One should also beware that the interface can reach sources during
evolution, thus breaking analyticity by forming a cusp, and after
this moment a solution ceases to exist.

\paragraph*{Logarithms}

In the paper \cite{KufVinog} Kufarev and Vinogradov have found a
logarithmic class of solutions of (\ref{LGE}), which was later
rediscovered and studied in detail by several authors
\cite{S59,BP,H86,ms,sm}.  This class appeared to be
particularly fruitful from both mathematical and physical points
of view: besides providing a significant extension from rational
solutions, the logarithmic ones are often free of finite time
singularities for an unstable exterior problem, which is the most
important for physics. Existence of these solutions for all times
allows to study the long time asymptotics, which is perhaps the
major goal of this research. These so-called multi-logarithmic
solutions have a form
\begin{equation}
\la{logr}
z = rw + \sum_{k=1}^N \alpha_k \log \left ( \frac{w}{a_k}-1 \right ),
\end{equation}
where $r\,\alpha_k,\,a_k$ are parameters (some of them are time
dependent), and $|a_k|<1$ for a conformal mapping from the
exterior of the unit disk. Using the method outlined above for
rational solutions one could easily figure out that (\ref{logr})
satisfies the LGE (\ref{LGE}) with all $\alpha_k$ to be constants
in time and the following time dependence of $r$ and $a_k$:
\begin{eqnarray}
\left\{
\begin{array}{l}
\la{eq12}
r^2+\sum_{k=1}^N\sum_{l=1}^N\,\alpha_k\,{\bar
\alpha_l}\log(1-a_k\,{\bar a_l}) = C +
Qt\\
\\
r/{\bar a_k} + \sum_{l=1}^N\,\alpha_l\,\log(\frac{1}{a_l{\bar
a_k}}-1) = \beta_k; \qquad  k=1,2,\ldots,N,
\end{array}
\right.
\end{eqnarray}
where $C$ and $\beta_k$ are constants of motion.  It is less
trivial to show that the solutions (\ref{logr}) may be free of finite time
singularities, but the following example illustrates it well:
let's impose a $\mathbb{Z}_N$ symmetry over the system (\ref{logr}) by setting
$\alpha_k = \alpha\,\exp{(2\pi i k /N)}$ and $a_k = a\,\exp{(2\pi
i k /N)}$, with positive $a$ and $\alpha$.  Then (\ref{eq12}) looks
significantly simpler:
\begin{eqnarray}
\left\{
\begin{array}{l}
\la{eq13}
r^2+N\alpha^2\sum_{k=1}^N\,\gamma_k\log(1-a^2\,\gamma_k) = C + Qt\\
\\
r/a +
\alpha\sum_{k=1}^N\,\gamma_k\,\log(\frac{1}{a^2\,\gamma_k}-1) =
\beta,
\end{array}
\right.
\end{eqnarray}
where $\gamma_k = e^{2\pi i k/N}$. Equating the derivative of (\ref{logr})
to zero, we find critical points, $b_k$.  As expected, $b_k = b\,\gamma_k$ and
\begin{equation}
b^N = a^N - \frac{\alpha N a^{N-1}}{r}.
\end{equation}
Assuming the initial $b$ to be positive, we see that ${\dot b}$ is
always positive, if $\dot a$ is, which is always the case, since
as follows from the second equation in (\ref{eq13})
\begin{equation}
{\dot r}=\left (\frac{r}{a}+2\alpha N\frac{a^{2N-2}}{1-a^{2N}} \right ){\dot a},
\end{equation}
and therefore ${\dot a}$ is positive, since ${\dot r}$ is. Let's
also notice that $a$ cannot reach $1$ since this would make the 
RHS of the second equation in (\ref{eq13}) infinite, which would
contradict to the fact that it is a finite constant.  Thus, from
\begin{equation}
0<b<a<1, \qquad {\dot a}>0, \qquad {\rm and} \qquad {\dot b}>0
\end{equation}
it follows that critical points and singularities of the conformal
map (\ref{logr}) will always stay inside the unit circle, which guarantees
existence of the solution (\ref{logr}) for all times.  This simple example
illustrates the fact that many of these solutions are free of
finite time-singularities (see details in \cite{ms,sm}), but the
interesting problem of comprehensive classification of initial
data for the solutions (\ref{logr}) which do not blow up in finite time
still is an open question.

\subsection{Mathematical structure of Laplacian growth}

\subsubsection{Conservation of harmonic moments}

The following remarkable property of the Laplacian growth was
found by S. Richardson in 1972 \cite{Richardson72} for a
point-like source Q at the origin, for which

$$ \nabla^2 \phi = \frac{Q}{2\pi}\,\delta^2(z). $$

He showed that all positive harmonic moments of the viscous
domain, $D(t)$, 
\begin{equation}
C_k = \int_{D(t)}\,z^k\,dx\,dy, \qquad k=1,2,\ldots
\end{equation}
do not change in time, while the zeroth moment, which is the area
of the growing bubble, changes linearly in time:
\begin{equation}
\frac{d C_k}{d t} = \oint_{\partial D(t)}\,z^k\,V_n\,\frac{dl}{\pi} =
\oint_{\partial D(t)}\,(p\,\partial_n z^k - z^k\,\partial_n p)\,
\frac{dl}{\pi},
\end{equation}
($dl$ is an element of arclength) because $V_n = -\partial_n\,p$
and $p=0$ along the boundary $\partial D(t)$.  By virtue of 
Gauss' theorem, it equals
\begin{equation}
\int_{D(t)} \nabla(p \nabla z^k - z^k \nabla
p)\,\frac{dx\,dy}{\pi} = Q\,\delta_{k,0}.
\end{equation}
This property may be used as the definition of the idealized
Laplacian growth problem, namely to find an evolution of the domain
whose area increases in  time, while all positive harmonic
moments do not change.

\subsubsection{LG and the Inverse Potential Problem}

One can easily notice that the harmonic moments are the
coefficients of the (negative) power expansion of the so called
Cauchy transform ${\cal C}_{D}(z)$ of the domain $D$, namely
\begin{equation}
\la{CT}
{\cal C}_{D}(z) = \frac{1}{\pi}\int_{D}\,\frac{dx'\,dy'}{z -
z'} = \sum_{k=0}^\infty\,\frac{C_k}{z^{k+1}}.
\end{equation}
Since the Cauchy transform, ${\cal C}_{D}(z)$, is
the derivative of the Newtonian potential $\Phi(z)$ created by
matter occupied the domain $D$ with a unit density,
\begin{equation}
\la{IPP}
\Phi(z) = \int_{D}\,\log|z-z'|\,\frac{\,dx'\,dy'}{\pi},
\end{equation}
we see a deep connection between the Laplacian growth with
the so-called inverse potential problem, asking to find a domain
$D$ occupied uniformly by matter which produces a given far field
Newtonian potential.  The harmonic moments in this context are
multipole moments of this potential. If the domain $D(t)$ grows in
accordance with the idealized Laplacian growth, then the potential
$\Phi(z)$ changes linearly in time, so (up to a constant):
\begin{equation}
\Phi(t,z)= \frac{Q t}{2 \pi}\,\log|z|,
\end{equation}
which is a potential of a point-like (increasing in time) mass at the origin.

\subsubsection{Laplacian growth in terms of the Schwarz function}

Let $F(x,y)=0$ define an analytic closed Jordanian contour
$\Gamma$ on the plane. Replacing the cartesian coordinates, $x$
and $y$, by complex ones, $z = x+iy$ and ${\bar z} = x-iy$, one
obtains a description of $\Gamma$ as
\begin{equation}
F \left ( \frac{z+{\bar z}}{2}, \frac{z-{\bar z}}{2i} \right ) = G(z,{\bar z}) =
0.
\end{equation}
Solving the last equation with respect to ${\bar z}$ one obtains:
\begin{equation}
{\bar z} = {S}(z),
\end{equation}
when $z \in \Gamma$.  The function ${S}(z)$ is
called the Schwarz function of the curve $\Gamma$ \cite{D}. It is the same mathematical object we
encountered in the previous chapter. 
This function  plays an outstanding role in the
theory of quadrature domains (see next chapter).  
It has the following Laurent expansion, valid at least in a strip around the curve $\Gamma$:
\begin{equation}
\la{SE}
{S}(z) = \sum_{k=0}^\infty\,\frac{C_k}{z^{k+1}}
+ \sum_{k=0}^\infty \, kt_k z^{k-1},
\end{equation}
where $t_k$ are the external harmonic moments defined as
\begin{equation}
t_k = \frac{1}{\pi k}\int_{D_-} \, \frac{dx\,dy}{z^k}, \qquad k =
1, 2, \ldots,
\end{equation}
where $D_-$ is the domain complimentary to the domain $D$. 

From (\ref{CT}) and (\ref{SE}) we obtain the connection between the 
Cauchy transform of a domain with the Schwarz function of its boundary:

\be
{\cal C}_D(z) = \oint_{\partial D}\,\frac{{S}(z')}{z-z'} \, \frac{d\,z'}{2\pi i}.
\ee

Rewriting the Laplacian growth dynamics in terms of the Schwarz
function, ${S}(z)$ \cite{Howison91} one obtains
\begin{equation}
\la{lgs}
\partial_t\,{S}(z) = 2\,\partial_z W,
\end{equation}
where $W = -p + i\phi = \log w$ is the complex potential defined
earlier. This last form of the Laplacian growth is very instructive.
In particular, it helps to understand the origin of
constants of integration in all exact solutions of the Laplacian
growth equation presented above as a result of direct integrating
efforts. Indeed, the RHS in the last equation is analytic in the
viscous domain $D(t)$ except a simple pole at the origin (we
consider an internal LG problem with a source at the origin).  In
order for the LHS to satisfy this condition, all the singularities
of ${S}(t,z)$ outside the interface should be constants of motion.
At zero the Schwarz function should have a simple pole with a residue
(which is the area of the domain $D(t)$) linearly changing in
time. This observation can be easily seen as an alternative proof of the
Richardson theorem, stated above.

\subsubsection{The correspondence of singularities}

The Schwarz function is connected to a conformal map $z = f(w)$
from the unit circle to the domain $D$ through the following
formula \cite{D}
\begin{equation}
{S}(z) = {\bar f} \left (\frac{1}{f^{-1}(z)} \right ),
\end{equation}
where $f^{-1}(z)$ is the inverse of the conformal map
$w = f^{-1}(z)$. This formula helps to derive a one-to-one
correspondence between singularities of ${S}(z)$ inside $D$
and $f(w)$ inside the unit circle: if near a singular point
$a$ the conformal map $f(w)$ diverges as
\begin{equation}
f(w) = \frac{A}{(w-a)^p},
\end{equation}
(here by convention $p=0$ stands for a logarithmic divergence),
then the Schwarz function ${S}(z)$ diverges near a point $b =
f(1/{\bar a})$ with the same power, $p$, as
\begin{equation}
{S}(z) = \frac{B}{(z-b)^p},
\end{equation}
where
\begin{equation}
A = \left [ \frac{\bar B}{(-a^2 {\bar f}')^p} \right ]_{w = 1/{\bar a}}.
\end{equation}
$B$ and $b$ are constants of motions as showed above, thus the
last formula together with the relation $b=f(1/{\bar a})$ and the
area linearly changing in time and expressed in terms of the
parameters of $f(w)$ constitute the whole time dynamics of
singularities of $f(t,w)$ \cite{Richardson72, Etingof}.  The 
reader can see the equivalence of these formulae with constants of integration obtained earlier when various classes of exact solutions 
were derived by direct integration.  

\subsubsection{A first classification of singularities}

As mentioned in the previous sections, existence of the singular limit was
established at the same time with the model \cite{PK, Galin}. It became
a fertile field of study in itself, and led to further developments of the
problem \cite{ST,Tanveer}. In a series of papers
\cite{S3, S4, S5, Howison86, Howison85, Hohlov-Howison94},
the possible boundary singularities were studied, as well as the problem of continuing the solutions for certain classes. It was found that, in the free-space set-up, the generic critical boundary features  a cusp at $(x_0, y_0)$, with local geometry of the type
\be
(x-x_0)^q \sim (y-y_0)^p, \quad (p, q) \mbox{ mutual } \mbox{primes.}
\ee
The most common cusp is characterized by $q=2, p=3$, but $q=2, p=5$
can also be obtained fairly easy by choosing proper initial conditions.
Very special situations, where a finite-angle geometry is assumed as
initial condition were also considered \cite{King}.

It was shown be several methods that dynamics can be continued through a
cusp of type $(2, 4k+1), k > 0$ \cite{Howison86, BAZW05}.

\subsubsection{Hydrodynamics of LG and the singularities of Schwarz function}

 As indicated above, the Schwarz function encodes information about the conserved moments $\{ t_k \}$, through its expansion at infinity \cite{MWZ}:
\be \la{expansion}
S(z) = \frac{t_0}{z} + \sum_{k > 0}t_k z^{k-1} + O(z^{-2}).
\ee
This function is useful when computing averages of integrable analytic
functions $f(z)$ over the domain $D_{+}$ (an interior domain):
\be \la{average}
\frac{1}{\pi}\int_{D_{+}} f(z) \dd x \dd y =
\sum_{k=1}^N\sum_{i=1}^{n_k}
c_{ik}f^{(i)}(z_k) + \sum_{m=1}^M \int_{\gamma_m} h_m(z)f(z) \dd z,
\ee
if the function $S$ has poles of order $n_k$ at $z=z_k$ and branch cuts $\gamma_m$
with jump functions $h_m(z)$. Applying formula (\ref{average}) for the characteristic function of the domain $f(z) = \chi_{D_{+}}(z)$ and taking a derivative with respect to $t_0$, we obtain the relation
\be \la{sources}
\frac{\dd \,\,\,\,}{\dd t_0}\left [ \sum_{k} {\rm{Res }} S(z_k) +
\sum_m \int h_m(z) \dd z\right ]= 1,
\ee
which shows that the singularity data of the Schwarz function in $D_+$ can be interpreted as giving the location and strength of fluid sources (isolated or line-distributed) \cite{Richardson72}. Identifying the 2D uniform measure with  another, singular (point or line-distributed) distribution, is referred to as {\emph{sweeping}} of a measure. We will repeatedly encounter this process in the next chapter. In the case when the Schwarz function is meromorphic in $D_+$ (it has only isolated poles as singular points), (\ref{average}) becomes
\be
\frac{1}{\pi}\int_{D_{+}} f(z) \dd x \dd y = \sum_{k=1}^N\sum_{i=1}^{n_k} c_{ik}f^{(i)}(z_k),
\ee
and the domain is called a {\emph{quadrature domain}} \cite{Bell03, Bell04, G1, G2, GSh}.
Generically, the Schwarz function may have branch cuts in $D_+$, in which case $D_+$ is called a
{\emph{generalized quadrature domain}} \cite{Sh}. This is the typical scenario for our problem. The 
rigorous theory of quadrature domains is outlined in the next chapter. 

The hydrodynamic interpretation of the Schwarz function arises from(\ref{lgs}), which is worthwhile to rewrite here
\begin{equation}
\la{darcy2}
\partial_t\,{S}(z) = \,\partial_z W,
\end{equation}
after rescaling by 2.
Let $C$ be some closed contour, boundary of a
domain $B$, and integrate equation (\ref{darcy2}) over it. We obtain
\be \la{re-im}
\p_t \oint_C S(z) \dd z =
\int \! \!\!\! \int_B \omega \, \dd x \dd y - \ii \int \!\!\!\!
\int_B \vec \nabla \vec v \, \dd x \dd y,
\ee
where $\omega = \p_y v_x - \p_x v_y$ is the vorticity field, and $\vec \nabla \vec v
= \p_x v_x + \p_y v_y$ is the divergence of velocity field. The real part of this identity shows if the flow has zero vorticity, we have
\be \la{boutroux}
{\rm{Re }} \,\, \p_t \oint S(z) \dd z = 0.
\ee
The imaginary part of (\ref{re-im}) illustrates again the interpretation of singularity
set of $S(z)$ as sources of {\it water} (which occupies $D_+$ in a canonical
Laplacian growth formulation, while the exterior domain, $D_-$, is occupied
by a viscous fluid, which we call {\it oil} \cite{MWZ}): assume that the contour $C$ in (\ref{re-im}) encircles
the droplet without crossing any other branch cuts, then the contour integral may be
performed using Cauchy's theorem, giving the total flux of water:
\be
\int \!\!\!\! \int_B \vec \nabla \cdot \vec v \, \dd x \dd y = Q = 1.
\ee

We note here that equation (\ref{darcy2}) implies existence of a closed form
\be \la{omega1}
\dd \Omega = S \dd z + W \dd t,
\ee
whose primitive $\Omega$ has for real part the {\emph{Baiocchi transform}} of $p$:
\be \la{baiocchi}
{\rm {Re }}\,\, \Omega  =  - \int_0^t p(z, \tau) \dd \tau.
\ee
One can see that ${\rm {Re }}\,\, \Omega$ coincides with the potential
$\Phi$ introduced earlier.
From the continuity equation for water $\dot \rho + \vec \nabla \vec v = 0$ and the
Darcy law for water (opposite to oil) $\vec v = \nabla p$, we obtain for the time
evolution of water density at a given point $z$,
\be \la{char-funct}
\dot \rho = -\Delta p \Rightarrow \rho(z,t) = \rho(z,0) - \Delta \int_0^t  p(z,\tau) \dd \tau.
\ee
Equation (\ref{char-funct}) may be immediately generalized in a weak sense, replacing the
water density by the characteristic function of the domain $D_+, \rho \to \chi_{D_+}$, which
shows that the Baiocchi transform Re $\Omega$ may be interpreted as the electrostatic potential giving the growth of the water domain.

Similarly, applying an antiholomorphic derivative to (\ref{darcy2}), we obtain
\be
\vec \nabla \cdot \vec v + \ii \omega = -\Delta p + \ii \Delta \phi,
\ee
so that the imaginary part of the form $\Omega$ can be considered an electrostatic potential for the time integral of vorticity at a given point $z$:
\be \la{vort-funct}
\Delta \, {\rm {Im}} \, \Omega(z, t) = \int_0^t \omega(z, \tau) \dd \tau.
\ee

\subsubsection{Variational formulation of Hele-Shaw dynamics}

Formula (\ref{average}) has another physical interpretation, which we explore in this section.
Besides hydrodynamics, it also allows to describe the droplet through a variational (minimization) formulation, which will become very relevant when considering the singular limit.

Consider the case when the Schwarz function has only simple poles $\{ z_k \}$ and
cuts at $\{ \gamma_m \}$, with residues Res $S(z_k)$ and jump functions $h_m(z)$, inside the droplet. A simple calculation shows that these singular points constitute electrostatic
sources for the potential Re $\Omega$:
\be \la{actual}
\Delta \,\, {\rm{Re}} \,\, \Omega (z) = \sum_k {\rm {Res}} \,\,S(z_k) \delta(z-z_k)
+ \sum_m \int_{\gamma_m} h_m(\zeta) \delta(z-\zeta) \dd \zeta.
\ee
If we apply (\ref{average}) to all positive powers $z^k, k \ge 0$, we conclude that the singular distribution $\{ z_k \}, \{ \gamma_m \}$ and the uniform distribution $\rho(z)
= \chi_{D_+}(z)$ have the same interior harmonic moments $v_k = \langle z^k \rangle,
k \ge 0$. Thus, they create the same electrostatic potential outside the droplet. It is therefore possible to substitute the actual singular distribution
$\{ z_k \}, \{ \gamma_m \}$ with the smooth, uniform distribution $\rho(z)$ in calculations
related to the exterior potential. Beyond the mathematical equivalence, however, this
fact has an important physical interpretation, whose full meaning will become apparent
in the critical limit: when one more quantum of water is pumped into the droplet, it
first appears as a new singular point of the Schwarz function (a $\delta$-function
singularity). After a certain time, though, the droplet adjusts to the new area (subject
to the constraints given by the fixed exterior harmonic moments), and reaches its new
shape (with uniform density of water inside). Therefore, we can say that the singular
distribution $\{ z_k \}, \{ \gamma_m \}$ represents the fast-time distribution of sources
of water, while the uniform distribution $\rho(z)$ is the long-time, equilibrium distribution
of the same amount of water. When the dynamics
becomes fully non-equilibrium (after the cusp formation), this equivalence breaks down, and
the correct distribution to work with is the set of poles (cuts) of the Schwarz function.
In that case, the issue becomes solving the Poisson problem $\Delta \, {\rm{Re}} \,
\Omega = \sum_k {\rm{Res}} S(z_k) \delta(z-z_k)$, and finding the actual (time-dependent) location of the distribution of charges $z_k(t)$, subject to usual conditions
for the electrostatic potential $\Delta \, {\rm{Re}} \, \Omega $.

In the equilibrium case, however, it is appropriate to work with the smooth distribution
$\rho(z)$. Since the actual electrostatic potential $\Delta \, {\rm{Re}} \, \Omega $ contains
the regular expansion $V(z) = \sum_k t_k z^k$, we also add it to the contribution due to the distribution $\rho(z)$. We obtain for the total potential:
\be \la{equiv}
\Phi(z, \bar z) = \int_{D_+}\rho(\zeta)\log|z-\zeta|^2 \dd^2 \zeta + V(z) + \overline{V(z)}.
\ee
Inside the droplet, this potential solves the Poisson problem $\Delta \, \Phi(z, \bar z) = \rho(z) = 1$, and on the boundary it creates the electric field $E(z) = \bar{\p} \Phi = z$.
This means that inside the droplet, this potential is actually equal to $|z|^2$. Therefore,
the problem of finding the actual shape of a droplet of area $t_0$ and harmonic moments
$\{ t_k \}$ can be stated as:
\begin{quotation}
\emph{Find the domain $D_+$ of area $t_0$ such that $\Phi(z,\bar z) = |z|^2$ on $D_+$.}
\end{quotation}
Since $\rho(z)$ is the characteristic function of $D_+$, we may also write this problem
in the variational form:
$$
\frac{\delta \,\,\,\,\,\,\,\,}{\delta \rho(z)} \int_{D_+}
\rho(z) \left [ |z|^2 - V(z) - \overline{V(z)} - \int_{D_+}\rho(\zeta) \log |z-\zeta|^2 \dd^2 \zeta \right ]
\dd^2 z = 0.
$$
This equation is simply the minimization condition for the total energy
of a distribution of charges $\rho(z)$, in the external potential $W(z, \bar z) = -|z|^2
+ V(z) + \overline{V(z)}$. Therefore, the equilibrium (long-time limit) distribution of
water has the usual interpretation of minimizing the total electrostatic energy of the
system. However, when the system is not in equilibrium, this criterion cannot be used 
to select the solution.

\section{Quadrature Domains} \la{fourth}

We have seen in the previous sections
that polynomial or rational conformal mappings from the disk have
as images planar domains which are relevant for the Laplacian
growth (with finitely many sources). The domains in
question were previously and independently studied by
mathematicians, for at least two separate motivations. First they
have appeared in the work of Aharonov and Shapiro, on extremal
problems of univalent function theory \cite{AS}. About the same
time, these domains have been isolated by Makoto Sakai in his
potential theoretic work \cite{S1}. These domains, known today as
{\it quadrature domains}, carry Gaussian type quadrature formulas
which are valid for several classes of functions, like integrable analytic, 
harmonic, and sub-harmonic functions. The geometric structure of
their boundary, qualitative properties of their boundary defining
function, and dynamics under the Laplacian growth law are well
understood. The reader can consult the recent collection of
articles \cite{qd} and the survey \cite{GP07}. The present section
contains a general view of the theory of quadrature domains, with
special emphasis of a matrix model realization of their defining
function.

This chapter is organized in the following way: after presenting the
theory of quadrature domains for subharmonic and analytic functions,
we give an overview of the (inverse) Markov problem of moments, followed by 
its analogue in two dimensions, which is based on the notion of exponential
transform in the complex plane. The following sections illustrate the
reconstruction algorithm for the shape of a droplet, and point to a few
essential properties of the problem for signed measures. 

\subsection{Quadrature domains for subharmonic functions}

Let $\varphi$ be a subharmonic function defined on an open subset
of the complex plane, that is $\Delta\varphi \geq 0$, in the sense
of distributions, or the submeanvalue property
$$
\varphi (a) \leq \frac{1}{|B(a,r)|}\int_{B(a,r)} \varphi\,{\rm dA}
$$
holds for any disc centered at $a$, of radius $r$, $B(a,r)$ contained in the
domain of definition of $\varphi$. Henceforth $dA$ denotes
Lebesgue planar measure. Thus, with $\Omega=B(a,r)$, $c=|B(a,r)|=
\pi r^2$ and $\mu =c \delta_a$ there holds
\begin{equation}\label{subharmqd}
\int \varphi \, d\mu \leq \int_{\Omega} \varphi\,{\rm dA}
\end{equation}
for all subharmonic functions $\varphi$ in $\Omega$. This set of
inequalities is encoded in the definition that $\Omega$ is a {\it
quadrature domain for subharmonic functions} with respect to $\mu$
\cite{S1}, and it expresses that $\Omega=B(a,r)$ is a {\emph{swept out}}
version of the measure $\mu=c\delta_a$. If $c$ increases the
corresponding expansion of $\Omega$ is a simple example of
Hele-Shaw evolution, or Laplacian growth, as we have seen in the
previous section.

The above can be repeated with finitely many points, i.e., with
$\mu$ of the form
\begin{equation}\label{mu}
\mu=c_1 \delta_{a_1}  +\dots +  c_n \delta_{a_n},
\end{equation}
$a_j\in\mathbb C$, $c_j>0$: there always exists a unique (up to
nullsets) open set $\Omega\subset \mathbb C$ such that
(\ref{subharmqd}) holds for all $\varphi$ subharmonic and
integrable in $\Omega$.
One can think of it as the union $\bigcup_{j=1}^n B(a_j, r_j)$,
$r_j =\sqrt{c_j/\pi}$, with all multiple coverings smashed out to
a singly covered set, $\Omega$. In particular, $\bigcup_{j=1}^n
B(a_j, r_j)\subset\Omega$.

The above sweeping process, $\mu\mapsto\Omega$, or better
 $\mu\mapsto\chi_\Omega\cdot \rm{(dA)}$, called
 {\it partial balayage} \cite{S1}, \cite{Gustafsson-Sakai94}, \cite{G4},
 applies to quite general measures $\mu\geq 0$ and can be
 defined in terms of a natural energy minimization:
 given $\mu$,
$\nu =\chi_\Omega\cdot \rm{(dA)} $ will be the unique solution of
$$
{\rm Minimize}_\nu \, ||\mu -\nu||_e^2 \quad {\rm{s.t.}} \quad \nu\leq \rm{dA},
\int\,d\nu =\int\,d\mu.
$$
Here $|| \cdot ||_e$ is the energy norm:
$$
|| \mu ||_e^2=(\mu,\mu)_e, \quad {\rm with}\quad 
(\mu, \nu)_e = \frac{1}{2\pi}\int \log\frac{1}{|z-\zeta|}\, d\mu
(z) d\nu (\zeta).
$$
If $\mu$ has infinite energy, like in (\ref{mu}), one minimizes
$-2(\mu,\nu)+||\nu||_e^2 $ instead of $||\mu -\nu||_e^2$, which
can always be given a meaning \cite{Saff-Totik}.

By choosing
$$
\varphi (\zeta) = \pm \log |z-\zeta|
$$
in equation (\ref{subharmqd}), the plus sign allowed for all
$z\in\mathbb C$, the minus sign allowed only for $z\notin\Omega$,
one gets the following statements for potentials:
\begin{equation}\label{potentials}
\left\{\begin{array}{l} U^\mu \geq U^\Omega\quad \rm{in\ all}\quad
\mathbb C,\\
 U^\mu = U^\Omega\quad \rm{outside}\,\, \Omega.
\end{array}\right.
\end{equation}
Here
$$
U^\mu (z)=\frac{1}{2\pi}\int \log\frac{1}{|z-\zeta|}\, d\mu
(\zeta)
$$
denotes the logarithmic potential of the measure $\mu$, and
$U^\Omega=U^{\chi_\Omega\cdot {\rm dA}}$. In particular, the
measures $\mu$ and $\chi_\Omega\cdot \rm{(dA)}$ are gravi-
equivalent outside $\Omega$. By an approximation argument,
(\ref{potentials}) is actually equivalent to (\ref{subharmqd}).

Let us consider now an integrable harmonic function $h$, defined
in the domain $\Omega$. Since both $\varphi = \pm h$ are
subharmonic functions, we find
\begin{equation}
\int_\Omega h dA = \int h d\mu = \sum_{j=1}^n c_j h(a_j).
\end{equation}
That is, a Gaussian type quadrature formula, with nodes $\{ a_j\}$
and weights $\{ c_j\}$ holds. We say in this case that $\Omega$ is
a {\it quadrature domain for harmonic functions}. Similarly, one
defines a quadrature domain for complex analytic functions, and it
is worth mentioning that the inclusions $\{$QD for subharmonic functions$\} 
 \subset $ $\{$QD for harmonic functions$\} \ \subset $ $\{$QD for analytic functions$\}$
are strict, see for details \cite{S1}.

Recall that for a given positive measure $\sigma$ on the line,
rapidly decreasing at infinity, the zeros of the $N$-th orthogonal
polynomial are the nodes of a Gauss quadrature formula, valid only
for polynomials of degree $2N-1$. The difference above is that the
same finite quadrature formula is valid, in the plane, for an
infinite dimensional space of functions. A common feature of the
two scenarios, which will be clarified in the sequel, is the link
between quadrature formulas (on the line or in the plane) and
spectral decompositions (of Jacobi matrices, respectively
hyponormal operators).

Let $K={\rm conv\,}{\rm supp\,}\mu$ be the convex hull of the
support of $\mu$, i.e., the convex hull of the points
${a_1,\dots,a_n}$. As mentioned, $\Omega$ can be thought of as
smashed out version of $\bigcup_{j=1}^n B(a_j, r_j)$. The geometry
of $\Omega$ which this enforces is expressed in the following
sharp result (\cite{Gustafsson-Sakai94}, \cite{Gustafsson-Sakai02a}, \cite{Gustafsson-Sakai02b}): 
assume that $\Omega$ satisfies (\ref{subharmqd}) for a measure
$\mu\geq 0$ of the form (\ref{mu}). Then:

\begin{enumerate}

\item[(i)] $\partial\Omega$ may have singular points (cusps, double
points, isolated points), but they are all located inside $K$.
Outside $K$, $\partial\Omega$ is smooth algebraic.

\end{enumerate}

\noindent For $z\in\partial\Omega\setminus K$, let $N_z$ denote
the inward normal of $\partial\Omega$ at $z$ (well defined by
(i)).

\begin{enumerate}

\item[(ii)] For each $z\in\partial\Omega\setminus K$, $N_z$ intersects $K$.

\item[(iii)] For $z,w\in \partial\Omega\setminus K$, $z\ne w$,
$N_z$ and $N_w$ do not intersect each other before they reach $K$.
Thus $\Omega\setminus K$ is the disjoint union of
the inward normals from $\partial\Omega\setminus K$.

\item[(iv)] There exist $r(z)>0$ for $z\in K \cap\Omega$ such that
$$
\Omega = \bigcup_{z\in K \cap\Omega} B(z,r(z)).
$$
\end{enumerate}

(Statement (iv) is actually a consequence of (iii).)\\

To better connect our discussion with the moving boundaries
encountered in the first part of this survey, we add the following
remarks. Since $\Omega$ is uniquely determined by $(a_j,c_j)$ one
can steer $\Omega$ by changing the $c_j$ (or $a_j$). Such
deformations are of Hele-Shaw type, as can be seen by the
following computation, which applies in more general situations:
Hele-Shaw evolution $\Omega (t)$ corresponding to a point source
at $a\in\mathbb{C}$ (``injection of fluid'' at $a$) means that
$\Omega(t)$ changes by $\partial\Omega (t)$ moving in the outward
normal direction with speed
$$
-\frac{\partial G_{\Omega (t)}(\cdot, a)}{\partial n}.
$$
Here $G_\Omega(z,a)$ denotes the Green function of the domain
$\Omega$. If $\varphi$ is subharmonic in a neighborhood of
$\overline{\Omega(t)}$ then, as a consequence of $G_\Omega(\cdot,
a)\geq 0$, $G_\Omega(\cdot, a)=0$ on $\partial\Omega$ and $-\Delta
G_\Omega(\cdot, a)=\delta_a$,
$$
\frac{d}{dt}\int_{\Omega(t)} \varphi \,{\rm dA} =
\int_{\partial \Omega(t)} ({\rm speed\ of\ } \partial\Omega(t) \,
{\rm in\ normal\ direction})\, \varphi\,ds
$$ $$
=-\int_{\partial\Omega(t)} \frac{\partial G_{\Omega (t)}(\cdot,
a)}{\partial n}\, \varphi\,ds =-\int_{\partial\Omega(t)}
\frac{\partial \varphi}{\partial n}\, G_{\Omega (t)}(\cdot, a)\,ds
$$
$$
-\int_{\Omega(t)}\varphi \,\Delta G_{\Omega (t)}(\cdot, a)\,{\rm
dA}
+\int_{\Omega(t)}G_{\Omega (t)}(\cdot, a)\, \Delta \varphi\,{\rm
dA} \geq \varphi (a).
$$

Hence, integrating from $t=0$ to an arbitrary $t>0$,
$$
\int_{\Omega(t)} \varphi\, {\rm dA} \geq \int_{\Omega(0)}
\varphi\, {\rm dA} +t\varphi (a),
$$
telling that if $\Omega(0)$ is a quadrature domain for $\mu$
then $\Omega(t)$ is a quadrature domain for $\mu + t\delta_a$.

We remark that quadrature domains for subharmonic functions can be
defined in any number of variables, but then much less of their
qualitative properties are known, see for instance \cite{qd}.

\subsection{Quadrature domains for analytic functions}

Critical for our study is the regularity and algebraicity of the
boundary of quadrature domains for analytic functions. This was
conjectured in the early works of Aharonov and Shapiro, and proved
in full generality by Gustafsson \cite{G1}. A description of the
possible singular points in the boundary of a quadrature domain
was completed by Sakai \cite{S3,S4,S5}.

Assume that the quadrature domain for analytic functions $\Omega$
has a sufficiently smooth boundary $\Gamma$.  Let us consider the
Cauchy transform of the area mass, uniformly distributed on
$\Omega$:
$$ C(z) = \frac{-1}{\pi} \int_\Omega \frac{d A(w)}{w-z}.$$
This is an analytic function on the complement of
$\overline{\Omega}$, which is continuous (due to the Lebesgue
integrability of the kernel) on the whole complex plane. In
addition, the quadrature identity implies
$$
C(z) = \sum_{j=1}^n \frac{c_j}{\pi(z-a_j)}, \ \  z \in {\mathbb C}
\setminus \overline{\Omega}.$$

From the Stokes formula,
$$ C(z) = \frac{-1}{2 \pi i} \int_\Gamma \frac{ \overline{w}
dw}{w-z}, \ \ z \in {\mathbb C} \setminus \overline{\Omega}.$$
Therefore, by standard arguments in function theory  one proves
that the continuous function $w \mapsto \overline{w}$ extends
meromorphically from $\Gamma$ to $\Omega$. The poles of this
meromorphic extension coincide with the quadrature nodes.

The converse also holds, in virtue of Cauchy's formula: if $f$ is
an integrable analytic function in $\Omega$, then
$$ \int_\Omega f dA = \int_\Gamma f(w) \overline{w} d w =
\sum_{j=1}^n c_j f(a_j).$$
Thus, we recover the following fundamental observation: {\emph{
if $\Omega$ is a bounded planar domain with
sufficiently smooth boundary $\Gamma$, then $\Omega$ is a
quadrature domain for analytic functions if and only if the
function $w \mapsto \overline{w}$ extends meromorphically from
$\Gamma$ to $\Omega$.}}

Note that above, and elsewhere henceforth, we do not assume that
the weights in the quadrature formula for analytic functions are
positive. In this way we recover the fact (already noted in the
previous chapters) that quadrature domains for analytic functions
are characterized by a meromorphic Schwarz function, usually
denoted $S(z)$. A second departure from the quadrature domains for
subharmonic functions is that the quadrature data $(a_j,c_j)$ {\it
do not determine} the quadrature domain for analytic functions.
Indeed, consider the annulus $A_{r,R} = \{ z, \ r<|z|<R\}.$ Then
$$ \int_{A_{r,R}} f dA = \pi(R^2-r^2) f(0),$$
for all analytic, integrable functions $f$ in $A_{r,R}$.

The question how weak the smoothness assumption on the boundary
$\Gamma$ can be to insure the use of the above arguments has a long
history by itself, and we do not enter into its details. Simply
the existence of the quadrature formula and the fact that the
boundary is a mere continuum implies, via quite sophisticated
techniques, the regularity of $\Gamma$. See for instance
\cite{S3,GP98}.

The Schwarz function is a central character in our story. It can also
be related to the logarithmic potentials introduced in the
previous subsection. More specifically,  given any measure $\mu$
as in (\ref{mu}) and any open set $\Omega$ containing ${\rm
supp\,}\mu$, define (as distributions in all $\mathbb{C}$)
$$
u=U^\mu-U^\Omega, \quad 
S(z)=\overline{z} -4\frac{\partial u}{\partial z}.
$$
Then
$$
\Delta u = \chi_\Omega- \mu, \quad \frac{\partial S}{\partial
\overline{z}}= 1- \chi_\Omega +\mu.
$$
Note that with $\mu$ of the form (\ref{mu}) $w$ is harmonic in
$\Omega$ except for poles at the points $a_j$ and that in
particular, $S(z)$ is meromorphic in $\Omega$.

It is clear from (\ref{potentials}) that $\Omega$ is a subharmonic
quadrature domain for $\mu$ if and only if $u\geq 0$ everywhere
and $u=0$ outside $\Omega$. Then also $\nabla u=0$ outside
$\Omega$. Similarly, the criterion for $\Omega$ being a quadrature
domain for harmonic functions is that merely $u = \nabla u =0$ on
$\mathbb C\setminus \Omega$. (The vanishing of the gradient is a
consequence of the vanishing of $u$, except at certain singular
points on the boundary.) To be a quadrature domain for analytic
functions it is enough that just the gradient vanishes, or better
in the complex-valued case, that $\frac{\partial u}{\partial z}
=0$ on $\mathbb C\setminus \Omega$ (or just on $\partial\Omega$).

Gustafsson's innovative idea, to use the Schottky double of the
domain, can be summarized as follows. Let $\Omega$ be a bounded
quadrature domain for analytic functions, with boundary $\Gamma$.
We consider a second copy $\tilde{\Omega}$ of $\Omega$, endowed
with the anti-conformal structure, and ``glue" them into a compact
Riemann surface
$$ X = \Omega \cup \Gamma \cup \tilde{\Omega}.$$
This (connected) Riemann surface carries two meromorphic
functions:
$$ f(z) = \left\{ \begin{array}{cc}
                 S(z), & z \in \Omega \\
                 \overline{z}, & z \in \tilde{\Omega}\\
                 \end{array}\right. , \ \ \
                 g(z) = \left\{ \begin{array}{cc}
                 z, & z \in \Omega \\
                 \overline{S(z)}, & z \in \tilde{\Omega}\\
                 \end{array}\right. .$$
Any pair of meromorphic functions on $X$ is algebraically
dependent, that is, there exists a polynomial $Q(z,w)$ with the
property $ Q(g,f)=0$, and in particular
$$ Q(z, S(z)) = Q(z, \overline{z}) = 0, \ \ z \in \Gamma.$$
The involution (flip from one side to its mirror symmetric) on $X$
yields the Hermitian structure of $Q$:
$$ Q(z,w) = \sum_{i,j}^n a_{ij} z^i w^j,\ \ a_{ij} =
\overline{a_{ji}}.$$ One also proves by elementary means of
Riemann surface theory that $Q$ is irreducible, and moreover, its
leading part is controlled by the quadrature identity data:
$$ Q(z,\overline{z}) - |P(z)|^2 = O(z^{n-1},
\overline{z}^{n-1}),$$ where
$$ P(z) = (z-a_1)(z-a_2)...(z-a_n).$$
This Riemann surface is the continuum limit of the spectral curve
(\ref{qc}), for $N \to \infty$. Following Gustafsson (\cite{G1}), we note a surprising result:  

{\it a) The boundary of a quadrature
domain for analytic functions is a real algebraic, irreducible
curve.

b) In every conformal class of finitely connected planar domains
there exists a quadrature domain.

c) Every bounded planar domain can be approximated in the
Haudorff distance by a sequence of quadrature domains.}

The last two assertions are proven in Gustafsson's influential
thesis \cite{G1}. Recently, considerable progress was made in the
construction of multiply connected quadrature domains, see
\cite{qd,Crowdy-Marshall03,Crowdy-Marshall03}.

It is important to point out that not every domain bounded by an
algebraic curve is an algebraic domain in the above sense. In
general, if a domain $\Omega\subset \mathbb C$ is bounded by an
algebraic curve $Q(z,\overline{z})=0$ ($Q$ a polynomial with
Hermitian symmetry), then one can associate two compact symmetric
Riemann surfaces to it: one is the Schottky double of $\Omega$ and
the other is the Riemann surface classically associated to the
complex curve $Q(z,w)=0$. For the latter the involution is given
by $(z,w)\mapsto (\overline{w},\overline{z})$. In the case of {\it
algebraic domains} (this is another circulating name for
quadrature domains for analytic functions), and only in that case,
the two Riemann surfaces canonically coincide: the lifting
$$
z\mapsto (z,S(z))
$$
from $\Omega$ to the locus of $Q(z,w)=0$ extends to the Schottky
double of $\Omega$ and then gives an isomorphism, respecting the
symmetries, between the two Riemann surfaces.

As a simple example, the Schottky double of the simply connected
domain
$$
\Omega=\{ z=x+iy \in {\mathbb{C}} : x^4+ y^4 <1 \}
$$
has genus zero, while the Riemann surface associated to the curve
$x^{4}+ y^{4} =1$ has genus $3$. Hence they cannot be identified,
and in fact $\Omega$ is not an algebraic domain.

Other ways of characterizing algebraic domains, by means of
rational embeddings into $n$ dimensional projective space, are
discussed in \cite{GP00}.

\subsection{Markov's moment problem}

 We pause for a while the main line
of our story, to connect the described phenomenology with a
classical, beautiful mathematical construct due to A. A. Markov, 
all gravitating around moment problems for bounded functions.

 The classical {\it $L$-problem of moments} (also known as {\it Markov's moment problem})
 offers a good theoretical
framework for reconstructing extremal measures $\mu$ from their
moments, or equivalently, from the germ at infinity of some of
their integral transforms. The material below is classical and can
be found in the monographs \cite{AK,KN}. We present only a
simplified version of the abstract $L$-problem, well adapted to
the main themes of this survey.

Let $K$ be a compact subset of ${\mathbb{R}}^n$ with interior points
and let $A \subset {\mathbb{N}}^n$ be a finite subset of multi-indices.
We are interested in the set $\Sigma_A$ of moment sequences $a(f)
= (a_\sigma(f))_{\sigma \in A}$:
$$ a_\sigma (f) = \int_K x^\sigma f(x) dx, \  \  \sigma \in A,$$
of all measurable functions $f: K \longrightarrow [0,1]$. Regarded
as a subset of ${\mathbb{R}}^{|A|}$, $\Sigma_A$ is a compact convex
set. An $L^1-L^\infty$ duality argument (known as the abstract
$L$-problem of moments) shows that every extremal point of
$\Sigma_A$ is a characteristic function of the form $\chi_{ \{ p < \gamma\} },$ 
where we denote:
$$  \{p < \gamma \} = \{ x \in K;\  p(x)< \gamma \}. $$
Above $\gamma$ is a real constant and $p$ is an $A$-polynomial
with real coefficients, that is $p(x) = \sum_{\sigma \in A}
c_\sigma x^\sigma$. Indeed, to find the special form of the
extremal functions $f$, one has to analyze when the inequality
$$ \int_K p(x) f(x) dx \leq \| p \|_1 \| f \|_\infty = \int_K
|p(x)| dx$$ is an equality. For a complete proof the reader can
consult Krein and Nudelman's monograph \cite{KN}.

As a consequence, the above description of the extremal points in
the moment set $\Sigma_A$ implies the following remarkable
uniqueness theorem due to Akhiezer and Krein:

{\it  For each characteristic
function $\chi$ of a level set in $K$ of an $A$-polynomial there
exists exactly one class of functions $f$ in $L^\infty(K)$
satisfying $a(f) = a(\chi)$. For a non-extremal point $a(f) \in
\Sigma_A$ there are infinitely many non-equivalent classes in
$L^\infty(K)$ having the same $A$-moments.}\\

Let us consider a simple example:
$$K = \{ (x,y); \ x^2 + y^2 \leq 1 \} \subset {\mathbb{R}}^2,$$
and
$$ \Omega_+ = \{ (x,y) \in K; \ x>0, \ y>0 \}, \ \ \Omega_- = \{ (x,y) \in K; \ x<0, \ y<0 \}.$$
The reader can prove by elementary means that the sets
$\Omega_\pm$ cannot be defined in the unit ball $K$ by a single
polynomial inequality. On the other hand, the set
$$ \Omega = \Omega_+ \cup \Omega_- = \{ (x,y); \ xy >0 \},$$
is defined  by a single equation of degree two.

Thus, no matter how the finite set of indices $A \subset {\mathbb
N}^2$ is chosen, there is a continuum $f_s, \ s \in {\mathbb{R}},$ of
essentially distinct measurable functions $f_s : K \longrightarrow
[0,1]$ possessing the same $A$-moments:
$$ \int_K x^{\sigma_1} y^{\sigma_2} f_s(x,y) dx dy = \int_{\Omega_+}
  x^{\sigma_1} y^{\sigma_2} dx dy, \ \ s \in {\mathbb{R}}, \ \sigma \in A.$$

On the contrary, if the set of indices $A$ contains $(0,0)$ and
$(1,1)$, then for every measurable function $f: K \longrightarrow
[0,1]$ satisfying
$$ \int_K x^{\sigma_1} y^{\sigma_2} f(x,y) dx dy = \int_{\Omega}
    x^{\sigma_1} y^{\sigma_2} dx dy, \  \ \sigma \in A,$$
we infer by Akhiezer and Krein's Theorem that $f = \chi_\Omega,$
almost everywhere.

On a more theoretical side, we can interpret Akhiezer and Krein's
Theorem in terms of geometric tomography, see \cite{Ga}. Fix
a unit vector $\omega \in {\mathbb{R}}^n, \ \| \omega \| =1,$ and let
us consider the parallel Radon transform of a function $f:K
\longrightarrow [0,1]$, along the direction $\omega$:
$$
(Rf)(\omega, s) = \int_{ \langle x, \omega \rangle = s} f(x) dx.$$
Accordingly, the $k$-th moment in the variable $s$ of the Radon
transform is, for a sufficiently large constant $M$:
\begin{equation}
\int_{-M}^M (Rf)(\omega,s ) s^k ds = \int_K \langle x, \omega
\rangle^k f(x) dx = 
\end{equation}
\begin{equation} \la{33}
\sum_{|\sigma| = k} \frac{|\sigma|!}{\sigma !} \int_K x^\sigma
\omega^\sigma f(x) dx = \sum_{|\sigma| = k}
\frac{|\sigma|!}{\sigma !} \omega^\sigma a_\sigma(f).
\end{equation}

Since there are $N(n,d) = C_{n+d}^n$ linearly independent
polynomials in $n$ variables of degree less than or equal to $d$,
a Vandermonde determinant argument shows, via the above formula,
that the same number of different parallel projections of the
"shade" function $f: K \longrightarrow [0,1]$, determine, via a
matrix inversion, all moments:
$$ a_\sigma(f), \ \ |\sigma| \leq d.$$ The converse also holds, by formula
(\ref{33}). These transformations are known and currently used in image
processing, see for instance \cite{GHMP} and the references cited
there.

In conclusion, Akhiezer and Krein's Theorem asserts then that in
the measurement process
$$f \mapsto ((Rf)(\omega_j,s))_{j=1}^{N(n,d)} \mapsto (a_\sigma(f))_{|\sigma| \leq d}$$
only black and white pictures, delimited by a single algebraic
equation of degree less than or equal to $d$, can be exactly
reconstructed. Even when these uniqueness conditions are met, the
details of the reconstruction from moments are delicate. We shall
see some examples in the next sections.

\subsubsection{  Markov's extremal problem and the phase shift} By
going back to the source and dropping a few levels of generality,
we recall Markov's original moment problem and some of its modern
interpretations. Highly relevant for our "quatization" approach to
moving boundaries of planar domains is the matrix interpretation
we will describe for Markov's moment. Again, this material is well
exposed in the monograph by Krein and Nudelman \cite{KN}.

Let us consider, for a fixed positive integer $n$, the $L$-moment
problem on the line:
$$
a_k = a_k(f) = \int_{\mathbb{R}} t^k f(t) dt, \ \ \ 0 \leq k \leq 2n,$$
where the unknown function $f$ is measurable, admits all moments
up to degree $2n$ and satisfies:$$ 0 \leq f \leq L, \  {\rm
a.e.}.$$

As noted by Markov, the next formal series transform is quite
useful for solving this question:
\beq 
\la{4.1}
{\rm exp} \left [\frac{1}{L} \left ( \frac{a_0}{z} + \frac{a_1}{z^2} + \ldots
\frac{a_{2n}}{z^{2n+1}} \right ) \right ] =
 1 + \frac{b_0}{z} + \frac{b_1}{z^2} + \ldots.
 \eeq
Remark that, although the series under the exponential is finite,
the resulting one might be infinite.

The following result is classical, see for instance \cite{AK} pp.
77-82. Its present form was refined by Akhiezer and Krein; partial
similar attempts are due, among others, to Boas, Ghizzetti,
Hausdorff, Kantorovich, Verblunsky and Widder, see \cite{AK,KN}.  

{\emph{(Markov) Let $a_0, a_1, \ldots, a_{2n}$ be a
sequence of real numbers and let $b_0, b_1, \ldots$ be its
exponential $L$-transform. Then there is an integrable function
$f, \ 0 \leq f \leq L,$ possessing the moments $a_k(f) = a_k , \ 0
\leq k \leq 2n,$ if and only if the Hankel matrix
$(b_{k+l})_{k,l=0}^n$ is non-negative definite. Moreover, the
solution $f$ is unique if and only if ${\rm det}
(b_{k+l})_{k,l=0}^n = 0.$ In this case the function $f/L$ is the
characteristic function of a union of at most $n$ bounded
intervals.}}

The reader will recognize above a concrete validation of the
abstract moment problem discussed in the previous section.

In order to better understand the nature of the $L$-problem, we
interpret below the exponential transform from two different and
complementary points of view. For simplicity we take the constant
$L$ to be equal to $1$ and consider only compactly supported
originals $f$, due to the fact that the extremal solutions have
anyway compact support. Let $\mu$ be a positive Borel measure on
${\mathbb{R}}$, with compact support. Its Cauchy transform
$$
F(z) = 1- \int_{\mathbb{R}} \frac{d\mu(t)}{t-z},
$$ provides an analytic function on ${\mathbb{C}} \setminus {\mathbb{R}}$
which is also regular at infinity, and has the normalizing value
$1$ there. The power expansion, for large values of $|z|$, yields
the generating moment series of the measure $\mu$:
$$
F(z) = 1 + \frac{b_0(\mu)}{z} + \frac{b_1(\mu)}{z^2} +
\frac{b_2(\mu)}{z^3} + \ldots.
$$

On the other hand,
$$
{\rm Im} F(z) = - {\rm Im} z  \int \frac{ d\mu(t)}{|t-z|^2},
$$
whence
$$ {\rm Im} F(z)\  {\rm Im} z < 0, \ \ z \in {\mathbb C} \setminus {\mathbb{R}}.$$
Thus the main branch of the logarithm ${\rm log} F(z)$ exists in
the upper half-plane and its imaginary part, equal to the argument
of $F(z)$, is bounded from below by $-\pi$ and from above by $0$.
According to Fatou's theorem, the non-tangential boundary limits
$$ f(t) = \lim_{\epsilon \rightarrow 0} \frac{-1}{\pi} {\rm Im}\  {\rm log} F(t+i\epsilon),$$
exist and produce a measurable function with values in the
interval $[0,1]$. According to Riesz-Herglotz formula for the
upper-half plane, we obtain:
$$
{\rm log} F(z) = - \int_{\mathbb{R}} \frac{f(t)dt}{t-z}, \ \ z \in
{\mathbb C} \setminus {\mathbb{R}}.$$ Or equivalently,
$$
F(z) = {\rm exp} \left [ - \int_{\mathbb{R}} \frac{f(t)dt}{t-z} \right ].$$

One step further, let us consider the Lebesgue space $L^2(\mu)$
and the bounded self-adjoint operator $A = M_t$ of multiplication
by the real variable. The vector $\xi = {\bf 1}$ corresponding to
the constant function $1$ is $A$-cyclic, and according to the
spectral theorem:
$$
\int_{\mathbb{R}} \frac{d\mu(t)}{t-z} = \langle (A-z)^{-1} \xi, \xi
\rangle, \ \ z \in  {\mathbb C} \setminus {\mathbb{R}}.$$

As a matter of fact an arbitrary function $F$ which is analytic on
the Riemann sphere minus a compact real segment, and which maps
the upper/lower half-plane into the opposite half-plane has one of
the above forms. These functions are known in rational
approximation theory as {\it Markov functions}.

In short, putting together the above comments we can state the
following result: the canonical representations:
$$ F(z) = 1- \int_{\mathbb{R}} \frac{d\mu(t)}{t-z} = 
 {\rm exp}( - \int_{\mathbb{R}} \frac{f(t)dt}{t-z}) =  1-  \langle (A-z)^{-1} \xi, \xi \rangle$$
establish constructive equivalences between the following classes:
\par
 \ \ \ \ \ a)\ Markov's functions F(z);

\  \  \  \  \   b)  Positive Borel measures $\mu$ of compact
support on ${\mathbb{R}}$;

\  \  \  \  \   c)   Functions $f \in L^\infty_{\rm comp} ({\bf
R})$ of compact support, $0 \leq f \leq 1;$

\  \  \  \  \  d) Pairs $(A,\xi)$ of bounded self-adjoint operators
with a cyclic vector $\xi$.

The extremal solutions correspond, in each case exactly, to:

\  \  \  \  \   a) Rational Markov functions $F$;

\  \  \  \  \   b) Finitely many point masses $\mu$;

\  \  \  \  \   c) Characteristic functions $f$ of finitely many
intervals;

\  \  \  \  \   d) Pairs $(A,\xi)$ acting on a finite dimensional
Hilbert space.
\bigskip

For a complete proof see for instance Chapter VIII of \cite{MP}
and the references cited there. The above dictionary is remarkable in many ways. Each of its terms
has intrinsic values. They were long ago recognized in moment
problems, rational approximation theory or perturbation theory of
self-adjoint operators.

For instance, when studying the change of the spectrum under a
rank-one perturbation $A \mapsto B= A - \xi\langle \cdot, \xi
\rangle$ one encounters the {\it perturbation determinant}:
$$ \Delta_{A,B}(z) = {\rm det} [ (A - \xi\langle \cdot, \xi \rangle - z)(A-z)^{-1}] =
1-  \langle (A-z)^{-1} \xi, \xi \rangle.$$ The above exponential
representation leads to the {\it phase-shift} function $f_{A,B}(t)
= f(t)$:
$$ \Delta_{A,B}(z) = {\rm exp}\left [  - \int_{\mathbb{R}} \frac{f_{A,B}(t)dt}{t-z} \right ].$$
The phase shift of, in general, a trace-class perturbation of a
self-adjoint operator has certain invariance properties; it
reflects by fine qualitative properties the nature of change in
the spectrum. The theory of perturbation determinants and of the
phase shift is nowadays well developed, mainly for its
applications to quantum physics, see \cite{Krein1953,Simon}.

The reader will recognize above an analytic continuation in the
complex plane of the real exponential transform
$$ F(x) = E_f(x) = {\rm exp} \left [  - \int_{\mathbb{R}} \frac{f(t)dt}{|t-x|} \right ], $$
assuming for instance that $x < M$ and the function $f$ is
supported by $[M,\infty)$.

To give the simplest, yet essential, example, we consider a
positive number $r$ and the various representations of the
function:
$$ F(z) = 1 + \frac{r}{z} = \frac{z+r}{z} =  1 - \int_{\mathbb{R}} \frac{ r d \delta_0(t)}{t-z} =
 {\rm exp} \left [ - \int_{-r}^0 \frac{dt}{t-z} \right ] =  {\rm det} [(-r -z)(-z)^{-1}].$$
In this case the underlying Hilbert space has dimension one and
the two self-adjoint operators are $A = 0$ and $A - \xi\langle
\cdot, \xi \rangle = -r$.

\subsubsection{The reconstruction algorithm in one real variable}
Returning to our main theme, and as a direct continuation of the
previous section, we are interested in the exact reconstruction of
the original $f:{\mathbb{R}} \longrightarrow [0,1]$ from a finite set
of its moments, or equivalently, from a Taylor polynomial of $E_f$
at infinity. The algorithm described in this section is the
diagonal Pad\'e approximation of the exponential transform of the
moment sequence. Its convergence, even beyond the real axis, is
assured by a famous result discovered by A. A.  Markov.

Let $a_0, a_1, \ldots, a_{2n}$ be a sequence of real numbers with
the property that its exponential transform:
$${\rm exp} \left [\frac{1}{L} \left (\frac{a_0}{z} + \frac{a_1}{z^2} +
  \ldots \frac{a_{2n}}{z^{2n+1}} \right ) \right ] =
   1 + \frac{b_0}{z} + \frac{b_1}{z^2} + \ldots,$$
produces a non-negative Hankel matrix $(b_{k+l})_{k,l=0}^n$.

According to Markov's Theorem, there exists at
least one bounded self-adjoint operator $A \in L(H)$, with a
cyclic vector $\xi$, such that:
$$
{\rm exp} \left [\frac{1}{L} \left (\frac{a_0}{z} + \frac{a_1}{z^2} + \ldots
\frac{a_{2n}}{z^{2n+1}} \right ) \right ] =  1 + \frac{\langle \xi, \xi
\rangle}{z} + \frac{\langle A\xi, \xi \rangle}{z^2} + \ldots
\frac{\langle A^{2n} \xi, \xi \rangle}{z^{2n+1}} +
O(\frac{1}{z^{2n+2}}).
$$

Let $k <n$ and $H_k$ be the Hilbert subspace spanned by the
vectors $\xi, A\xi, \ldots, A^{k-1}\xi$. Suppose that ${\rm dim}
H_k =k$, which is equivalent to saying that ${\rm det}
(b_{i+j})_{i,j=0}^{k-1} \neq 0$. Let $\pi_k$ be the orthogonal
projection of $H$ onto $H_k$ and let $A_k = \pi_k A \pi_k$. Then
$$ \langle A_k^{i+j} \xi, \xi \rangle = \langle A_k^i \xi, A_k^j \xi \rangle =\langle A^i \xi, A^j \xi \rangle = \langle A^{i+j} \xi, \xi \rangle,$$
whenever $0 \leq i, j \leq k-1.$ In other terms, for large values
of $|z|$:
$$ \langle (A-z)^{-1} \xi, \xi \rangle = \langle (A_k-z)^{-1} \xi, \xi \rangle + O(\frac{1}{z^{2k+1}}).
$$

By construction, the vector $\xi$ remains cyclic for the matrix
$A_k \in L(H_k)$. Let $q_k(z)$ be the minimal polynomial of $A_k$,
that is the monic polynomial of degree $k$ which annihilates
$A_k$. In particular,
$$ q_k(z) \langle (A_k-z)^{-1} \xi, \xi \rangle =
 \langle (q_k(z) - q_k(A_k)) (A_k-z)^{-1} \xi, \xi \rangle = p_{k-1}(z)$$
 is a polynomial of degree $k-1$.

The two observations yield:
$$ q_k(z) \langle (A-z)^{-1} \xi, \xi \rangle = 
q_k(z) \langle (A_k-z)^{-1} \xi, \xi \rangle +
O(\frac{1}{z^{k+1}}) =  p_{k-1}(z) + O(\frac{1}{z^{k+1}}).$$

The resulting rational function $R_k(z) =
\frac{p_{k-1}(z)}{q_k(z)}$ is characterized by the property:
$$ 1 + \frac{b_0}{z} + \frac{b_1}{z^2} + \ldots = 1+ R_k(z) + O(\frac{1}{z^{2k+1}});
$$ it is known as the {\it Pad\'e approximation} of
order $(k-1,k)$, of the given series.

A basic observation is now in order: since $b_0, b_1, \ldots,
b_{2k+1}$ is the power moment sequence of a positive measure,
$q_k$ is the associated orthogonal polynomial of degree $k$ and
$p_k$ is a second order orthogonal polynomial of degree $k-1$. In
particular their roots are simple and interlaced. We prove only
the first assertion, the second one being of a similar nature. Indeed, let $\mu$ be the spectral measure of $A$
localized at the vector $\xi$.
Then, for $j<k$,
$$ \int_{\mathbb{R}} t^j q_k(t) dt = \langle A^j \xi, q_k(A) \xi
\rangle = \langle A_k^j \xi, q_k(A_k) \xi \rangle = 0.$$

Assume now that we are in the extremal case ${\rm det}
(b_{i+j})_{i,j=0}^n = 0$ and that $n$ is the smallest integer with
this property, that is ${\rm det} (b_{i+j})_{i,j=0}^{n-1} \neq 0$.
Since
$$ b_{i+j} = \langle A^i \xi, A^j \xi \rangle,$$
this means that the vectors $\xi, A\xi, \ldots, A^n\xi$ are
linearly dependent. Or equivalently that $H_n = H$ and
consequently $A_n = A$.

According to the dictionary established above, this
is another proof that the extremal case of the truncated moment
1-problem with data $a_0, a_1, \ldots, a_{2n}$ admits a single
solution. The unique function $f : {\mathbb{R}} \longrightarrow [0,1]$
with this string of moments will then satisfy:
$$  {\rm exp} \left [ - \int_{\mathbb{R}} \frac{f(t)dt}{t-z} \right ] = 1 + R_n(z) =  1 - \sum_{i=1}^n \frac{r_i}{a_i - z} ={\rm det} [(A- \xi\langle  \cdot, \xi \rangle - z)(A-z)^{-1}] = $$
$$ \prod_{i=1}^n \frac{b_i-z}{a_i-z},$$
where the spectrum of the matrix $A$ is $\{a_1, \ldots, a_n\},$
that of the perturbed matrix $B = A- \xi\langle  \cdot, \xi
\rangle$ is $b_1, \ldots, b_n$ and $r_i$ are positive numbers.
Again, one can easily prove that $b_1 < a_1 < b_2 < a_2 < \ldots <
b_n < a_n$. By the last example considered, we infer:
$$ f = \sum_{i=1}^n \chi_{[b_i,a_i]},$$
or equivalently
$$ f = \frac{1}{2} \left [1- {\rm sign} \frac{p_{k-1} + q_k}{q_k} \right ].$$

The above computations can therefore be put into a (robust)
reconstruction algorithm of all extremal functions $f$. The
Hilbert space method outlined above has other benefits, too. We
illustrate them with a proof of another celebrated result due to
A. A. Markov, and related to the convergence of the mentioned
algorithm, in the case of non-extremal functions.

{\it  Let $\mu$ be a positive measure,
compactly supported on the real line and let $F(z) = \int_{\mathbb{R}}
(t-z)^{-1} d\mu(t)$ be its Cauchy transform. Then the diagonal
Pad\'e approximation $R_n(z) = p_{n-1}(z)/q_n(z)$ converges to
$F(z)$ uniformly on compact subsets of ${\mathbb C} \setminus {\mathbb
R}$.}

This is the basic argument proving the statement: let $A$ be 
the multiplication operator with the real
variable on the Lebesgue space $H= L^2(\mu)$ and let $\xi = {\bf
1}$ be its cyclic vector. The subspace generated by $\xi,
A\xi,...,A^{n-1}\xi$ will be denoted as before by $H_n$ and the
corresponding compression of $A$ by $A_n = \pi_n A \pi_n$.

If there exists an integer $n$ such that $H=H_n,$ then the
discussion preceding the theorem shows that $F=R_n$ and we have
nothing else to prove. Assume the contrary, that is the measure
$\mu$ is not finite atomic.

Let $p(t)$ be a polynomial function, regarded as an element of
$H$. Then $$(A-A_n)p(t) = tp(t) - (\pi_n A \pi_n) p(t) = tp(t) -
tp(t) = 0$$ provided that ${\rm deg} (p) < n$. Since $\|A_n\| \leq
\|A\|$ for all $n$, and by Weierstrass Theorem, the polynomials
are dense in $H$, we deduce:
$$ \lim_{n \rightarrow \infty} \| (A-A_n) h \| =
0, \ \ h \in H.$$

Fix a point $a \in {\mathbb C} \setminus {\mathbb{R}}$ and a vector $h \in
H$. Then
$$ \lim_{n \rightarrow \infty} \| [(A-a)^{-1} - (A_n-a)^{-1}]h \| =  \lim_{n \rightarrow \infty}
      \| (A_n-a)^{-1}(A-A_n)(A-a)^{-1} h \| \leq $$
$$  \lim_{n \rightarrow \infty} \frac{1}{| {\rm Im}\  a|} \| (A-A_n)(A-a)^{-1} h \| = 0.$$
A repeated use of the same argument shows that, for every $k \geq
0$,
$$ \lim_{n \rightarrow \infty} \| [(A-a)^{-k} - (A_n-a)^{-k}]h \| = 0.$$

Choose a radius $r < |{\rm Im}\  a| \leq \| (A_n -a)^{-1} \|^{-1}
$, so that the Neumann series
$$ (A_n -z)^{-1} = (A_n - a - (z-a))^{-1} = \sum_{k=0}^\infty (z-a)^k (A_n - a)^{-k-1} $$
converges uniformly and absolutely, in $n$ and $z$, in the disk
$|z-a| \leq r$. Consequently, for a fixed vector $h \in H$,
$$ \lim_{n \rightarrow \infty} \| (A_n -z)^{-1} h - (A-z)^{-1} h \| = 0,$$
uniformly in $z,\ |z-a|\leq r$. In particular,
$$ \lim_{n \rightarrow \infty}
R_n(z) = \langle (A_n-z)^{-1} \xi, \xi \rangle = \lim_{n
\rightarrow \infty} \langle (A_n-z)^{-1} \xi, \xi \rangle = $$ $$
\langle (A-z)^{-1} \xi, \xi \rangle = F(z),$$ uniformly in $z, \
|z-a| \leq r$.

Details and a generalization of the above operator theory approach
to Markov theorem can be found in \cite{Put02}.

\subsection{The exponential transform in two dimensions}

We return now to two real dimensions, and establish an analog of
the matrix model for Markov's moment problem. Fortunately this is
possible due to the import of some key results in the theory of
semi-normal operators. We expose first the analog of Markov's
exponential transform, and second, we will make a digression into
semi-normal operator theory, with the aim at realizing the
exponential transform in terms of (infinite) matrices, and
ultimately of reconstructing planar shapes from their moments.

 The case of two
real variables is special, partly due to the existence of a
complex variable in ${\mathbb{R}}^2$ . Let $g : {\mathbb C} \longrightarrow
[0,1]$ be a measurable function and let $dA(\zeta)$ stand for the
Lebesgue area measure. The {\it exponential transform} of $g$, is
by definition the transform: \begin{equation}
 E_g(z) = {\rm exp}
(-\frac{1}{\pi} \int_{\mathbb C} \frac{ g(\zeta) dA(\zeta)} {|\zeta -
z|^2}),\ \  z \in {\mathbb C} \setminus {\rm supp}( g).
\end{equation}
 This
expression invites to consider a polarization in $z$:
$$ E_g(z, w) = {\rm exp} (-\frac{1}{\pi} \int_{\mathbb{C}} \frac{ g(\zeta) dA(\zeta)}
{(\zeta - z)(\overline{\zeta}-\overline{w})}),\  \   z, w \in {\mathbb
C} \setminus {\rm supp} (g). \eqno{(6.1)}$$ The resulting function
$E_g(z,w)$ is analytic in $z$ and antianalytic in $w$, outside the
support of the function $g$. Note that the integral converges for
every pair $(z,w) \in {\mathbb{C}}^2$ except the diagonal $z=w$.
Moreover, assuming by convention ${\rm exp} (- \infty) =0$, a
simple application of Fatou's Theorem reveals that the function
$E_g(z,w)$ extends to the whole ${\mathbb{C}}^2$ and it is separately
continuous there. Details about these and other similar
computations are contained in \cite{MP}.

As before, the exponential transform contains, in its power
expansion at infinity, the moments
$$
a_{mn}= a_{mn}(g) = \int_{\mathbb{C}} z^m \overline{z}^n g(z) dA(z), \
\  m,n \geq 0.$$ According to Riesz Theorem these data determine
$g$. We will denote the resulting series by: \begin{equation}
 {\rm exp}\left [ \frac{-1}{\pi} \sum_{m,n=0}^\infty \frac{a_{mn}}{z^{n+1}
\overline{w}^{m+1}} \right ] =
 1- \sum_{m,n=0}^\infty \frac{b_{mn}}{z^{n+1} \overline{w}^{m+1}}.
 \end{equation}

The exponential transform of a uniformly distributed mass on a
disk is simple, and in some sense special, this being the building
block for more complicated domains. A direct elementary
computation leads to the following formulas for the unit disk
${\bf D}$, cf. \cite{GP98}:
$$ E_{\bf D}(z,w) = \cases {
                      1- \frac{1}{z \overline{w}},  \ \ z,w \in \overline{\bf D}^c,\cr
              1- \frac{\overline{z}}{\overline{w}}, \ \ z \in {\bf D},\ w \in \overline{\bf D}^c, \cr
        1- \frac{{w}}{{z}}, \ \ w \in {\bf D},\ z \in \overline{\bf D}^c,\cr
    \frac{|z-w|^2}{1-z\overline{w}}, \ \ z,w \in {\bf D}.\cr}
            $$
Remark that $E_{\bf D}(z) = E_{\bf D}(z,z)$ is a rational function
and its value for $|z|>1$ is $1- \frac{1}{|z|^2}$. The
coefficients $b_{mn}$ of the exponential transform are in this
case particularly simple: $b_{00} = 1$ and all other values are
zero.

Once more, an additional structure of the exponential transform in
two variables comes from operator theory. More specifically, for
every measurable function $g: {\mathbb{C}} \to [0,1]$ of
compact support there exists a unique irreducible, linear bounded
operator $T \in L(H)$ acting on a Hilbert space $H$, with rank-one
self-commutator $[T^\ast, T] = \xi \otimes \xi = \xi \langle
\cdot, \xi \rangle$, which factors $E_g$ as follows:
\begin{equation} \label{factorE} E_g(z,w) = 1 - \langle (T^\ast -
\overline{w})^{-1}\xi, (T^\ast - \overline{z})^{-1}\xi \rangle, \
\ z,w \in {\rm supp} (g)^c. \end{equation} As a matter of fact,
with a proper extension of the definition of localized resolvent
$(T^\ast - \overline{w})^{-1}\xi$ the above formula makes sense on
the whole ${\mathbb{C}}^2$. The function $g$ is called the {\it
principal function} of the operator $T$. The next section will
contain a brief incursion into this territory of operator theory.

Let $g : {\mathbb{C}} \to [0,1]$ be a measurable function
and let $E_g(z,w)$ be its polarized exponential transform. We
retain from the above discussion the fact that the kernel:
$$ 1- E_g(z,w), \ \ \ z,w \in {\mathbb{C}},$$
is positive definite. Therefore the distribution $H_g(z,w)  = -
\frac{\partial}{\partial \overline{z}} \frac{\partial}{\partial
{w}} E_g(z,w)$ has compact support and it is positive definite, in
the sense:
$$ \int_{{\mathbb{C}}^2} H_g(z,w) \phi(z) \overline{\phi(w)} dA(z) dA(w) \geq 0, \ \ \phi \in C^\infty({\mathbb{C}}).$$
If $g$ is the characteristic function of a bounded domain $\Omega
\subset {\mathbb{C}}$, then it is elementary to see that the
distribution $H_\Omega(z,w)  = H_g(z,w)$ is given on $\Omega
\times \Omega$ by a smooth, jointly integrable function which is
analytic in $z \in \Omega$ and antianalytic in $w \in \Omega$, see
\cite{GP98}.

In particular, this gives the useful representation:
$$
E_\Omega(z,w) = 1- \frac{1}{\pi^2} \int_{\Omega^2}
\frac{H_\Omega(u,v) dA(u) dA(v)}{(u-z)(\overline{v} -
\overline{w})}, \ \ z,w \in \overline{\Omega}^c,$$ where the
kernel $H_\Omega$ is positive definite in $\Omega \times \Omega$.

The example of the disk considered in this section suggests that
the exterior exponential transform of a bounded domain
$E_\Omega(z,w)$ may extend analytically in each variable inside
$\Omega$. This is true whenever $\partial \Omega$ is real analytic
smooth. In this case there exists an analytic function $S$ defined
in a neighborhood of $\partial \Omega$, with the property:
$$ S(z) = \overline{z}, \ \ z \in \partial \Omega.$$
The anticonformal local reflection with respect to $\partial
\Omega$ is then the map $z \mapsto \overline{S(z)}$; for this
reason $S(z)$ is called the {\it Schwarz function} of the real
analytic curve $\partial \Omega$, introduced earlier in this text. 
Let $\omega$ be a relatively
compact subdomain of $\Omega$, with smooth boundary, too, and such
that the Schwarz function $S(z)$ is defined on a neighborhood of
$\Omega \setminus \omega$. A formal use of Stokes' Theorem yields:
$$ 1- E_\Omega(z,w) = \frac{1}{4 \pi^2} \int_{\partial \Omega}
\int_{\partial \Omega} H_\Omega(u,v) \frac{\overline{u}du}{u-z}
\frac{v d \overline{v}}{\overline{v}-\overline{w}} = $$
$$ \frac{1}{4 \pi^2} \int_{\partial \omega}
\int_{\partial \omega} H_\Omega(u,v) \frac{\overline{u}du}{u-z}
\frac{v d \overline{v}}{\overline{v}-\overline{w}}.$$

But the latter integral is analytic/antianalytic for $z,w \in
\overline{\omega}^c$. A little more work with the above Cauchy
integrals leads to the following remarkable formula for the
analytic extension of $E_\Omega(z,w)$ from $z,w \in
\overline{\Omega}^c$ to $z,w \in \overline{\omega}^c$:
$$
F(z,w) = \cases{ E(z,w), \ \ z,w \in \Omega^c, \cr
                 (z-\overline{S(w)})(S(z) - \overline{w})
                 H_\Omega(z,w), \ \ z,w \in \Omega \setminus
                 \overline{\omega}. \cr}
                 $$
The study outlined above of the analytic continuation phenomenon
of the exponential transform $E_\Omega(z,w)$ led to a proof of a
priori regularity of boundaries of domains which admit analytic
continuation of their Cauchy transform. The most general result of
this type was obtained by different means by Sakai. We
simply state the result, giving in this way a little more insight into the proof
of the regularity of the boundaries of quadrature domains.

{\it {Let $\Omega$ be a bounded planar domain
with the property that its Cauchy transform $$ \hat{\chi}_\Omega
(z) = \frac{-1}{\pi} \int_\Omega \frac{dA(w)}{w-z}, \ \ z \in
\overline{\Omega}^c $$ extends analytically across $\partial
\Omega$. Then the boundary $\partial \Omega$ is real analytic.}}
\par

\bigskip

Moreover, Sakai has classified the possible singular points of the
boundary of such a domain. For instance angles not equal to $0$ or
$\pi$ cannot occur on the boundary.

\subsection{Semi-normal operators}

A normal operator is modelled via the spectral theorem as
multiplication by the complex variable on a vector valued Lebesgue
$L^2$-space. The interplay between measure theory and the
structure of normal operators is well known and widely used in
applications. One step further, there are by now well understood
functional models, and a complete classification for classes of
close to normal operators. We record below a few aspects of the
theory of semi-normal operators with trace class self-commutators.
They will be serve as Hilbert space counterparts for the study of
moving boundaries in two dimensions. The reader is advised to
consult the monographs \cite{MP,Xia} for full details.

Let $H$ be a separable, complex Hilbert space and let $T \in
\mathcal L(H)$ be a linear bounded operator. We assume that the
self-commutator $[T^\ast , T] = T^\ast T - T T^\ast$ is
trace-class, and call $T$ semi-normal. If $[T^\ast , T] \geq 0,$
then T is called {\it hypo-normal}. For a pair of polynomials
$p(z,\overline{z}), q(z,\overline{z})$ one can choose (at random)
an ordering in the functional calculus $p(T,T^\ast), q(T,T^\ast)$,
for instance putting all adjoins to the left of all other
monomials. The functional
$$
(p,q) \rightarrow {\rm trace} [p(T,T^\ast), q(T,T^\ast)]
$$
is then well-defined, independent of the ordering in the
functional calculus, and possesses the algebraic identities of the
Jacobian $\frac{ \partial(p,q)}{\partial(\overline{z}, z)}$.  A
direct (algebraic) reasoning will imply the existence of a
distribution $u_T \in \mathcal D'(\mathbb C) $ satisfying
$$
{\rm trace} [p(T,T^\ast), q(T,T^\ast)] = u_T \left [\frac{
\partial(p,q)}{\partial(\overline{z}, z)} \right ],
$$
see \cite{HH}. The distribution $u_T$ exists in any number of
variables (that is for tuples of self-adjoint operators subject to
a trace class multi-commutator condition) and it is known as the
{\it Helton-Howe functional}.

Dimension two is special because of a theorem of J. D. Pincus
which asserts that $u_T = \frac{1}{\pi} g_T \,{\rm dA}$, that is
$u_T$ is given by an integrable function function $g_T$, called
the {\it principal function} of the operator $T$, see
\cite{Pincus,CareyPincus}.

The analogy between the principal function and the phase shift
(the density of the measure appearing in Markov's moment problem
in one variable) is worth mentioning in more detail. More
precisely, if $B = A- K$ is a trace-class, self-adjoint
perturbation of a bounded self-adjoint operator $A \in L(H)$, then
for every polynomial $p(z)$, Krein's {\it trace formula} holds:
$$
{\rm tr}[ p(B) - p(A)] =  \int_{\mathbb{R}} p'(t) f_{A,B}(t) dt,
$$ where $f_{A,B}$ is the corresponding phase-shift function, \cite{Krein1953}.
It is exactly this link between Hilbert space operations and
functional expressions which bring the two scenarios very close.
Taking one step further, exactly as in the one variable case, the moments
of the principal function can be interpreted in terms of the
Hilbert space realization, as follows:
$$
m k \int z^{m-1} \overline{z}^{k-1} g_T(z) {\rm dA} = $$ $${\rm
trace} [T^{\ast k}, T^m], \ \  k,m \geq 1.
$$
In general, the principal function can be regarded as a
generalized Fredholm index of $T$, that is, when the left hand
side below is well defined, we have
$$
{\rm ind} (T-\lambda) = - g_T(\lambda).
$$
Moreover $g_T$ enjoys the functoriality properties of the index,
and it is obviously invariant under trace class perturbations of
$T$. Moreover, in the case of a fully non-normal operator $T$,
$$ {\rm supp} g_T = \sigma (T),$$
and various parts of the spectrum $\sigma(T)$ can be interpreted
in terms of the behavior of $g_T$, see for details \cite{MP}.

To give a simple, yet non-trivial, example we proceed as follows.
Let $\Omega$ be a planar domain bounded by a smooth Jordan curve
$\Gamma$.  Let $H^2(\Gamma)$ be the closure of complex polynomials
in the space $L^2(\Gamma, ds)$, where $ds$ stands for the arc
length measure along $\Gamma$ (the so-called {\it Hardy space}
attached to $\Gamma$). The elements of $H^2(\Gamma)$ extend
analytically to $\Omega$. The multiplication operator by the
complex variable, $T_z f = z f, \ \ f \in H^2(\Gamma),$ is
obviously linear and bounded. The regularity assumption on
$\Gamma$ implies that the commutator $[T_z, T_z^\ast]$ is trace
class. Moreover, the associated principal function is the
characteristic function of $\Omega$, so that the trace formula
above becomes:
$$
{\rm trace} [p(T_z,T_z^\ast), q(T_z,T_z^\ast)] =
\frac{1}{\pi} \int_\Omega \frac{
\partial(p,q)}{\partial(\overline{z}, z)} {\rm dA}, \ \ p,q \in
\mathbb C[z,\overline{z}].
$$
See for details \cite{MP,Xia}.

A second, more interesting (generic example this time) can be
constructed as follows. Let $u(t), v(t)$ be real valued, bounded
continuous functions on the interval $[0,1]$. Consider the
singular integral operator, acting on the Lebesgue space
$L^2([0,1],dt)$ by the formula:
$$ (Tf)(t) = tf(t) +i[u(t)f(t) + \frac{1}{\pi} \int_0^1 \frac{
v(t)v(s) f(s) ds}{s-t}.$$ Then it is easy to see that the
self-commutator $[T^\ast,T]$ is rank one. The principal function
$g_T$ will be in this case the characteristic function of the
closure of the domain $G$ given by the constraints
$$ G = \{ (x,y) \in {\mathbb R}^2; \ \  |y-u(x)| \leq v(x)^2, \ x\in
[0,1]\}.$$ Based on a refinement of this example, in general every
{\it hyponormal operator} with trace class self-commutator can be
represented by such a singular integral model, with matrix valued
functions $u,v$, acting on a direct integral of Hilbert spaces
over $[0,1]$; in which case the principal function relates
directly to Krein's phase shift, by the following remarkable
formula due to Pincus \cite{Pincus}:
$$ g_T(x,y) = f_{u(x)-v(x)^\ast v(x), u(x)+v(x)^\ast v(x)}(y).$$

The case of rank-one self-commutators is singled out in the
following key classification result:

{\it There exists a bijective correspondence $T
\mapsto g_T$ between irreducible hyponormal operators $T$, with
rank-one self-commutator, and bounded measurable functions with
compact support in the complex plane.}

An invariant formula, relating the moments of the principal
function $g$ to the Hilbert space operator $T, \ [T^\ast,T] = \xi
\langle \cdot, \xi \rangle, $, satisfying $g_T = g, a.e.$ is
furnished by the determinantal formula:
$$  {\rm exp} (-\frac{1}{\pi} \int_{\mathbb{C}} \frac{ g(\zeta) dA(\zeta)}
{(\zeta - z)(\overline{\zeta}-\overline{w})}) =  {\rm det} [
(T^\ast - \overline{w})^{-1}(T-z)(T^\ast -
\overline{w})(T-z)^{-1}]= $$ $$ 1 - \langle (T^\ast -
\overline{w})^{-1}\xi, (T^\ast - \overline{z})^{-1}\xi \rangle, \
\ z,w \in {\rm supp} (g)^c.$$ This formula explains the positivity
property of the exponential transform, alluded to in the previous
section.

The bijective correspondence between classes $g \in L^\infty_{\rm
comp} ({\mathbb{C}}), \ 0 \leq g \leq 1$ and irreducible operators $T$
with rank-one self-commutator was exploited in \cite{Put96,Put98}
for solving the $L$-problem of moments in two variables. The
theory of the principal function has inspired and played a basic
role in the foundations of modern non-commutative geometry
(specifically the cyclic cohomology of operator algebras) and
non-commutative probability.

We have to stress the fact that the above bijective correspondence
between``shade functions" $g_T$ and irreducible hyponormal operators
 $T$ with rank-one self-commutator can in principle transfer {\it any}
dynamic $g(t)$ into a Hilbert space operator dynamic $T(t)$. However,
the details of the evolution law of $T(t)$ even in the case of
elliptic growth are not trivial, nor make the integration
simpler. We will see some relevant low degree examples in the next section.

\subsubsection{Applications: Laplacian growth}

To give a single abstract illustration, consider a growing family of bounded
planar domains $D(t)$ with smooth boundary:
$$ D(t) \subset D(s),\ \ {\rm whenever}\ t<s.$$
The evolution of the exponential transforms
$$ E_{D(t)}(z,w) = \exp \left [\frac{-1}{\pi} \int_{D(t)}
        \frac{dA(\zeta)}{(\zeta-z)(\bar \zeta -\bar w)} \right ] ,
        $$
is governed by the differential equation (in the standard vector calculus notation)
$$ \frac{d}{dt} E_{D(t)}(z,w) = \frac{-1}{\pi} E_{D(t)}(z,w)                     \int_{\partial D(t)} \frac{V_n
 d\ell(\zeta)}{(\zeta-z)(\overline{\zeta}-\overline{w})}.$$
 Any evolution law at the level of the pair $(T(t),\xi(t))$
 will have the form
 $$ \frac{d}{dt} E_{D(t)}(z,w) =  \langle (T^\ast(t) - \overline{w})^{-1})
 T^\ast (t) (T^\ast(t) - \overline{w})^{-1}) \xi(t),
 (T^\ast(t) - \overline{z})^{-1}\xi(t)\rangle -
 $$
 $$
 \langle (T^\ast(t) - \overline{w})^{-1}\xi'(t),(T^\ast(t) -
 \overline{z})^{-1}\xi(t)\rangle +
 $$ 
 $$
 \langle (T^\ast(t) - \overline{w})^{-1}\xi(t), (T^\ast(t) -
 \overline{z})^{-1} T^\ast (t) (T^\ast(t) -
 \overline{z})^{-1}\xi(t)\rangle -
 $$
 $$
 \langle (T^\ast(t) - \overline{w})^{-1})\xi(t),
 (T^\ast(t) - \overline{z})^{-1})\xi'(t)\rangle.
 $$
 A series of simplification in the case of elliptic growth are
 immediate: for instance $\| \xi(t)\|$ is proportional to the area
 of $D(t)$, whence we can choose the vector of the form
 $$ \xi(t) = t \xi(0).$$  Second, the higher harmonic moments
 are preserved by the evolution, whence the Cauchy transform/resolvent
 $$\frac{d}{dt} \pi \langle \xi(t), (T^\ast(t) -
 \overline{z})^{-1}\xi(t)\rangle = \frac{d}{dt} \int_{D(t)}
 \frac{dA(\zeta)}
 {\zeta - z} = - \frac{c}{z},$$
 gives full information about the first row and first column in the matrix
 representation of $T^\ast(t)$ in the  basis obtained by orthonormalizing
 the sequence $\xi(t), T^\ast(t)\xi(t), T^{\ast 2}\xi(t),...$. The reader
 can consult the article [122] for more details about computations
 related to the above ones.

\subsection{Linear analysis of quadrature domains} If we would infer
from the one-variable picture a good class of extremal domains for
Markov's $L$-problem in two variables we would choose the disjoint
unions of disks, as immediate analogs of disjoint unions of
intervals. In reality, the nature of the complex plane is much
more complicated, but again, fortunately for our survey, the class
of quadrature domains plays the role of extremal solutions in two
real dimensions.

Recall from our previous sections that a bounded domain $\Omega$
of the complex plane is called a {\it quadrature domain} (always
henceforth  for analytic functions) if there exists a finite set
of points $a_1, a_2, \ldots, a_d \in \Omega$, and real weights
$c_1, c_2, \ldots, c_d$, with the property:
$$ \int_\Omega f(z) dA(z) = c_1 f(a_1) + c_2 f(a_2) + \ldots +
c_d f(a_d),\ \  f \in AL^1(\Omega)$$ where the latter denotes the
space of all integrable analytic functions in $\Omega$. In case
some of the above points coincide, a derivative of $f$ can
correspondingly be evaluated.

Let $\Omega$ be a bounded planar domain with moments
$$ a_{mn} = a_{mn}(\Omega) = \int_\Omega z^m \overline{z}^n dA(z),
\ \ m,n \geq 0.$$ The exponential transform produces the sequence
of numbers $b_{mn} = b_{mn}(\Omega), \ \ m,n \geq 0.$ Let $T$
denote the irreducible hyponormal operator with rank-one
self-commutator $[T^\ast,T] = \xi \langle \cdot, \xi \rangle.$ In
virtue of the factorization (\ref{factorE}),
$$ b_{mn} = \langle T^{\ast m}\xi, T^{\ast n}\xi \rangle, \ \ m,n
\geq 0.$$ Hence the matrix $(b_{mn})_{m,n=0}^\infty$ turns out to
be non-negative definite. The following result identifies a part
of the extremal solutions of the $L$-problem of moments as the
class of quadrature domains:

{\it  A bounded planar domain $\Omega$ is a
quadrature domain if and only if there exists a positive integer
$d \geq 1$ with the property}
$ {\rm det} (b_{mn}(\Omega))_{m,n=0}^d = 0.$

For a proof see \cite{Put96}. The vanishing condition in the
statement is equivalent to the fact that the span $H_d$ of the
vectors $\xi, T^\ast \xi, T^{\ast 2}\xi, \ldots$ is finite
dimensional (in the Hilbert space where the associated hyponormal
operator $T$ acts). Thus, if $\Omega$ is a quadrature domain with
corresponding hyponormal operator $T$, and $T_d$ is the
compression of $T$ to the $d$-dimensional subspace $H_d$, then:
$$ E_\Omega(z,w) = 1 - \langle (T_d^\ast - \overline{w})^{-1}\xi, (T_d^\ast - \overline{z})^{-1}\xi
\rangle, \ \ z,w \in \overline{\Omega}^c.$$ In particular this
proves that the exponential transform of a quadrature domain is a
rational function. As a matter of fact a more precise statement
can easily be deduced:

{\it  Let $\Omega$ be the quadrature domain defined above. Then
$$ E_\Omega(z,w) = \frac{Q(z,w)}{P(z) \overline{P(w)}}, \ \ z,w
\in \overline{\Omega}^c.$$}

This result offers an efficient characterization of quadrature
domains in terms of a finite set of their moments (see the
reconstruction section below) and it opens a natural
correspondence between quadrature domains and certain classes of
finite rank matrices. We only describe a few results in this
direction. For more details see \cite{GP98,GP00,Put96}.

In the conditions of the above result, let $\Omega$ be a
quadrature domain with associated hyponormal operator $T$; let
$H_0 = \bigvee_{k\geq 0} T^{\ast k} \xi$ and let $p$ denote the
orthogonal projection of the Hilbert space $H$ (where $T$ acts)
onto $H_0$. Denote $C_0 = p T p$ (the compression of $T$ to the
$d$-dimensional space $H_0$) and $D_0^2 = [T^\ast, T]$. Then the
operator $T$ has a two block-diagonal structure:
$$
T= \left( \begin{array} {ccccc}
           {C_0}&0&0&0&\ldots\\
            {D_1}&{C_1}&0&0&\ldots\\
           0&{D_2}&{C_2}&0&\ldots\\
           0&0&{D_3}&{C_3}&\ldots\\
           \vdots& & \vdots& & \ddots\\
           \end{array} \right),
$$
where the entries are all $d \times d$ matrices, recurrently
defined by the system of equations:
$$
\left\{ \begin{array}{c}
          [{{C_{k}}^\ast}, {C_{k}}]
          +{{D_{k+1}}^\ast}{D_{k+1}}= {D_k}{{D_k}^\ast}\\
          {{C_{k+1}}^\ast}{D_{k+1}}
          = {D_{k+1}}{{C_k}^\ast},\hspace{.2in} k \geq 0.\\
          \end{array} \right.
$$
Note that $D_k >0$ for all $k$. This decomposition has an array of
consequences:
\begin{enumerate}
\item The spectrum of $C_0$ coincides with the quadrature nodes of
$\Omega$;

\item $\Omega = \{ z; \| (C_0^\ast - \overline{z})^{-1}\xi \| >1\}
$ (up to a finite set);

\item The quadrature identity becomes
$$
\int_\Omega f(z) {\rm dA}(z) = \pi \langle f(C_0)\xi, \xi\rangle,
$$
for $f$ analytic in a neighborhood of $\overline{\Omega}$;

\item The Schwarz function of $\Omega$ is
$$
S(z) = \overline{z} - \langle \xi,  (C_0^\ast -
\overline{z})^{-1}\xi\rangle  + \langle \xi, (T^\ast -
\overline{z})^{-1}\xi \rangle ,$$ where $ z \in \Omega$.

\end{enumerate}

To give the simplest and most important example, let $\Omega =
\mathbf D$ be the unit disk (which  is a quadrature domain of
order one) . Then the associated operator is the unilateral shift
$T = T_z$ acting on the Hardy space $H^2(\partial \mathbf D)$.
Denoting by $z^n$ the orthonormal basis of this space we have $T
z^n = z^{n+1}, \ \ n \geq 0,$ and $[T^\ast , T] = 1\langle \cdot,
1\rangle$ is the projection onto the first coordinate $1 = z^0$.
The space $H_0$ is one dimensional and $C_0 = 0$. This will
propagate to $C_k =0$ and $D_k =1$ for all $k$. Thus the matricial
decomposition of $T$ becomes the familiar realization of the shift
as an infinite Jordan block.

In view of the  linear algebra realization outlined in the
preceding section we obtain more information about the defining
equation of the quadrature domain. For instance:
$$
\frac{Q(z,\overline{w})}{P(z)\overline{P(w)}} = 1 - \langle
(C_0^\ast - \overline{w})^{-1}\xi, (C_0^\ast -
\overline{z})^{-1}\xi \rangle,
$$
which yields
$$
Q(z,{z}) = |P(z)|^2 - \sum_{k=0}^{d-1} |Q_k(z)|^2,
$$
where $Q_k$ is a polynomial of degree $k$ in $z$, see \cite{GP00}.

Thus the exponential transform of a quadrature domain contains
explicitly the irreducible polynomial $Q$ which defines the
boundary and the polynomial $P$ which vanishes at the quadrature
nodes. By putting together all these remarks we obtain a
strikingly similar picture to that of a single variable. 
More specifically, if $\Omega$ is a quadrature domain
with $d$ nodes, as given above, and associated hyponormal operator
$T$, then:
$$ E_\Omega(z,w) = \frac{Q(z,w)}{P(z) \overline{P(w)}} = 1 - \langle (T_d^\ast - \overline{w})^{-1}\xi, (T_d^\ast - \overline{z})^{-1}\xi
\rangle = $$
$$ \frac{1}{\pi^2} \sum_{i,j=1}^d
 H_\Omega(a_i,a_j) \frac{c_i}{a_i-z} \frac{
\overline{c_j}}{\overline{a_j}-\overline{w}}, \ \ z,w \in
\overline{\Omega}^c. $$

In particular we infer, assuming that all nodes are simple:
$$
-\pi^2 \frac{Q(a_i,a_j)}{P'(a_i) \overline{P'(a_j)}} = c_i
\overline{c_j} H_\Omega(a_i, \overline{a_j}), \ \ 1 \leq i,j \leq
d.$$ For details see \cite{Put96,GP00}.

The interplay between these additive, multiplicative and Hilbert
space decompositions of the exponential transform gives an exact
reconstruction algorithm of a quadrature domain from its moments.
The next section will be devoted to this algorithm.

Before ending the present section we consider an illustration of
the above formulas. Let $\Omega = \cup_{i=1}^d D(a_i, r_i)$ be a
union of $d$ pairwise disjoint disks. This is a quadrature domain
with data:
$$ P(z) = (z-a_1) \ldots (z-a_d),$$
$$ Q(z,w) = [(z-a_1)(\overline{w} - \overline{a_1}) - r_1^2]
\ldots [(z-a_d)(\overline{w} - \overline{a_d}) - r_d^2].$$ The
associated matrix $T_d$ is also computable, involving a sequence
of square roots of matrices, but we do not need here its precise
form. Whence the exponential transform is, for large values of
$|z|, |w|$:
$$ E_\Omega(z,w) = \prod_{i=1}^d [1- \frac{r_i^2}{(z-a_i)(\overline{w} -
\overline{a_i})}] =  1+ \sum_{i,j=1}^d \frac{Q(a_i,\overline{a_j})}{P'(a_i) \overline{P'(a_j)}}
\frac{r_i}{a_i-z} \frac{ r_j}{\overline{a_j}-\overline{w}}.$$

The essential positive definiteness of the exponential transform
of an arbitrary domain can be deduced, via an approximation
argument, from the positivity of the matrix $(-Q(a_i,
\overline{a_j}))_{i,j=1}^d$, where $Q$ is the defining equation of
a disjoint union of disks. We note that $(-Q(a_i,
\overline{a_j}))_{i,j=1}^d \geq 0$ is only a necessary condition
for the disks $D(a_i, r_i), \ 1 \leq i \leq d,$ to be disjoint.
Exact computations for $d=2$ immediately show that this matrix can
remain positive definite even the two disks overlap a little.
However, if two disks overlap, then, by adding an external disk,
even far away, this prevents the new $3 \times 3$ matrix to be
positive definite.

We end this section with two examples, covering the totality of
quadrature domains of order two.

{\bf Quadrature domains with a double node.} Let $z={w^2}+bw$ be
the conformal mapping of the disk $|w| <1$, where $b \geq 2$. Then
$z$ describes a quadrature domain $\Omega$ of order $2$, whose
boundary has the equation:$$
Q(z,\overline{z})=|z{|^4}-(2+{b^2})|z{|^2}-{b^2}z-{b^2}\overline{z}+1-{b^2}=0.$$

The Schwarz function of $\Omega$ has a double pole at $z=0$,
whence the associated $2 \times 2$-matrix $C_0$ is
nilpotent. Moreover, we know that:$$ |z{|^4} \|
({C_0^\ast}-\overline{z}{)^{-1}}\xi {\|^2} =
|z{|^4}-P(z,\overline{z}).$$ Therefore $$ \|
({C_0^\ast}+\overline{z}) \xi{\|^2}=
(2+{b^2})|z{|^2}+{b^2}z+{b^2}\overline{z} +{b^2}-1,$$ or
equivalently: $\|\xi{\|^2}=2+{b^2}, \langle {C_0^\ast}\xi, \xi
\rangle= {b^2}$ and $\| {C_0^\ast}\xi {\|^2}={b^2}-1.$

Consequently the linear data of the quadrature domain $\Omega$
are:$$ {C_0^\ast}=\left( \begin{array}{cc}
              0& \frac{{b^2}-1}{({b^2}-2{)^{1/2}}}\\
              0&0
              \end{array}\right) ,\hspace{.3in}
\xi=\left( \begin{array}{c}
           \frac{b^2}{({b^2}-1{)^{1/2}}}\\
           (\frac{{b^2}-2}{{b^2}-1}{)^{1/2}}
           \end{array} \right) . $$

{\bf  Quadrature domains with two distinct nodes.} Assume that the
nodes are fixed at $\pm 1$. Hence $P(z) = z^2 -1$. The defining
equation of the quadrature domain $\Omega$ of order two with these
nodes is: $$ Q(z, \overline{z}) = (|z+1|^2 - r^2)(|z-1|^2 -r^2)
-c,$$ where $r$ is a positive constant and $c \geq 0$ is chosen so
that either $\Omega$ is a union of two disjoint open disks (in
which case $c=0$), or $Q(0,0)=0$, see \cite{G2}. A short
computation yields:
$$ Q(z,\overline{z}) = z^2
\overline{z}^2 - 2r z \overline{z} -z^2 - \overline{z}^2
+\alpha(r), $$ where $$ \alpha(r) = \left\{ \begin{array}{lcc}
                  (1-r^2)^2 , & & r<1\\
                  0 ,& & r \geq 1.
                  \end{array}
                  \right. $$

One step further, we can identify the linear data from the
identity:
\begin{equation}
|P(z)|^2 (1- \| (C_0^\ast -\overline{z})^{-1} \xi \|^2) =
Q(z,\overline{z}).
\end{equation}
Consequently,
$$ \xi = \left( \begin{array}{c}
                 \sqrt{2} r \\
                 0
                 \end{array} \right) ,
                 \
         C_0^\ast = \left( \begin{array}{cc}
                   0 & \frac{\sqrt{2} r}{\sqrt{1-\alpha(r)}} \\
                   \frac{\sqrt{1-\alpha(r)}}{\sqrt{2}r} & 0
                   \end{array}
                   \right) .$$

This simple computation illustrates the fact that, although the
process is affine in $r$, the linear data of the growing domains
have discontinuous derivatives at the exact moment when the
connectivity changes.

\subsection{Signed measures, instability, uniqueness}

Contrary to the uniqueness of a quadrature domain for subharmonic
functions with a prescribed quadrature measure, quadrature domains
for harmonic or analytic functions are not determined by the
quadrature nodes and weights. This is an intriguing global
phenomenon which has haunted mathematicians for many decades. we
briefly record below some significant discoveries in this
direction.

 Consider  quadrature domains for
harmonic test functions and real-valued measures (\ref{mu}). As to
the relationship between the geometry of $\Omega$ and the location
of ${\rm supp\,}\mu$ there are then drastic differences between
the cases of having all $c_j>0$ respectively having no
restrictions on the signs of $c_j$. This is clearly demonstrated
in the following theorem due to M.~Sakai \cite{Sakai98},
\cite{Sakai99a}. The second part of the theorem is discussed (and
proved) in some other forms also in \cite{G1}, \cite{Gustafsson96a}, \cite{Bell03}, \cite{Bell04}, 
\cite{Zabrodin}, for
example:

 {\it Let $r$ and $R$ be  positive numbers, $R\geq
2r$. Consider measures $\mu$ of the form (\ref{mu}) with $c_j$
real and related to $r$ and $R$ by
\begin{equation}\label{supp}
{\rm supp\,}\mu \subset B(0,r),
\end{equation}
\begin{equation}\label{R}
\sum_{j=1}^n c_j = \pi R^2.
\end{equation}

\begin{enumerate}

\item[(i)] If $\mu\geq 0$, then any quadrature domain $\Omega$
for harmonic functions for $\mu$ is also a quadrature domain for
subharmonic functions. Hence the previous result  applies,
and in addition
$$
B(0,R-r)\subset \Omega\subset B(0,R+r).
$$

\item[(ii)] With $\mu$  not necessarily $\geq 0$, and with no restrictions on
$\sum_{j=1}^n |c_j|$ and $n$, any bounded domain containing
$B(0,r)$ and having area $\pi R^2$ can be uniformly approximated
by quadrature domains for harmonic functions for measures $\mu$
satisfying (\ref{supp}), (\ref{R}).
\end{enumerate}}

With $\mu$ a signed measure of the form (\ref{mu}) we still have
$\sum_{j=1}^n c_j =|\Omega|$, but $\sum_{j=1}^n |c_j|$ may be much
larger. In view of the theorem, the ratio
$$
\rho=\frac{\sum_{j=1}^n c_j}{\sum_{j=1}^n |c_j|} =\frac{\int
d\mu}{\int|d\mu|}
$$
($0<\rho \leq 1$) might give an indication of  how strong is the
coupling between the geometry of ${\rm supp\,}\mu$ and the
geometry of $\Omega$.

As mentioned, a quadrature domain for harmonic functions is not
always uniquely determined by its measure $\mu$. Still there is
uniqueness at the infinitesimal level: if
\begin{equation}\label{qi}
\sum_{j=1}^n c_j \varphi (a_j) = \int_\Omega \varphi\, {\rm dA}
\end{equation}
and (for example) the $a_j$ are kept fixed, then one can always
increase the $c_j$ (indefinitely) and get a unique evolution of
$\Omega$ (Hele-Shaw evolution). If $\partial\Omega$ has no
singularities then one can also decrease the $c_j$ slightly and
have a unique evolution (backward Hele-Shaw, which is ill-posed).
Thus it makes sense to write
$$
\Omega = \Omega (c_1, \dots, c_n)
$$
for $c_j$ in some interval around the original values. Note
however that decreasing the $c_j$ makes the ratio $\rho$ decrease,
indicating a loss of control or stability.

In the simply connected case, $\Omega$ will be the image of the
unit disc ${\mathbf D}$ under a rational conformal map $f=f_{(c_1,
\dots ,c_n)}: {\mathbf D}\to \Omega(c_1, \dots ,c_n)$. This
rational function is simply the conformal pull-back of the
meromorphic function $(z,S(z))$ on the Schottky double of $\Omega$
to the Schottky double of ${\mathbf D}$, the latter being
identified with the Riemann sphere. It follows that the poles of
$f$ are the mirror points (with respect to the unit circle) of the
points $f^{-1} (a_j)$. When the $c_j$ increase then the $|f^{-1}
(a_j)|$ decrease (this follows by an application of Schwarz' lemma
to $f^{-1}_{{\rm larger\,} c_j} \circ f_{{\rm original\,} c_j}$),
hence the poles of $f$ move away from the unit circle. Conversely,
the poles of $f$ approach the unit circle as the $c_j$ decrease,
also indicating a loss of stability.

For decreasing $c_j$ the evolution $\Omega(c_1, \dots ,c_n)$
always breaks down by singularity development of $\partial\Omega$
or $\partial\Omega$ reaching some of the points $a_j$ (see e.g.
\cite{Hohlov-Howison94}, \cite{Gustafsson-Vasiliev06}) before
$\Omega$ is empty, except in the case that $\Omega(c_1, \dots
,c_n)$ is a quadrature domain for subharmonic functions.  In the
latter case the $c_j$ (necessarily positive) can be decreased down
to zero, and $\Omega$ will be empty in the limit $c_1=\dots
=c_n=0$. However, it may happen that $\Omega(c_1, \dots ,c_n)$
breaks up into components under the evolution.

Assume now that $\Omega$ is simply connected. Then the analytic
and harmonic functions are equivalent as test classes for
(\ref{qi}). In the limit case that all the points $a_j$ coincide,
say $a_1=\dots=a_n=0$, then (\ref{qi}) corresponds to
\begin{equation}\label{qi0}
\sum_{j=1}^n c_j\varphi^{(j-1)} (0) =\int_\Omega \varphi\, {\rm
dA}
\end{equation}
for $\varphi$ analytic. The $c_j$ (allowed to be complex) now have
a slightly different meaning than before. In fact, they are
essentially the analytic moments of $\Omega$:
$$
 c_j =\frac{1}{(j-1) !} \int_\Omega z^{j-1} dA
\quad(j= 1,\dots , n).
$$
The higher order moments vanish, and the conformal map $f=f_{(c_1,
\dots ,c_n)}: {\mathbf D}\to \Omega(c_1, \dots ,c_n)$ (normalized
by $f(0)=0$, $f'(0)>0$) is a polynomial of degree $n$. A precise
form of the local bijectivity of the map $(c_1, \dots, c_n)\mapsto
\Omega (c_1, \dots, c_n)$ has been established by O.~Kouznetsova
and V.~Tkachev \cite{Kouznetsova-Tkachev2004}, \cite{Tkachev2005}, 
who proved an  explicit formula for the (nonzero) Jacobi
determinant of the map from the coefficients of $f$ to the moments
$(c_1,\dots, c_n))$. This formula was conjectured (and proved in
some special cases) by C.~Ullemar \cite{Ullemar80}.

On the global level, it does not seem to be known whether
(\ref{qi0}), or (\ref{qi}), with a given left member, can hold for
two different simply connected domains and all analytic $\varphi$.

Leaving the realm of quadrature domains, an explicit example of
two different simply connected domains having the same analytic
moments has been given by M.~Sakai \cite {Sakai78}. The idea of
the example is that a disc and a concentric annulus of the same
area have equal moments. If the disc and annulus are not
concentric, then the union of them (if disjoint) will have the
same moments as the domain obtained by interchanging their roles.
Arranging everything carefully, with removing and adding some
common parts, two different Jordan domains having equal analytic
moments can be obtained. Similar examples were known earlier by
A.~Celmin{s} \cite {Celmins57}, and probably even by
P.~S.~Novikov.
 On the positive side, a classical theorem of Novikov
\cite{Novikoff38} asserts that domains which are starshaped with
respect to one and the same point are uniquely determined by their
moments. See \cite{Zalcman87} for further discussions.

Returning now to quadrature domains, there is definitely no
uniqueness for harmonic and analytic test classes if multiply
connected domains are allowed. If $\Omega$ has connectivity $m+1$
($m\geq 1$), i.e., has $m$ ``holes'', then there is generically an
$m$-parameter family $\Omega (t_1,\dots,t_m)$ of domains such that
$\Omega (0,\dots,0)=\Omega$ and
$$
\frac{\partial}{\partial t_j} \int_{\Omega (t_1,\dots,t_m)}
\varphi\,{\rm dA} =0 \quad (j=1,\dots ,m)
$$
for every $\varphi$ analytic in a neighborhood of the domains.
These deformations are Hele-Shaw evolutions, driven not by Green
functions but by ``harmonic measures'', i.e., regular harmonic
functions which take (different) constant boundary values on the
components of $\partial\Omega$.

It follows that multiply connected quadrature domains for analytic
functions for a given $\mu$ occur in continuous families. It even
turns out \cite{Gustafsson90}, \cite{Sjodin04} that {\em any
two} algebraic domains for the same $\mu$ can be deformed into
each other through families as above. Thus there is a kind of
uniqueness at a higher level: given any $\mu$ there is at most one
connected family of algebraic domains belonging to it.

For harmonic quadrature domains there are no such continuous
families (choosing $\varphi (z) =\log |z-a|$ in (\ref{qi}) with
$a\in \mathbb{C}\setminus\overline{\Omega}$ in the holes stops
them), but one can still construct examples with a discrete set of
different domains for the same $\mu$. It is for example possible
to imitate the example with a disc and an annulus with quadrature
domains for measures $\mu$ of the form (\ref{mu}), with
$a_j=e^{2\pi j/n}$ ($n\geq 3$) and $c_1=\dots =c_n=c>0$ suitably
chosen. However, it seems very
difficult to imitate the full Sakai construction, with ``removing
and adding some common parts'', in the context of quadrature
domains. Therefore it is not at all easy to  construct different
simply connected quadrature domains for the same $\mu$.

We end this section with the simplest example of a continuous
class of quadrature domains with the same quadrature data.

{\bf Three points, non-simply connected quadrature domains and the
non-uniqueness phenomenon.}\  Quadrature domains (for analytic
functions) with at most two nodes, as in the above examples, are
uniquely determined by their quadrature data and are simply
connected. For three nodes and more it is no longer so. The
following example, taken from \cite{G2}, with three nodes and
symmetry under rotations by $2\pi/3$, illustrates the general
situation quite well. More details on the present example are
given in \cite{G2}, and similar examples with more nodes are
studied in  \cite{Crowdy-Marshall03}.

Let the quadrature nodes and weights be $a_j=\omega^j$ and
$c_j=\pi r^2$ respectively ($j=1,2,3$), where $\omega=e^{2\pi
i/3}$ and where $r>0$ is a parameter. Considering first the
strongest form of quadrature property, namely for subharmonic
functions, as in (\ref{subharmqd}), (\ref{mu}), the situation is
in principle easy: $\Omega$ is for any given $r>0$ uniquely
determined up to nullsets and can be viewed as a swept out version
of the quadrature measure $\mu=\sum_{j=1}^3 c_j \delta_{a_j}$ or
as the union of the discs $B(a_j,r)$ with (possible) multiple
coverings smashed out.

For $0<r\leq \frac{\sqrt{3}}{2}$ the above discs are disjoint,
hence $\Omega =\cup_{j=1}^3 B(a_j,r)$. For $r$ larger than
$\frac{\sqrt{3}}{2}$ but smaller than a certain critical value
$r_0$ (which seems to be difficult to determine explicitly)
$\Omega$ is doubly connected with a hole containing the origin,
while for $r\geq r_0$ the hole will be filled in so that  $\Omega$
is a simply connected domain. The above quadrature domains (or
open sets) are actually uniquely determined even within nullsets,
except in the case $r=r_0$ when both $\Omega$ and $\Omega\setminus
\{ 0\}$ satisfy (\ref{subharmqd}).

Consider next the general class of quadrature domains for analytic
functions (algebraic domains). For $0<r\leq \frac{\sqrt{3}}{2}$
only the disjoint discs qualify, as before. However, for any
$r>\frac{\sqrt{3}}{2}$ there is a whole one-parameter family of
domains $\Omega$ satisfying the quadrature identity for analytic $\varphi$.
These are defined by the polynomials
\begin{equation}\label{polynomial}
Q(z,\overline{z}) = z^3 \overline{z}^3 - z^3 -\overline{z}^3
-3r^2z^2\overline{z}^2-
\end{equation}
$$
3\tau (\tau^3 - 2r^2\tau +1)z\overline{z} +\tau^3 (2\tau^3 -
3r^2\tau +1),
$$
where $\tau>0$ is a free parameter, independent of the quadrature
data. When completed as to nullsets, the quadrature domains in
question are more precisely
$$
\Omega (r,\tau)= {\rm int clos}\{z\in \mathbb{C}:
Q(z,\overline{z})<0\}.
$$

The interpretation of the parameter $\tau$ is that on each radius
$\{z=t\omega^{j+\frac{1}{2}}: t>0 \}$, $j=1,2,3$, there is exactly
one singular point of the algebraic curve $Q(z,\overline{z})=0$,
and $\tau=|z|$ for that point. This singular point is either a
cusp on $\partial\Omega$ or an isolated point of
$Q(z,\overline{z})=0$, a so-called {\it special point}. Special
points are those points $a\in\Omega$ for which the quadrature
identity  admits the (integrable) meromorphic
function $\varphi (z)=\frac{1}{z-a}$. Equivalently,
$\Omega\setminus\{a\}$ remains to be a quadrature domain for
integrable analytic functions.

For $\frac{\sqrt{3}}{2}<r<2^{-\frac{1}{6}}$ the quadrature domains
for analytic functions are exactly the domains $\Omega(r,\tau)$
(with possible removal of special points) for $\tau$ in an
interval $\tau_1(r)\leq \tau\leq\tau_2(r)$, where $\tau_1(r)$,
$\tau_2(r)$ satisfy $0<\tau_1(r)<\frac{1}{2}<\tau_2(r)$, and more
precisely can be defined as the positive zeros of the polynomial
$4\tau^3-4r^2\tau +1$. (see \cite{G2} for further explanations and
proofs). The domains $\Omega(r,\tau)$ are doubly connected with a
hole containing the origin. When $\tau$ increases the hole shrinks
and both boundary components move towards the origin. For
$\tau=\tau_2(r)$ there are three cusps on the outer boundary
component which stop further shrinking of the hole, and for
$\tau=\tau_1(r)$ there are three cusps on the inner boundary
component which stop the expansion of the hole.

For exactly one parameter value, $\tau=\tau_{\rm subh}(r)$,
$\Omega(r,\tau)$ is a quadrature domain for subharmonic functions
(and so also for harmonic functions). This $\tau_{\rm subh}(r)$
can be determined implicitly by evaluating the quadrature identity
for $\varphi(z)=\log |z|$, which gives the equation
$$
\int_{\Omega(r,\tau_{\rm subh}(r))} \log |z| \,{\rm dA}(z)=0.
$$

For $r=\frac{\sqrt{3}}{2}$, $\tau_1 (r)=\tau_{2}(r)=\frac{1}{2}$,
and as $r$ increases, $\tau_1(r)$ decreases and $\tau_2(r)$
increases. What happens when $r= 2^{-\frac{1}{6}}$ is that for
$\Omega(r,\tau_2(r))$, i.e., for the domain with cusps on the
outer component, the hole has shrunk to a point (the origin).
Hence, for $r=2^{-\frac{1}{6}}$, $\Omega(r,\tau_2(r))$ is simply
connected, while $\Omega(r,\tau)$ for $\tau_1(r)\leq
\tau<\tau_2(r)$ remain doubly connected.

For all $\frac{\sqrt{3}}{2}<r\leq 2^{-\frac{1}{6}}$,
$\tau_1(r)<\tau_{\rm subh}(r)<\tau_2(r)$ because a subharmonic
quadrature domain cannot have the type of cusps which appear for
$\tau=\tau_1(r), \tau_2(r)$ (see \cite{S4}, \cite{S5}). It follows
that the critical value $r=r_0$, when $\Omega(r,\tau_{\rm
subh}(r))$ becomes simply connected, is larger that
$2^{-\frac{1}{6}}$.

For $r\geq  2^{-\frac{1}{6}}$ the quadrature domains for analytic
functions are the domains $\Omega(r,\tau)$ (with possible deletion
of special points), with $\tau$ in an interval
$\tau_1(r)\leq\tau\leq\tau_3(r)$. Here $\tau_1(r)$ is the same as
before (i.e., corresponds to cusps on the inner boundary), while
$\tau_3(r)$ is the value of $\tau$ for which the hole at the
origin degenerates to just the origin itself (which for $r>
2^{-\frac{1}{6}}$ occurs before cusps have developed on the outer
boundary). The origin then is a special point, and one concludes
from (\ref{polynomial}) that $\tau=\tau_3(r)$ is the smallest
positive zero of the polynomial $2\tau^3 -3r^2\tau+1$. For
$r=2^{-\frac{1}{6}}$, $\tau_3(r)=\tau_2(r)= 2^{-\frac{2}{3}}$.

For $2^{-\frac{1}{6}}\leq r<r_0$ we have $\tau_1(r)<\tau_{\rm
subh}(r)<\tau_3(r)$, while for $r\geq r_0$, $\tau_{\rm
subh}(r)=\tau_3(r)$. Since $\Omega(r,\tau_3(r))$ is simply
connected and is a quadrature domain for analytic functions it is
also a quadrature domain for harmonic functions. It follows that
in the interval  $2^{-\frac{1}{6}}\leq r<r_0$ there are (for each
$r$) two different quadrature domains for harmonic functions,
namely $\Omega(r,\tau_{\rm subh}(r))$ and $\Omega(r,\tau_{3}(r))$
(doubly respectively simply connected).

In summary, we have for each $r>\frac{\sqrt{3}}{2}$ a
one-parameter family of algebraic domains $\Omega(r,\tau)$, for
exactly one parameter value ($\tau=\tau_{\rm subh}(r)$) this is a
quadrature domain for subharmonic functions, and for each $r$ in a
certain interval ($2^{-\frac{1}{6}}\leq r<r_0$) there are two
different quadrature domains for harmonic functions
($\Omega(r,\tau_{\rm subh}(r))$ and $\Omega(r,\tau_{3}(r))$).

\section{Other physical applications of the operator theory formulation}

The preceding chapters provide a review of the relationships between 
the theory of normal random matrices, where evolution is defined by 
increasing the size of the matrix (a discrete time), its continuum (or infinite
size) limit - Laplacian growth -  and the general theory of semi-normal 
operators whose spectrum approximates generic domains. The exposition 
reflects, to some extent, the parallel historical development of
the two non-commutative generalizations of Laplacian growth (random
matrix theory and semi-normal operator theory). It is quite natural, at this
point, to investigate the direct relationships between these two theories. 
However, this is a task of a magnitude which would require a separate 
review at the very least. We will therefore contend ourselves with exposing
only a few of these relations, via their applications to physical problems. 

The first application has to do with refined asymptotic expansions which 
characterize Laplacian growth in the critical case, before formation of a 
$(2, 3)$ cusp. As we will see, to obtain this limit, one must take a ``double-scaling limit"
by fine-tuning two parameters of the random matrix ensemble. Alternatively, this
procedure is equivalent to a special choice of Pad\'e approximants 
in the operator theory approach. 

The second application described in this section is a very brief introduction of the notion of 
free, non-commutative random variables, and its relevance in open problems 
of strongly interacting quantum models, particularly in the 2D metal-insulator 
transition and the determination of ground state for 2D spin models. The review
concludes with this cursory exposition. 

\subsection{Cusps in Laplacian growth: Painlev\'e equations}

In this section, we exploit the formalism built up to now, in order to address a 
problem of great significance both at the mathematical and physical levels: 
what happens when a planar domain evolving under Laplacian growth approaches 
a generic (2, 3) cusp? We have already seen that a $classical$ solution does not
exist, in that no singly-connected domain with uniform density would satisfy the
conditions of the problem. However, since we now have alternative formulations 
of Laplacian growth via the $balayage$ of the uniform measure, we may generalize
the problem and ask whether there is $any$ equilibrium measure, dropping the 
uniformity (and indeed, the two-dimensional support) of the classical solution. 
By analogy with the 1D situation, we seek a solution in the sense of Saff and Totik, 
where the support and density of the equilibrium measure are given by the proper
weighted limit of orthogonal polynomials. In order to obtain this limit, we must organize 
the evolution equations of the wavefunction such as to extract the correct scaling 
limit, for $N \to \infty$.

\subsubsection{Universality in the scaling region at critical points -- a conjecture}

Detailed analysis of critical Hermitian ensembles indicates that the 
behavior of orthogonal polynomials in a specific region including the
critical point (the {\it {scaling region}}), upon appropriate scaling
of the degree $n$, is essentially independent of the bulk features of the ensemble. This $universality$ property (a common working hypothesis in the physics of critical phenomena) is expected 
to occur for critical NRM ensembles as well -- and is indeed easy to 
verify in critical Gaussian models, $2|t_2|=1$. Analytically, it means that by suitable
scaling of the variables $z, n$: 
$$
n \to \infty, \,\,\, \hbar \to 0, \,\,\, n\hbar = t_0, \,\,\,
t_0 = t_c - \hbar^{\delta} \nu, \,\,\, z = z_c + \hbar^{\epsilon} \zeta,
$$  
where $z_c$ is the location of the critical point and $t_c$ the critical
area, the wave function $\Psi_n(z)$ will reveal a universal part 
$\phi(\nu,\zeta)$ which depends exclusively on the local singular 
geometry $x^p \sim y^q$ ($p, q$ mutual primes) of the complex curve at the critical point. This conjecture is a subject of active research. Its main consequence is that in order to describe the 
scaling behavior for a certain choice of $p, q$, 
it is possible to replace a given ensemble with another which leads to
the same type of critical point, though they may be very different at 
other length scales. 

\subsubsection{Scaling at critical points of normal matrix ensembles} \label{painl}

In the remainder of the section we analyze the regularization of Laplacian
Growth for a critical point of type $p=3, q=2$, by discretization of the 
conformal map as described in the previous paragraph. For simplicity, we
start from the conformal map corresponding to the potential $V(z) = t_3z^3$, which is the simplest model leading to the specified type of cusp. It 
should be noted that the analysis will be identical for any monomial potential $V(z) = t_nz^n, n \ge 3$; for every such map, $n$ singular 
points of type $p=3, q=2$ will form simultaneously on the boundary. The 
critical boundary corresponding to $n=3$ is shown in Figure~\ref{critical}. 

\paragraph{The scaling limit from the string equation}  

We start from the Lax pair corresponding to the potential $V(z)=t_3z^3$:
\be
L\psi_n = r_n \psi_{n+1} + u_n \psi_{n-2}, \,\, 
L^{\dag}\psi_n = r_{n-1}\psi_{n-1} + \bar u_{n+2}\psi_{n+2}.
\ee
The string equation (\ref{string}) $[L^{\dag}, \, L] = \hbar$ 
translates into 
\begin{eqnarray} \nonumber
(r^2_n +|u|_n|^2 - r^2_{n-1} - |u|_{n+2}^2) \psi_n  +  (r_n\bar u_{n+3} - r_{n+2}\bar u_{n+2}) \psi_{n+3} \\
+(r_{n-3}u_n - r_{n-1}u_{n-1}) \psi_{n-3} = \hbar \psi_n.
 \end{eqnarray}
Identifying the coefficients gives
\be \la{ar}
\left ( r_n^2 - |u|^2_{n+2} - |u|^2_{n+1} \right ) - \left ( r_{n-1}^2 - |u|^2_{n+1} - |u|^2_{n} \right ) = \hbar, 
\ee
and
\be \la{co}
\frac{\bar u_{n+2}}{r_nr_{n+1}} = \frac{\bar u_{n+3}}{r_{n+1}r_{n+2}} = 3t_3.
\ee
Equation (\ref{ar}) gives the quantum area formula
\be
r_n^2 - (|u|_{n+2}^2 + |u|_{n+1}^2) = n\hbar,
\ee
which together with the conservation law (\ref{co}) leads to the discrete Painlev\'e equation
\be
r_n^2 \left [ 1 - 9|t_3|^2 (r_{n-1}^2  + r_{n+1}^2) \right ] = n\hbar.
\ee

\noindent In the continuum limit, the equation becomes
\be
r^2 - 18|t_3|^2 r^4 = t_0.
\ee
The critical (maximal) area is given by 
\be
\frac{dt_0}{dr^2} = 0, \,\,  36|t_3|^2r^2_c = 1. 
\ee
Choosing $r_c =1$ gives $6|t_3| = 1$ and $t_c = \frac{1}{2}$. It also follows that
\be
u_n = \frac{r_{n-2}r_{n-1}}{2}, \,\, z_c = \frac{3}{2}.
\ee
Introduce the notations
\be
N\hbar = t_c, \,\, n\hbar = t_0 = t_c +\hbar^{4a} \nu, \,\, r_n^2 = 1 - \hbar^{2a}u(\nu) , 
\,\, z = \frac{3}{2} + \hbar^{2a}\zeta,    
\ee
where $a=\frac{1}{5}.$ We get $\p_n = \hbar^a \p_\nu$ and
\be
r^2_{n+k} = 1 - \hbar^{2a}u - k \hbar^{3a}\dot u(\nu) - \frac{k^2}{2}\hbar^{4a}\ddot{u}, 
\ee
where dot signifies derivative with respect to $\nu$. 
The scaling limit of the quantum area formula becomes
\be
(1-\hbar^{2a}u)\left [\frac{1}{2} + \hbar^{2a}\frac{u}{2} + \hbar^{4a} \frac{\ddot u}{4} \right ] = \frac{1}{2} + \hbar^{4a}\nu,
\ee
giving at order $\hbar^{4a}$ the Painlev\'e I equation
\be \la{painleve}
\ddot u - 2 u^2 = 4 \nu.
\ee
Rescaling $u \to c_2 u$, $\nu \to c_1 \nu$ gives the standard form
\be
\ddot u - 3u^2 = \nu,
\ee
for $c_2 = 4c_1^3, 8c_1^5 = 3$.

\paragraph{Painlev\'e I as compatibility equation}

Inspired by the Saff-Totik approach, we construct the wavefunctions based 
on monic polynomials,  ($\mbox{Pol}$ is the polynomial part)
\be
\phi_n = \prod_{i=0}^{n-1}r_i \psi_n, \,\, 
\mbox{Pol } \phi_n(z) = z^n + O(z^{n-1}),
\ee
and rewrite the equations for the Lax pair as
\be
L \phi_n = \phi_{n+1} + \frac{r^2_{n-2}r^2_{n-1}}{2} \phi_{n-2}, \,\, 
L^{\dag} \phi_n = r^2_{n-1}\phi_{n-1} + \frac{\phi_{n+2}}{2}.
\ee
Notice that using the shift operator $\mathcal{W}$, the system can also be written
\be
L = \mathcal{W} + \frac{1}{2}\left ( r^2_{n-1}\mathcal{W}^{-1} \right )^2, \,\, 
L^{\dag} = r^2_{n-1}\mathcal{W}^{-1} + \frac{1}{2} \mathcal{W}^2. 
\ee
Introduce the scaling $\psi$  function through 
\be
\phi_n(z) = e^{\frac{z^2}{2\hbar}} \psi(\zeta, \nu).
\ee
The action of Lax operators on $\psi$ gives the representation
\be
L = \frac{3}{2} + \hbar^{2a}\zeta, \, \, L^{\dag} = z + \hbar \p_\zeta = \frac{3}{2} + \hbar^{2a} \zeta + \hbar^{3a} \p_\zeta.
\ee
Therefore, the action of $\zeta$ is given by the sum of equations at order $\hbar^{2a}$:
\be
3 + 2 \hbar^{2a} \zeta  = \mathcal{W} + \frac{1}{2} \mathcal{W}^2 + r^2_{n-1}\mathcal{W}^{-1}
+ \frac{1}{2}\left ( r^2_{n-1}\mathcal{W}^{-1} \right )^2, 
\ee
and the action of $\p_\zeta$ by their difference:
\be
\hbar^{3a}\p_\zeta  = - \mathcal{W} + \frac{1}{2} \mathcal{W}^2 + r^2_{n-1}\mathcal{W}^{-1}
- \frac{1}{2}\left ( r^2_{n-1}\mathcal{W}^{-1} \right )^2. 
\ee
Equivalently, we can write
\be
\hbar^{2a}\zeta = \frac{1}{2} \left [ \left ( \mathcal{W}+1 \right )^2 + 
\left ( r^2_{n-1}\mathcal{W}^{-1} + 1 \right )^2  \right ] - 4,
\ee
\be
\hbar^{3a}\p_\zeta
= \frac{1}{2} \left [ \left ( \mathcal{W}-1 \right )^2 - 
\left ( r^2_{n-1}\mathcal{W}^{-1} - 1 \right )^2  \right ].
\ee
Expanding the shift operator in $\hbar$ leads to
\be
\mathcal{W} = 1 + \hbar^{a} \p_\nu + \hbar^{2a}\frac{\p^2_\nu}{2} + \hbar^{3a}\frac{\p^3_\nu}{6},
\ee
and
\be
r^2_{n-1}\mathcal{W}^{-1} = 1 - \hbar^{a}\p_\nu + \hbar^{2a}\left (\frac{\p^2_\nu}{2}-u \right) + 
\hbar^{3a}\left ( -\frac{\p^3_\nu}{6}  + u \p_\nu + \dot u \right ).
\ee
Substituting into the equations for $\zeta, \p_\zeta$ gives the system of equations
\be
\ddot \psi = \frac{2(\zeta + u)}{3}\psi, \,\,\,\, \psi' = \frac{\dot u}{6} \psi +  \frac{2\zeta - u}{3}\dot \psi, 
\ee
where primed variables are differentiated with respect to $\zeta$. The equations can be written in matrix form as
\be
\Psi ' = \Lambda \Psi, \,\, \dot \Psi = Q \Psi, \,\, \Psi = 
\left (
\begin{array}{c}
\psi \\
\dot \psi
\end{array}
\right ),
\ee
where
\be
\Lambda = 
\left (
\begin{array}{cc}
\frac{\dot u}{6} & \frac{2\zeta - u}{3} \\
\frac{\ddot u}{6} + \frac{2(\zeta + u)(2\zeta - u)}{9} \,\,\,\, & -\frac{\dot u}{6}
\end{array}
\right ), \,\,
Q = 
\left (
\begin{array}{cc}
0 & 1 \\
\frac{2(\zeta + u)}{3} \,\,\, & 0
\end{array}
\right ).
\ee
The compatibility equations 
\be
\dot \Lambda - Q' = [Q, \,\Lambda]
\ee
yield the Painlev\'e equation derived in the previous section:
\be
\dot \Lambda = 
\left (
\begin{array}{cc}
\frac{\ddot u}{6} & -\frac{ \dot u}{3} \\
\frac{\stackrel{\ldots}{u}}{6} + \frac{2\zeta \dot u -4u \dot u }{9} \,\,\,\, & -\frac{\ddot u}{6}
\end{array}
\right ), \,\,
Q' = 
\left (
\begin{array}{cc}
0 & 0\\
\frac{2}{3} \, & 0
\end{array}
\right ),
\ee
and
\be
[Q, \,\, \Lambda ] =
\left (
\begin{array}{cc}
\frac{\ddot u}{6} & -\frac{ \dot u}{3} \\
\frac{2(\zeta+u) \dot u }{9} \,\,\,\, & -\frac{\ddot u}{6}
\end{array}
\right ).
\ee
Thus,
\be
0 = \dot \Lambda - Q' - [Q, \,\Lambda] =
\left (
\begin{array}{cc}
0 & 0\\
\frac{\stackrel{\ldots}{u}}{6} - \frac{6u \dot u}{9} - \frac{2}{3} \,\,\,\,\,\, & 0
\end{array}
\right ).
\ee
The only non-trivial element of the matrix gives
\be
\stackrel{\ldots}{u} -4u\dot u - 4 = 0,
\ee
$i.e.$ the Painlev\'e equation derived in the previous section.

\subsubsection{Conclusions} The derivations presented above indicate that, 
in the vicinity of a (2, 3) cusp, the refined asymptotics for Laplacian growth 
are based on the behavior of the Baker-Akhiezer function for the Painlev\'e I
equation. This fact allows to properly define the evolution of the domain 
beyond the critical time, by identifying the support of the measure with the 
support of the zeros of this function. This is a work in progress which will be
reported elsewhere. 

It is also interesting to note that the double-scaling limit required to derive
the refined asymptote mirrors an earlier result, due in its original form to 
 Stahl \cite{Stahl}, and related to orthogonal polynomials in \cite{Saf}. It 
 describes an approximation of the Cauchy transform of a planar domain
via a special sequence of Pad\'e approximants (in the spirit of section 
5.3.2), which by exponentiation would translate into the double-scaling
limit presented in this section.

\subsection{Non-commutative probability theory and 2D quantum models}

We conclude this review with a brief presentation of outstanding problems in
two-dimensional quantum models, where the use of random matrix theory 
led to important results, and (perhaps most importantly) pointed out to the
need for a probability theory for {\emph{non-commutative}} random variables. 
In turn, such a theory is intimately related to the semi-normal operator approach
presented in the previous chapter. 

\subsubsection{Metal-insulator transition in two dimensions}

The details of the transition from conductive to insulating behavior for a system
of interacting 2D electrons, in the presence of disorder, referred to as metal-insulator
transition (or MIT) are not well understood, despite decades of research.  Here 
we give a very sketchy description of this problem, in order to illustrate the
mathematical essence of the model and of the difficulties, and we refer the reader 
to one of the several excellent monographs on the subject \cite{RevModPhys}. The fact that a system 
of electrons may ``jam", i.e. behave like an insulator, because of either strong interactions 
(Mott transition) or strong disorder (Anderson transition) has been known for roughly 
half a century. However, creating a theoretical model which could incorporate both 
interactions and disorder in a proper fashion, was difficult to achieve. The foundation 
for our current formulation of this problem was laid by Wegner \cite{Wegner}, and 
later improved by Efetov \cite{Efetov}. A very clear exposition of this formulation can
be found in the synopsis \cite{Zirnbauer}.

In its simplest formulation, the model consists of a lattice in $d-$dimensions (which
may be taken to be $\mathbb{Z}^d$), where to each vertex corresponds an $n-$dimensional
vector space of states (also called orbitals), and with hamiltonian 
$$
H = H_0 + H_d, \quad H_0 = \sum_{n, \langle x, y \rangle} t_{x,y} |x, n \rangle \langle y, n|, 
\quad H_d = \sum_{x, i, j} f^{ij} | x, i\rangle \langle x, j |, 
$$ 
where the state $|n, x\rangle$ depends on position $x$ and orbital $n$, $H_0$  refers to the 
interaction between adjacent vertices and $H_d$ implements the disorder component, 
via the random matrix $f^{ij}$, which can be Hermitian, Orthogonal, Symplectic, etc. based on 
symmetries of the system. Efetov's idea was to use supersymmetry to incorporate 
interactions and disorder on the same footing; the method was later extended to 
 implement the ``Hermitization" of non-Hermitian 
random matrices with non-Gaussian weights, appearing in the same physical
context \cite{Feinberg}. For rotationally invariant measures, the authors showed 
that the distribution of eigenvalues can be either a disc or an annulus, and that
there is a phase transition between the two, as a function of model
parameters. 

The difficulties related to this formulation of the problem are due to the fact that 
the transition cannot be described within the established models of phase transitions. 
In all these models, the state of the system is obtained by minimization of a proper 
thermodynamic potential (for instance, free energy), or equivalently, finding the 
points of extrema of action in a path-integral approach (via a saddle-point condition). 
``Proper" phase transitions are characterized by potentials that are globally convex, 
so that the minimization problem is well-defined. However, the supersymmetric 
formulation of MIT does not lead to a true extremum, but rather a saddle-point, due 
to the non-compact, hyperbolic geometry structure of the effective theory ($SU(1,1)$
in the simplest case).  The interested reader can find a detailed exposition of this
phenomenon in \cite{Spencer}, for example. In the case where the system has a 
finite scale, it can be shown, following Efetov, that only the zero modes of the theory
are important, which leads to an effective simplification in computing the multipoint 
correlation functions. However, the full model is still not solved for the 2D case, due
to the difficulties pointed out above. 

In a nutshell, we may summarize the problem as non-tractable using the standard 
statistical physics formulation of phase transitions. In that sense, the situation is
similar to another famous unsolved physical model, the disordered spin problem 
in the presence of magnetic field, where determination of the ground state is a task
of exponential complexity (with respect to the size of the system).  The phase transition
where the system goes from an ordered state to a state with local order but no long-range
order (a spin glass) is equally intractable as MIT, for the reasons explained.

Interestingly enough, both problems may be {\emph{approximately}} studied using a
physicist's approach notorious for its lack of control: the replica-symmetry breaking 
(RSB) \cite{Parisi}. We mention it here mainly because of its statistical interpretation. 

Starting from the elementary observation $\log Z = \lim_{n \to 0}(Z^n - 1)/n$, it is tempting 
to replace averages (over disorder) computed from the thermodynamic potential $\langle \log Z \rangle$,
with averages computed with $\langle Z^n \rangle$, because of the implicit assumption
that repeated products of the random variable $Z$ will {\emph{self-average}} (an implicit
application of the Central Limit Theorem). By extension, one may assume that averages 
of products of operators, projected on special states, $\langle 0 |\phi_1 \phi_2 \ldots \phi_k | 0 \rangle$
(correlation functions), may also be computed by the same argument. 

In this (statistical inference) approach, the failure of standard descriptions of phase transitions 
is related to reducing correlation functions of products of operators, to their {\emph{projections}} 
onto selected states. At the critical point, such projections do not have the expected convergence
properties.  It is therefore natural to ask whether one may use other (weaker) criteria to 
determine the critical point. In particular, is it possible to define statistical inference for the 
operators themselves, rather than special projections? 

The answer is affirmative, and such a theory was constructed almost in parallel with the 
MIT and spin glass models described above.  

\subsubsection{Non-commutative probability theory and free random variables}

The basic elements in the probability theory for non-commutative operators \cite{Voiculescu}
are the following: $\mathcal{A}$, a non-commutative (operator) algebra over $\mathbb{C}$, $1 \in \mathcal{A}$;
a functional $\phi: \mathcal{A} \to \mathbb{C}, \phi(1) = 1$, called expectation functional. 
 
Quantum mechanics offers specific examples: 

\begin{itemize}
\item Example 1: $\mathcal{A}=$ bounded operators over Hilbert space of states $\mathcal{H}$, 
$\xi \in \mathcal{H}, ||\xi||=1$, the ground state, and 
$$
\phi(A) = \langle \xi | A |\xi \rangle. 
$$ 
\item Example 2:  $\mathcal{A}=$ von Neumann algebra over $\mathcal{H}$, and functional $\phi = $ Tr.  
\end{itemize}

In order to develop inference methods within this theory, it is necessary to define 
the equivalent of independent variables in commutative probability. Such variables 
are called {\emph{free}}, and satisfy the following property: $A_1, A_2, \ldots, A_k$ 
are free if $\phi(A_i) = 0,$ and 
$$
\phi(A_{i_1} A_{i_2} \ldots A_{i_k}) = 0, \quad A_{i_j} \ne A_{i_{j+1}}.
$$

Using these tools, generalizations of standard results in large sample theory 
are possible. We mention a few: 

\begin{itemize}
\item The ``Gaussian" distribution (limit distribution for Central Limit Theorem) 
in free probability theory is given by operators with eigenvalues obeying the 
semi-circle distribution (Wigner-Dyson) $\rho(\lambda) = \sqrt{a^2-\lambda^2}$; 
\item Similarly, the Poisson distribution has as free correspondent the operators 
with eigenvalues distribution according  to the Marchenko-Pastur (elliptical law),
 $\rho(\lambda) = \sqrt{(\lambda -a )(b-\lambda)}$; 
\item The free Cauchy distribution is the Cauchy distribution itself. 
\end{itemize}
Likewise, there is a notion of free Fisher entropy, Cram\'er-Rao bound, etc. 

As announced earlier, the relation between this theory, random matrices and 
operator theory for 2D spectral support is two-fold: on one hand, we have the 
important result that random matrices, in the large size limit, become free 
non-commutative random variables. Thus, inference in free non-commutative 
probability may be approximated using ensembles of random matrices, which
explains the success of this concept in the physics of disordered quantum systems.

On the other hand, the limit distributions specified above (via Wigner-Dyson, 
Marchenko-Pastur laws, and their 2D counterparts), are described through spectral 
data. Taking this as a starting point, it is relevant to construct sequences of operators
which approximate the spectrum, which points directly to the methods of section 5. 

As a last remark, an early attempt to employ non-commutative probability theory 
in MIT was reported in \cite{Speicher}. It is likely that the application of this 
generalized inference method will help elucidate open questions like the ones
discussed in this section.

\section*{Acknowledgments}

The authors wish to thank P Wiegmann, K Efetov and E Saff for useful discussions
regarding parts of this project, and A Zabrodin for comments on the final manuscript.
Research of M.M. and R. T.  was carried out under the auspices of the
National Nuclear Security Administration of the U.S. Department of
Energy at Los Alamos National Laboratory under Contract No. DE
C52-06NA25396. M.M and M.P. were supported by the LANL LDRD project 
20070483ER. M.P. was partially supported by the Natl. Sci. Foundation
grant DMS-0701094. R.T. acknowledges support from the LANL 
LDRD Directed Research grant on {\it Physics of Algorithms}. 

\section*{References}

\def\cprime{$'$} \def\cprime{$'$} \def\cprime{$'$} \def\cprime{$'$}
  \def\cprime{$'$}


\begin{thebibliography}{100}

\bibitem{Wigner1}
E.P. Wigner.
\newblock On the statistical distribution of the widths and spacings of nuclear
  resonance levels.
\newblock {\em Proc. Cambridge Philos. Soc.}, 47:790, 1950.

\bibitem{Wigner2}
E.~P. Wigner.
\newblock Characteristic vectors of bordered matrices with infinite dimensions.
\newblock {\em Ann. of Math. (2)}, 62:548--564, 1955.

\bibitem{Wigner3}
E.~P. Wigner.
\newblock Characteristic vectors of bordered matrices with infinite dimensions.
  {II}.
\newblock {\em Ann. of Math. (2)}, 65:203--207, 1957.

\bibitem{Dyson1}
F.~J. Dyson.
\newblock Statistical theory of the energy levels of complex systems. {I}.
\newblock {\em J. Mathematical Phys.}, 3:140--156, 1962.

\bibitem{Dyson2}
F.~J. Dyson.
\newblock Statistical theory of the energy levels of complex systems. {II}.
\newblock {\em J. Mathematical Phys.}, 3:157--165, 1962.

\bibitem{Dyson3}
F.~J. Dyson.
\newblock Statistical theory of the energy levels of complex systems. {III}.
\newblock {\em J. Mathematical Phys.}, 3:166--175, 1962.

\bibitem{Dyson4}
F.~J. Dyson.
\newblock The threefold way. {A}lgebraic structure of symmetry groups and
  ensembles in quantum mechanics.
\newblock {\em J. Mathematical Phys.}, 3:1199--1215, 1962.

\bibitem{tHoft}
G.~'t~Hooft.
\newblock Planar diagram theory for strong interactions.
\newblock {\em Nucl. Phys. B}, 72(3):461, 1974.

\bibitem{IZPB}
E.~Br{\'e}zin, C.~Itzykson, G.~Parisi, and J.~B. Zuber.
\newblock Planar diagrams.
\newblock {\em Comm. Math. Phys.}, 59(1):35--51, 1978.

\bibitem{Kazakov}
V.~A. Kazakov.
\newblock Ising model on a dynamical planar random lattice: exact solution.
\newblock {\em Phys. Lett. A}, 119(3):140--144, 1986.

\bibitem{Kostov}
I.~K. Kostov.
\newblock Matrix models as conformal field theories.
\newblock In {\em Applications of random matrices in physics}, volume 221 of
  {\em NATO Sci. Ser. II Math. Phys. Chem.}, pages 459--487. Springer,
  Dordrecht, 2006.

\bibitem{Serban}
M.~Bocquet, D.~Serban, and M.~R. Zirnbauer.
\newblock Disordered 2d quasiparticles in class {D}: {D}irac fermions with
  random mass, and dirty superconductors.
\newblock {\em Nuclear Phys. B}, 578(3):628--680, 2000.

\bibitem{Pastur}
L.~A. Pastur.
\newblock The distribution of eigenvalues of the {S}chr\"odinger equation with
  a random potential.
\newblock {\em Funkcional. Anal. i Prilo\v zen.}, 6(2):93--94, 1972.

\bibitem{Wegner}
F.~Wegner.
\newblock Disordered electronic system as a model of interacting matrices.
\newblock {\em Phys. Rep.}, 67(1):15--24, 1980.
\newblock Common trends in particle and condensed matter physics (Proc. Winter
  Adv. Study Inst., Les Houches, 1980).

\bibitem{Efetov}
K.~B. Efetov.
\newblock Supersymmetry and theory of disordered metals.
\newblock {\em Adv. in Phys.}, 32(1):53--127, 1983.

\bibitem{Zirnbauer}
P.~Heinzner, A.~Huckleberry, and M.~R. Zirnbauer.
\newblock Symmetry classes of disordered fermions.
\newblock {\em Comm. Math. Phys.}, 257(3):725--771, 2005.

\bibitem{Fyodorov}
Y.~V. Fyodorov.
\newblock Complexity of random energy landscapes, glass transition, and
  absolute value of the spectral determinant of random matrices.
\newblock {\em Phys. Rev. Lett.}, 92(24):240601, 4, 2004.

\bibitem{Altshuler}
B.~L. Al{\cprime}tshuler, V.~E. Kravtsov, and I.~V. Lerner.
\newblock Statistics of mesoscopic fluctuations and scaling theory.
\newblock In {\em Localization in disordered systems (Bad Schandau, 1986)},
  volume~16 of {\em Teubner-Texte Phys.}, pages 7--17. Teubner, Leipzig, 1988.

\bibitem{Verbaarschot}
J.~Verbaarschot.
\newblock The spectrum of the {D}irac operator near zero virtuality for {$N\sb
  c=2$} and chiral random matrix theory.
\newblock {\em Nuclear Phys. B}, 426(3):559--574, 1994.

\bibitem{Akemann}
G.~Akemann, Y.~V. Fyodorov, and G.~Vernizzi.
\newblock On matrix model partition functions for {QCD} with chemical
  potential.
\newblock {\em Nuclear Phys. B}, 694(1-2):59--98, 2004.

\bibitem{MIMO}
D.~Tse and P.~Viswanath.
\newblock On the capacity of the multiple antenna broadcast channel.
\newblock In {\em Multiantenna channels: capacity, coding and signal processing
  (Piscataway, NJ, 2002)}, volume~62 of {\em DIMACS Ser. Discrete Math.
  Theoret. Comput. Sci.}, pages 87--105. Amer. Math. Soc., Providence, RI,
  2003.

\bibitem{Schutz1}
A.~R{\'a}kos and G.~M. Sch{\"u}tz.
\newblock Current distribution and random matrix ensembles for an integrable
  asymmetric fragmentation process.
\newblock {\em J. Stat. Phys.}, 118(3-4):511--530, 2005.

\bibitem{Schutz2}
A.~R{\'a}kos and G.~M. Sch{\"u}tz.
\newblock Bethe ansatz and current distribution for the {TASEP} with
  particle-dependent hopping rates.
\newblock {\em Markov Process. Related Fields}, 12(2):323--334, 2006.

\bibitem{Cuernavaca}
P.~Deift J.~Baik, A.~Borodin and T.~Suidan.
\newblock A model for the bus system in {C}uernavaca ({M}exico).
\newblock {\em Journal of Physics A: Mathematical and General},
  39(28):8965--8975, 2006.

\bibitem{Szego}
G.~Szeg{\H{o}}.
\newblock {\em Orthogonal polynomials}.
\newblock American Mathematical Society, Providence, R.I., third edition, 1967.
\newblock American Mathematical Society Colloquium Publications, Vol. 23.

\bibitem{Saff-Totik}
E.~B. Saff and V.~Totik.
\newblock {\em Logarithmic potentials with external fields}, volume 316 of {\em
  Grundlehren der Mathematischen Wissenschaften [Fundamental Principles of
  Mathematical Sciences]}.
\newblock Springer-Verlag, Berlin, 1997.

\bibitem{BEH1}
M.~Bertola, B.~Eynard, and J.~Harnad.
\newblock Partition functions for matrix models and isomonodromic tau
  functions.
\newblock {\em J. Phys. A.}, 36:3067, 2003.

\bibitem{BEH2}
M.~Bertola, B.~Eynard, and J.~Harnad.
\newblock Differential systems for biorthogonal polynomials appearing in
  2-matrix models and the associated {R}iemann-{H}ilbert problem.
\newblock {\em Communications in Mathematical Physics}, 243:193, 2003.

\bibitem{Its}
P.~Bleher and A.~Its.
\newblock Double scaling limit in the random matrix model: the
  {R}iemann-{H}ilbert approach.
\newblock {\em Comm. Pure Appl. Math.}, 56(4):433--516, 2003.

\bibitem{Deift}
P.~A. Deift.
\newblock {\em Orthogonal polynomials and random matrices: a
  {R}iemann-{H}ilbert approach}, volume~3 of {\em Courant Lecture Notes in
  Mathematics}.
\newblock New York University Courant Institute of Mathematical Sciences, New
  York, 1999.

\bibitem{WZ03}
P.~Wiegmann and A.~Zabrodin.
\newblock Large scale correlations in normal and general non-hermitian matrix
  ensembles.
\newblock {\em J. Phys. A.}, 36:3411, 2003.

\bibitem{Teodorescu04}
R.~Teodorescu, E.~Bettelheim, O.~Agam, A.~Zabrodin, and P.~Wiegmann.
\newblock Normal random matrix ensemble as a growth problem.
\newblock {\em Nuclear Phys. B}, 704(3):407--444, 2005.

\bibitem{Mehta}
M.~L. Mehta.
\newblock {\em Random matrices}, volume 142 of {\em Pure and Applied
  Mathematics (Amsterdam)}.
\newblock Elsevier/Academic Press, Amsterdam, third edition, 2004.

\bibitem{Dumitriu}
I.~Dumitriu and A.~Edelman.
\newblock Matrix models for beta ensembles.
\newblock {\em J. Math. Phys.}, 43(11):5830--5847, 2002.

\bibitem{Ginibre}
J.~Ginibre.
\newblock Statistical ensembles of complex, quaternion, and real matrices.
\newblock {\em J. Mathematical Phys.}, 6:440--449, 1965.

\bibitem{Girko}
V.~L. Girko.
\newblock The elliptic law.
\newblock {\em Teor. Veroyatnost. i Primenen.}, 30(4):640--651, 1985.

\bibitem{MarPas}
L.~A. Pastur.
\newblock The spectrum of random matrices.
\newblock {\em Teoret. Mat. Fiz.}, 10(1):102--112, 1972.

\bibitem{Rider}
B.~Rider.
\newblock Deviations from the circular law.
\newblock {\em Probab. Theory Related Fields}, 130(3):337--367, 2004.

\bibitem{Sosh}
A.~Soshnikov and Y.~V. Fyodorov.
\newblock On the largest singular values of random matrices with independent
  {C}auchy entries.
\newblock {\em J. Math. Phys.}, 46(3):033302, 15, 2005.

\bibitem{Zakh}
I.~Zakharevich.
\newblock A generalization of {W}igner's law.
\newblock {\em Comm. Math. Phys.}, 268(2):403--414, 2006.

\bibitem{Zaboronsky}
L-L. Chau and O.~Zaboronsky.
\newblock On the structure of normal matrix model.
\newblock {\em Communications in Mathematical Physics}, 196:203, 1998.

\bibitem{Dikey}
L.~A. Dickey.
\newblock {\em Soliton equations and {H}amiltonian systems}, volume~26 of {\em
  Advanced Series in Mathematical Physics}.
\newblock World Scientific Publishing Co. Inc., River Edge, NJ, second edition,
  2003.

\bibitem{Aratyn}
H.~Aratyn.
\newblock Integrable {L}ax hierarchies, their symmetry reductions and
  multi-matrix models [arxiv.org:hep-th/950321], 1995.

\bibitem{BEHdual}
M.~Bertola, B.~Eynard, and J.~Harnad.
\newblock Duality of spectral curves arising in two-matrix models.
\newblock {\em Theor. Math. Phys.}, 134:32, 2003.

\bibitem{Eynard03}
M.~Bertola, B.~Eynard, and J.~Harnad.
\newblock Differential systems for biorthogonal polynomials appearing in
  2-matrix models and the associated {R}iemann-{H}ilbert problem.
\newblock {\em Communications in Mathematical Physics}, 243:193, 2003.

\bibitem{Krichev-red}
I.~M. Krichever.
\newblock The {$\tau$}-function of the universal {W}hitham hierarchy, matrix
  models and topological field theories.
\newblock {\em Comm. Pure Appl. Math.}, 47(4):437--475, 1994.

\bibitem{Wasow}
W.~Wasow.
\newblock {\em Linear turning point theory}, volume~54 of {\em Applied
  Mathematical Sciences}.
\newblock Springer-Verlag, New York, 1985.

\bibitem{Gustafsson90}
B.~Gustafsson.
\newblock On quadrature domains and an inverse problem in potential theory.
\newblock {\em J. Analyse Math.}, 55:172--216, 1990.

\bibitem{Oxf98}
K.A. Gillow and S.D. Howison.
\newblock {\em A bibliography of free and moving boundary problems for
  {H}ele-{S}haw and {S}tokes flow}.
\newblock http://www.maths.ox.ac.uk/howison/Hele-Shaw/, 1998.

\bibitem{MWZ}
M.~Mineev-Weinstein, P.B. Wiegmann, and A.~Zabrodin.
\newblock Integrable structure of interface dynamics.
\newblock {\em Physical Review Letters}, 84:5106, 2000.

\bibitem{ST}
P.~G. Saffman and G.~Taylor.
\newblock The penetration of a fluid into a porous medium or {H}ele-{S}haw cell
  containing a more viscous liquid.
\newblock {\em Proc. Roy. Soc. London. Ser. A}, 245:312--329. (2 plates), 1958.

\bibitem{Ristroph}
L.~Ristroph, M.~Thrasher, M.~B. Mineev-Weinstein, and H.~L. Swinney.
\newblock Fjords in viscous fingering: Selection of width and opening angle.
\newblock {\em Physical Review E (Statistical, Nonlinear, and Soft Matter
  Physics)}, 74(1):015201, 2006.

\bibitem{Pelce}
(ed)~P. Pelce.
\newblock {\em Dynamics of Curved Fronts}.
\newblock Academic (Boston), 1988.

\bibitem{Nakaya}
U.~Nakaya.
\newblock {\em Snow Crystals}.
\newblock Harvard University Press., Cambridge, 1954.

\bibitem{Langer}
J.~S. Langer.
\newblock Eutectic solidification and marginal stability.
\newblock {\em Phys. Rev. Lett.}, 44(15):1023--1026, 1980.

\bibitem{Gollub}
Y.~Sawada, A.~Dougherty, and J.~P. Gollub.
\newblock Dendritic and fractal patterns in electrolytic metal deposits.
\newblock {\em Phys. Rev. Lett.}, 56(12):1260--1263, 1986.

\bibitem{BenJacob}
E.~Ben-Jacob.
\newblock {\em Ann. \'Ecole Norm.}, 30:265--375, 1913.

\bibitem{DLA81}
T.~A. Witten and L.~M. Sander.
\newblock Diffusion-limited aggregation, a kinetic critical phenomenon.
\newblock {\em Phys. Rev. Lett.}, 47(19):1400--1403, 1981.

\bibitem{Zubarev}
N.~M. Zubarev and O.~V. Zubareva.
\newblock Exact solutions for equilibrium configurations of charged conducting
  liquid jets.
\newblock {\em Physical Review E}, 71(1):016307, 2005.

\bibitem{Bear}
J.~Bear.
\newblock {\em Dynamics of fluids in porous media}.
\newblock Elsevier (New York), 1972.

\bibitem{Hele-Shaw}
H.~S.~S. Hele-Shaw.
\newblock {\em Nature}, 58(1489):34--36, 1898.

\bibitem{Lamb}
H.~Lamb.
\newblock {\em Hydrodynamics}.
\newblock Cambridge Mathematical Library. Cambridge University Press,
  Cambridge, sixth edition, 1993.

\bibitem{LL}
L.~D. Landau and E.~M. Lifshits.
\newblock {\em Teoreticheskaya fizika. {T}om {VI}}.
\newblock ``Nauka'', Moscow, third edition, 1986.
\newblock Gidrodinamika. [Fluid dynamics].

\bibitem{PK}
P.~Ya. Polubarinova-Kochina.
\newblock {\em Dokl. Acad. Nauk SSSR}, 47:254--7, 1945.

\bibitem{Galin}
L.~A. Galin.
\newblock {\em Dokl. Acad. Nauk SSSR}, 47(1-2):250--3, 1945.

\bibitem{Kuf}
P.~P. Kufarev.
\newblock {\em Dokl. Acad. Nauk SSSR}, 57:335--48, 1947.

\bibitem{bs84}
B.~Shraiman and D.~Bensimon.
\newblock Singularities in nonlocal interface dynamics.
\newblock {\em Phys. Rev. A}, 30(5):2840--2842, 1984.

\bibitem{Halsey}
T.~C. {Halsey}.
\newblock {Diffusion-limited aggregation: A model for pattern formation}.
\newblock {\em Physics Today}, 53:36--41, 2000.

\bibitem{Procaccia}
F.~Barra, B.~Davidovitch, A.~Levermann, and I.~Procaccia.
\newblock {Laplacian Growth and Diffusion Limited Aggregation: Different
  Universality Classes}.
\newblock {\em Phys. Rev. Lett.}, 87(13):134501, 2001.

\bibitem{HLS-1}
O~{Praud} and H.~L. {Swinney}.
\newblock {Fractal dimension and unscreened angles measured for radial viscous
  fingering}.
\newblock {\em Phys. Rev.z E}, 72(1):011406, July 2005.

\bibitem{DH}
B.~Derrida and V.~Hakim.
\newblock Needle models of {L}aplacian growth.
\newblock {\em Phys. Rev. A}, 45(12):8759--8765, 1992.

\bibitem{Peterson}
M.~A. Peterson and J.~Ferry.
\newblock Spontaneous symmetry breaking in needle crystal growth.
\newblock {\em Phys. Rev. A}, 39(5):2740--2741, 1989.

\bibitem{CahnHill}
J.~W. Cahn and J.~E. Hilliard.
\newblock Free energy of a nonuniform system. i. interfacial free energy.
\newblock {\em The Journal of Chemical Physics}, 28(2):258--267, 1958.

\bibitem{Gollub-Langer}
J.~P. Gollub and J.~S. Langer.
\newblock Pattern formation in nonequilibrium physics.
\newblock {\em Rev. Mod. Phys.}, 71(2):S396--S403, 1999.

\bibitem{Khachaturyan}
Y.~U. {Wang}, Y.~M. {Jin}, A.~M. {Cuiti{\~n}o}, and A.~G. {Khachaturyan}.
\newblock {Phase field microelasticity theory and modeling of multiple
  dislocation dynamics}.
\newblock {\em Applied Physics Letters}, 78, 2001.

\bibitem{Matsushita}
M.~{Matsushita} and H.~{Fujikawa}.
\newblock {Diffusion-limited growth in bacterial colony formation}.
\newblock {\em Physica A Statistical Mechanics and its Applications},
  168:498--506, 1990.

\bibitem{SpinDec}
A.~Onuki.
\newblock {\em {Phase Transition Dynamics}}.
\newblock Cambridge University Press, 2002.

\bibitem{caginalp}
G.~{Caginalp} and X.~{Chen}.
\newblock {Phase field equations in the singular limit of sharp interface
  problems}.
\newblock {\em Institute for Mathematics and Its Applications}, 43, 1992.

\bibitem{Langer-1}
J.~S. Langer.
\newblock Models of pattern formation in first-order phase transitions.
\newblock In {\em Directions in condensed matter physics}, volume~1 of {\em
  World Sci. Ser. Dir. Condensed Matter Phys.}, pages 165--186. World Sci.
  Publishing, Singapore, 1986.

\bibitem{TS}
G.~Taylor and P.~G. Saffman.
\newblock A note on the motion of bubbles in a {H}ele-{S}haw cell and porous
  medium.
\newblock {\em Q J Mechanics Appl Math}, 12(3):265--279, 1959.

\bibitem{Libch}
G.~L. Vasconcelos.
\newblock Analytic solution for two bubbles in a {H}ele-{S}haw cell.
\newblock {\em Phys. Rev. E}, 62(3):R3047--R3050, 2000.

\bibitem{Glicksman}
J.~S. Langer and H.~M\"uller-Krumbhaar.
\newblock Mode selection in a dendritelike nonlinear system.
\newblock {\em Phys. Rev. A}, 27(1):499--514, 1983.

\bibitem{Hopper}
R.~W. Hopper.
\newblock Capillarity-driven plane stokes flow exterior to a parabola.
\newblock {\em Q J Mechanics Appl Math}, 46(2):193--210, 1993.

\bibitem{Rich}
S.~Richardson.
\newblock Two-dimensional {S}tokes flows with time-dependent free boundaries
  driven by surface tension.
\newblock {\em European J. Appl. Math.}, 8(4):311--329, 1997.

\bibitem{Crowdy}
D.~{Crowdy} and S.~{Tanveer}.
\newblock {A Theory of Exact Solutions for Annular Viscous Blobs}.
\newblock {\em Journal of Nonlinear Science}, 8:375--400, 1998.

\bibitem{Mineev90}
M.~B. Mineev.
\newblock A finite polynomial solution of the two-dimensional interface
  dynamics.
\newblock {\em Physica D}, 43(2-3):288--292, 1990.

\bibitem{Howison91}
S.~D. Howison.
\newblock Complex variable methods in {H}ele-{S}haw moving boundary problems.
\newblock {\em European J. Appl. Math.}, 3(3):209--224, 1992.

\bibitem{KufVinog}
Yu.~P. Vinogradov and P.~P. Kufarev.
\newblock On some particular solutions of the problem of filtration.
\newblock {\em Doklady Akad. Nauk SSSR (N.S.)}, 57:335--338, 1947.

\bibitem{S59}
P.~G. Saffman.
\newblock Exact solutions for the growth of fingers from a flat interface
  between two fluids in a porous medium or {H}ele-{S}haw cell.
\newblock {\em Quart. J. Mech. Appl. Math.}, 12:146--150, 1959.

\bibitem{BP}
D.~Bensimon and P.~Pelc\'e.
\newblock Tip-splitting solutions to a {S}tefan problem.
\newblock {\em Phys. Rev. A}, 33(6):4477--4478, 1986.

\bibitem{H86}
S.~D. Howison.
\newblock Fingering in {H}ele-{S}haw cells.
\newblock {\em J. Fluid Mech.}, 167:439--453, 1986.

\bibitem{ms}
M.~B. Mineev-Weinstein and S.~P. Dawson.
\newblock Class of nonsingular exact solutions for {L}aplacian pattern
  formation.
\newblock {\em Phys. Rev. E}, 50(1):R24--R27, 1994.

\bibitem{sm}
S.~P. Dawson and M.~Mineev-Weinstein.
\newblock Long-time behavior of the {N}-finger solution of the {L}aplacian
  growth equation.
\newblock {\em Physica D}, 73(4):373--387, 1994.

\bibitem{Richardson72}
S.~Richardson.
\newblock {Hele {S}haw flows with a free boundary produced by the injection of
  fluid into a narrow channel}.
\newblock {\em Journal of Fluid Mechanics}, 56:609--618, 1972.

\bibitem{D}
P.~J. Davis.
\newblock {\em The {S}chwarz function and its applications}.
\newblock The Mathematical Association of America, Buffalo, N. Y., 1974.
\newblock The Carus Mathematical Monographs, No. 17.

\bibitem{Etingof}
A.~N. Varchenko and P.~I. Etingof.
\newblock {\em Why the boundary of a round drop becomes a curve of order four},
  volume~3 of {\em University Lecture Series}.
\newblock American Mathematical Society, Providence, RI, 1992.

\bibitem{Tanveer}
S.~Tanveer.
\newblock The effect of surface tension on the shape of a {H}ele-{S}haw cell
  bubble.
\newblock {\em Phys. Fluids}, 29(11):3537--3548, 1986.

\bibitem{S3}
M.~Sakai.
\newblock Regularity of a boundary having a {S}chwarz function.
\newblock {\em Acta Math.}, 166(3-4):263--297, 1991.

\bibitem{S4}
M.~Sakai.
\newblock Regularity of boundaries of quadrature domains in two dimensions.
\newblock {\em SIAM J. Math. Anal.}, 24(2):341--364, 1993.

\bibitem{S5}
M.~Sakai.
\newblock Regularity of free boundaries in two dimensions.
\newblock {\em Ann. Scuola Norm. Sup. Pisa Cl. Sci. (4)}, 20(3):323--339, 1993.

\bibitem{Howison86}
S.~D. Howison.
\newblock Cusp development in {H}ele-{S}haw flow with a free surface.
\newblock {\em SIAM J. Appl. Math.}, 46(1):20--26, 1986.

\bibitem{Howison85}
S.~D. Howison, J.~R. Ockendon, and A.~A. Lacey.
\newblock Singularity development in moving-boundary problems.
\newblock {\em Quart. J. Mech. Appl. Math.}, 38(3):343--360, 1985.

\bibitem{Hohlov-Howison94}
Y.~E. Hohlov and S.~D. Howison.
\newblock On the classification of solutions to the zero-surface-tension model
  for {H}ele-{S}haw free boundary flows.
\newblock {\em Quart. Appl. Math.}, 51(4):777--789, 1993.

\bibitem{King}
J.~R. King, A.~A. Lacey, and J.~L. V{\'a}zquez.
\newblock Persistence of corners in free boundaries in {H}ele-{S}haw flow.
\newblock {\em European J. Appl. Math.}, 6(5):455--490, 1995.
\newblock Complex analysis and free boundary problems (St.\ Petersburg, 1994).

\bibitem{BAZW05}
E.~Bettelheim, O.~Agam, A.~Zabrodin, and P.~Wiegmann.
\newblock Singular limit of {H}ele-{S}haw flow and dispersive regularization of
  shock waves.
\newblock {\em Physical Review Letters}, 95:244504, 2005.

\bibitem{Bell03}
S.~R. Bell.
\newblock Quadrature domains and kernel function zipping.
\newblock {\em Ark. Mat.}, 43(2):271--287, 2005.

\bibitem{Bell04}
S.~R. Bell.
\newblock The {B}ergman kernel and quadrature domains in the plane.
\newblock In {\em Quadrature domains and their applications}, volume 156 of
  {\em Oper. Theory Adv. Appl.}, pages 61--78. Birkh\"auser, Basel, 2005.

\bibitem{G1}
B.~Gustafsson.
\newblock Quadrature identities and the {S}chottky double.
\newblock {\em Acta Appl. Math.}, 1(3):209--240, 1983.

\bibitem{G2}
B.~Gustafsson.
\newblock Singular and special points on quadrature domains from an algebraic
  geometric point of view.
\newblock {\em J. Analyse Math.}, 51:91--117, 1988.

\bibitem{GSh}
B.~Gustafsson and H.~S. Shapiro.
\newblock What is a quadrature domain?
\newblock In {\em Quadrature domains and their applications}, volume 156 of
  {\em Oper. Theory Adv. Appl.}, pages 1--25. Birkh\"auser, Basel, 2005.

\bibitem{Sh}
H.~S. Shapiro.
\newblock {\em The {S}chwarz function and its generalization to higher
  dimensions}.
\newblock University of Arkansas Lecture Notes in the Mathematical Sciences, 9.
  John Wiley \& Sons Inc., New York, 1992.

\bibitem{AS}
D.~Aharonov and H.~S. Shapiro.
\newblock Domains on which analytic functions satisfy quadrature identities.
\newblock {\em J. Analyse Math.}, 30:39--73, 1976.

\bibitem{S1}
M.~Sakai.
\newblock {\em Quadrature domains}, volume 934 of {\em Lecture Notes in
  Mathematics}.
\newblock Springer-Verlag, Berlin, 1982.

\bibitem{qd}
P.~Ebenfelt, B.~Gustafsson, D.~Khavinson, and M.~Putinar, editors.
\newblock {\em Quadrature domains and their applications}, volume 156 of {\em
  Operator Theory: Advances and Applications}.
\newblock Birkh\"auser Verlag, Basel, 2005.

\bibitem{GP07}
B.~Gustafsson and M.~Putinar.
\newblock Analytic continuation of the exponential transform from convex
  cavities.
\newblock {\em J. Math. Anal. Appl.}, 328(2):995--1006, 2007.

\bibitem{Gustafsson-Sakai94}
B.~Gustafsson and M.~Sakai.
\newblock Properties of some balayage operators, with applications to
  quadrature domains and moving boundary problems.
\newblock {\em Nonlinear Anal.}, 22(10):1221--1245, 1994.

\bibitem{G4}
B.~Gustafsson.
\newblock Lectures on balayage.
\newblock In {\em Clifford algebras and potential theory}, volume~7 of {\em
  Univ. Joensuu Dept. Math. Rep. Ser.}, pages 17--63. Univ. Joensuu, Joensuu,
  2004.

\bibitem{Gustafsson-Sakai02a}
B.~Gustafsson and M.~Sakai.
\newblock Sharp estimates of the curvature of some free boundaries in two
  dimensions.
\newblock {\em Ann. Acad. Sci. Fenn. Math.}, 28(1):123--142, 2003.

\bibitem{Gustafsson-Sakai02b}
B.~Gustafsson and M.~Sakai.
\newblock On the curvature of the free boundary for the obstacle problem in two
  dimensions.
\newblock {\em Monatsh. Math.}, 142(1-2):1--5, 2004.

\bibitem{GP98}
B.~Gustafsson and M.~Putinar.
\newblock An exponential transform and regularity of free boundaries in two
  dimensions.
\newblock {\em Ann. Scuola Norm. Sup. Pisa Cl. Sci. (4)}, 26(3):507--543, 1998.

\bibitem{Crowdy-Marshall03}
D.~Crowdy and J.~Marshall.
\newblock Constructing multiply connected quadrature domains.
\newblock {\em SIAM J. Appl. Math.}, 64(4):1334--1359 (electronic), 2004.

\bibitem{GP00}
B.~Gustafsson and M.~Putinar.
\newblock Linear analysis of quadrature domains. {II}.
\newblock {\em Israel J. Math.}, 119:187--216, 2000.

\bibitem{AK}
N.~I. Aheizer and M.~Krein.
\newblock {\em Some questions in the theory of moments}.
\newblock Translations of Mathematical Monographs, Vol. 2. American
  Mathematical Society, Providence, R.I., 1962.

\bibitem{KN}
M.~G. Kre{\u\i}n and A.~A. Nudel{\cprime}man.
\newblock {\em The {M}arkov moment problem and extremal problems}.
\newblock American Mathematical Society, Providence, R.I., 1977.
\newblock Translations of Mathematical Monographs, Vol. 50.

\bibitem{Ga}
R.~J. Gardner.
\newblock {\em Geometric tomography}, volume~58 of {\em Encyclopedia of
  Mathematics and its Applications}.
\newblock Cambridge University Press, Cambridge, 1995.

\bibitem{GHMP}
B.~Gustafsson, C.~He, P.~Milanfar, and M.~Putinar.
\newblock Reconstructing planar domains from their moments.
\newblock {\em Inverse Problems}, 16(4):1053--1070, 2000.

\bibitem{MP}
M.~Martin and M.~Putinar.
\newblock {\em Lectures on hyponormal operators}, volume~39 of {\em Operator
  Theory: Advances and Applications}.
\newblock Birkh\"auser Verlag, Basel, 1989.

\bibitem{Krein1953}
M.G. Krein.
\newblock On a trace formula in perturbation theory.
\newblock {\em Mat. Sbornik}, 33:597--626, 1953.

\bibitem{Simon}
B.~Simon.
\newblock Spectral analysis of rank one perturbations and applications.
\newblock In {\em Mathematical quantum theory. II. Schr\"odinger operatorss},
  volume~8 of {\em CRM Proc. Lecture Notes}, pages 109--149. Amer. Math. Soc,
  Providence, 1995.

\bibitem{Put02}
M.~Putinar.
\newblock On a diagonal {P}ad\'e approximation in two complex variables.
\newblock {\em Numer. Math.}, 93(1):131--152, 2002.

\bibitem{Xia}
D.~Xia.
\newblock {\em Spectral theory of hyponormal operators}, volume~10 of {\em
  Operator Theory: Advances and Applications}.
\newblock Birkh\"auser Verlag, Basel, 1983.

\bibitem{HH}
J.~W. Helton and R.~E. Howe.
\newblock Traces of commutators of integral operators.
\newblock {\em Acta Math.}, 135(3-4):271--305, 1975.

\bibitem{Pincus}
J.~D. Pincus.
\newblock Commutators and systems of singular integral equations. {I}.
\newblock {\em Acta Math.}, 121:219--249, 1968.

\bibitem{CareyPincus}
R.~W. Carey and J.~D. Pincus.
\newblock An exponential formula for determining functions.
\newblock {\em Indiana Univ. Math. J.}, 23:1031--1042, 1973/74.

\bibitem{Put96}
M.~Putinar.
\newblock Extremal solutions of the two-dimensional {$L$}-problem of moments.
\newblock {\em J. Funct. Anal.}, 136(2):331--364, 1996.

\bibitem{Put98}
M.~Putinar.
\newblock Extremal solutions of the two-dimensional {$L$}-problem of moments.
  {II}.
\newblock {\em J. Approx. Theory}, 92(1):38--58, 1998.

\bibitem{Sakai98}
M.~Sakai.
\newblock Sharp estimates of the distance from a fixed point to the frontier of
  a {H}ele-{S}haw flow.
\newblock {\em Potential Anal.}, 8(3):277--302, 1998.

\bibitem{Sakai99a}
M.~Sakai.
\newblock Linear combinations of harmonic measures and quadrature domains of
  signed measures with small supports.
\newblock {\em Proc. Edinburgh Math. Soc. (2)}, 42(3):433--444, 1999.

\bibitem{Gustafsson96a}
B.~Gustafsson.
\newblock A distortion theorem for quadrature domains for harmonic functions.
\newblock {\em J. Math. Anal. Appl.}, 202(1):169--182, 1996.

\bibitem{Zabrodin}
A.~V. Zabrodin.
\newblock The {W}hitham hierarchy in growth problems.
\newblock {\em Teoret. Mat. Fiz.}, 142(2):197--217, 2005.

\bibitem{Gustafsson-Vasiliev06}
B.~Gustafsson and A.~Vasil{\cprime}ev.
\newblock {\em Conformal and potential analysis in {H}ele-{S}haw cells}.
\newblock Advances in Mathematical Fluid Mechanics. Birkh\"auser Verlag, Basel,
  2006.

\bibitem{Kouznetsova-Tkachev2004}
O.~S. Kuznetsova and V.~G. Tkachev.
\newblock Ullemar's formula for the {J}acobian of the complex moment mapping.
\newblock {\em Complex Var. Theory Appl.}, 49(1):55--72, 2004.

\bibitem{Tkachev2005}
V.~G. Tkachev.
\newblock Ullemar's formula for the moment map. {II}.
\newblock {\em Linear Algebra Appl.}, 404:380--388, 2005.

\bibitem{Ullemar80}
C.~Ullemar.
\newblock A uniqueness theorem for domains satisfying a quadrature identity for
  analytic functions.
\newblock {\em Royal Institute of Technology}, TRITA-MAT-1980-37, 1980.

\bibitem{Sakai78}
M.~Sakai.
\newblock A moment problem on {J}ordan domains.
\newblock {\em Proc. Amer. Math. Soc.}, 70(1):35--38, 1978.

\bibitem{Celmins57}
A.~Celmin{s}.
\newblock Direkte {V}erfahren zur {A}uswertung von {S}chweremessungen bei
  zweidimensionaler {M}assenverteilung.
\newblock {\em Geofis. Pura Appl.}, 38:81--122, 1957.

\bibitem{Novikoff38}
P.S. Novikoff.
\newblock On uniqueness for the inverse problem of potential theory.
\newblock {\em C.R. (Dokl.) Acad. Sci. URSS (N.S.)}, 18:165--168, 1938.

\bibitem{Zalcman87}
L.~Zalcman.
\newblock Some inverse problems of potential theory.
\newblock In {\em Integral geometry (Brunswick, Maine, 1984)}, volume~63 of
  {\em Contemp. Math.}, pages 337--350. Amer. Math. Soc., Providence, RI, 1987.

\bibitem{Sjodin04}
T.~Sj{\"o}din.
\newblock Quadrature identities and deformation of quadrature domains.
\newblock In {\em Quadrature domains and their applications}, volume 156 of
  {\em Oper. Theory Adv. Appl.}, pages 239--255. Birkh\"auser, Basel, 2005.

\bibitem{Stahl}
H.~Stahl.
\newblock Beitr\"age zum {P}roblem der {K}onvergenz von
  {P}ad\'eapproximierenden.
\newblock {\em Ph. D. Thesis, Technical University Berlin}, 1976.

\bibitem{Saf}
E.B. Saff.
\newblock Incomplete and orthogonal polynomials.
\newblock {\em Approximation Theory}, IV:219, 1981.

\bibitem{RevModPhys}
D.~Belitz and T.~R. Kirkpatrick.
\newblock The {A}nderson-{M}ott transition.
\newblock {\em Rev. Mod. Phys.}, 66(2):261--380, 1994.

\bibitem{Feinberg}
J.~Feinberg and A.~Zee.
\newblock Non-{H}ermitian random matrix theory: method of {H}ermitian
  reduction.
\newblock {\em Nuclear Phys. B}, 504(3):579--608, 1997.

\bibitem{Spencer}
T.~Spencer and M.~R. Zirnbauer.
\newblock Spontaneous symmetry breaking of a hyperbolic sigma model in three
  dimensions.
\newblock {\em Comm. Math. Phys.}, 252(1-3):167--187, 2004.

\bibitem{Parisi}
G.~Parisi.
\newblock Toward a mean field theory for spin glasses.
\newblock {\em Phys. Lett. A}, 73(3):203--205, 1979.

\bibitem{Voiculescu}
D.~V. Voiculescu, K.~J. Dykema, and A.~Nica.
\newblock {\em Free random variables}, volume~1 of {\em CRM Monograph Series}.
\newblock American Mathematical Society, Providence, RI, 1992.

\bibitem{Speicher}
P.~Neu and R.~Speicher.
\newblock Rigorous mean-field model for coherent-potential approximation:
  {A}nderson model with free random variables.
\newblock {\em J. Statist. Phys.}, 80(5-6):1279--1308, 1995.

\end{thebibliography}
\end{document}